\documentclass[%
 reprint,
superscriptaddress,
amsmath,amssymb,
 prb,
]{revtex4-2}

\usepackage{graphicx}
\usepackage{dcolumn}
\usepackage{bm}
\usepackage{float}
\usepackage{chemformula}
\usepackage{appendix, comment}
\usepackage{booktabs}
\usepackage[colorlinks=true,linkcolor=blue,anchorcolor=red,citecolor=blue,urlcolor=blue]{hyperref}
\usepackage[mathlines]{lineno}
\newcommand{\RNum}[1]{\uppercase\expandafter{\romannumeral #1\relax}}

\begin{document}
\title{Floquet engineering of topological phase transitions in quantum spin Hall $\alpha$-$T_{3}$ system}

\author{Kok Wai Lee}
\affiliation{Science, Mathematics and Technology, Singapore University of Technology and Design, Singapore 487372, Singapore}

\author{Mateo Jalen Andrew Calderon}
\affiliation{Engineering Product Development, Singapore University of Technology and Design, Singapore 487372, Singapore}

\author{Xiang-Long Yu}
\affiliation{School of Science, Shenzhen Campus of Sun Yat-sen University, Shenzhen 518107, China.}

\author{Ching Hua Lee}
\email{phylch@nus.edu.sg}
\affiliation{Department of Physics, National University of Singapore, Singapore 117542}

\author{Yee Sin Ang}
\email{yeesin\_ang@sutd.edu.sg}
\affiliation{Science, Mathematics and Technology, Singapore University of Technology and Design, Singapore 487372, Singapore}

\author{Pei-Hao Fu}
\email{peihao\_fu@sutd.edu.sg}
\affiliation{Science, Mathematics and Technology, Singapore University of Technology and Design, Singapore 487372, Singapore}

\begin{abstract}
Floquet engineering of topological phase transitions driven by a high-frequency time-periodic field is a promising approach to realizing new topological phases of matter distinct from static states. 
Here, we theoretically investigate Floquet engineering topological phase transitions in the quantum spin Hall $\alpha$-$T_{3}$ system driven by an off-resonant circularly polarized light. 
In addition to the quantum spin (anomalous) Hall insulator phase with multiple helical (chiral) edge states, spin-polarized topological metallic phases are observed, where the bulk topological band gap of one spin sub-band overlaps with the other gapless spin sub-band. 
Moreover, with a staggered potential, the topological invariants of the system depend on whether the middle band is occupied because of the breaking of \textcolor{black}{symmetry with respect to the center of energy-momentum plane}.
Our work highlights the significance of Floquet engineering in realizing new topological phases in $\alpha$-$T_{3}$ lattices.
\end{abstract}
\maketitle

\section{Introduction}\label{introduction}
The study of topological phases of matter is a central focus in condensed matter physics owing to their fundamentally rich physical properties and potential future device applications \cite{RevModPhys.82.3045, RevModPhys.83.1057, ren2016topological, Chan_2024}.  
The quantum spin Hall insulator (QSHI) represents the paradigmatic topological phase, which was first proposed in the Kane-Mele model for the honeycomb lattice \cite{PhysRevLett.95.226801} and experimentally observed in \ch{HgTe/(Hg, Cd)Te}  \cite{doi:10.1126/science.1148047, konig2008quantum, brune2012spin, doi:10.1126/science.1174736}, \ch{InAs/GaSb} quantum wells \cite{PhysRevLett.107.136603}, 
charge-neutral monolayer graphene \cite{young2014tunable} and monolayer \ch{WTe_{2}} \cite{fei2017edge, doi:10.1126/science.aan6003}.
The band topology of QSHI is characterized by the spin Chern number, $\mathcal{C}_{s}$ which corresponds to the number of helical edge states in the sample boundary \cite{PhysRevApplied.21.054057, PhysRevLett.101.246807, PhysRevB.85.241402, Lü_2024, PhysRevB.110.195409}. 
Recently, the physics of the Kane-Mele model was generalized from the honeycomb lattice to the $\alpha$-$T_{3}$ lattice \cite{PhysRevB.103.075419, PhysRevLett.81.5888, PhysRevLett.112.026402, PhysRevB.92.245410, Biswas, Tamang_Lakpa, PhysRevB.110.165426} 
where the latter, with a tuning parameter $\alpha$ ($0 \le \alpha \le 1$),  interpolates between the honeycomb ($\alpha = 0$) and dice ($\alpha = 1$) lattices.
As a result, 
%
%
the $\alpha$-$T_{3}$ lattice is found to undergo a phase transition from a QSHI with $\mathcal{C}_{s}=1$ ($\alpha=0$) to $\mathcal{C}_{s}=2$ ($\alpha=1$) because of the additional quasi-flat band.
Being characterized by a higher spin Chern number, the quantum spin Hall $\alpha$-$T_{3}$ system which supports multiple helical edge states and a quasi-flat band has potential applications in the study of strongly correlated systems \cite{Lin_Fu, pizzi2023light}, quantum geometry \cite{PhysRevLett.131.240001, PhysRevB.106.165133}, spintronics \cite{RevModPhys.76.323, di2023flat}, 
entanglement entropy \cite{xue2024topologicallyprotectednegativeentanglement} and fractional Chern insulators \cite{PhysRevB.93.155155, PhysRevB.96.165150, PhysRevResearch.3.L032070, PhysRevB.86.165129, PhysRevLett.109.186805}.

New topological phases can be generated from more experimentally established phases by driving a topological transition.
One representative example is the realization of a quantum anomalous Hall insulator (QAHI) from QSHI by breaking time-reversal symmetry. 
As a result, the QAHI has been experimentally observed in the chromium-doped \ch{(Bi, Sb)_{2}Te_{3}} magnetic topological insulator \cite{Qi-Kun} after its prediction from the Haldane model \cite{Haldane}. 
Similarly, QAHI is characterized by the charge Chern number, $\mathcal{C}$ corresponding to the number of chiral edge states \cite{PhysRevLett.71.3697, PhysRevLett.49.405}.
Apart from introducing a magnetic order, the QSHI-QAHI phase transition can be achieved by applying an off-resonant polarized light which is another 
%
%
method for breaking time-reversal symmetry \cite{oka2019floquet}. 
Although the illuminated system depends physically on the time-dependent optical driving, it possesses an effective time-independent description through the Floquet approach \cite{PhysRevX.4.031027, PhysRevA.68.013820, PhysRevA.7.2203, PhysRev.138.B979}.
%
%
Hence, Floquet-engineered topological phase transitions have been studied extensively in realizing higher Chern insulators from higher-order topological insulators \cite{PhysRevB.109.085148}, 
Floquet topological Anderson insulator \cite{PhysRevLett.114.056801}, four-dimensional Floquet topological insulator
\cite{PhysRevB.109.125303, liu2024tuning}, Weyl semimetals \cite{PhysRevB.94.121106, PhysRevB.106.195115, Wang_2014, FU20173499,Fu_2022, PhysRevLett.123.206601} with potential applications in realizing odd-frequency superconducting pairs \cite{PhysRevB.103.104505}, 
spin-polarized tunable photocurrents \cite{berdakin2021spin}, Floquet engineering Higgs dynamics \cite{PhysRevB.109.134517}, anomalous Josephson effect \cite{PhysRevB.105.064503}, Josephson diodes \cite{PhysRevApplied.21.054057},
non-Abelian fractional quantum Hall states \cite{PhysRevLett.121.237401},
axion insulator \cite{PhysRevB.108.075435}, 
nonlinear Hall effect \cite{qin2024light},
Floquet semimetal with exotic topological linkages \cite{PhysRevLett.121.036401} and half-integer quantized conductance \cite{PhysRevB.97.165142}.

\begin{figure}[tbp]
    \centering
    \includegraphics[width = 0.485\textwidth]{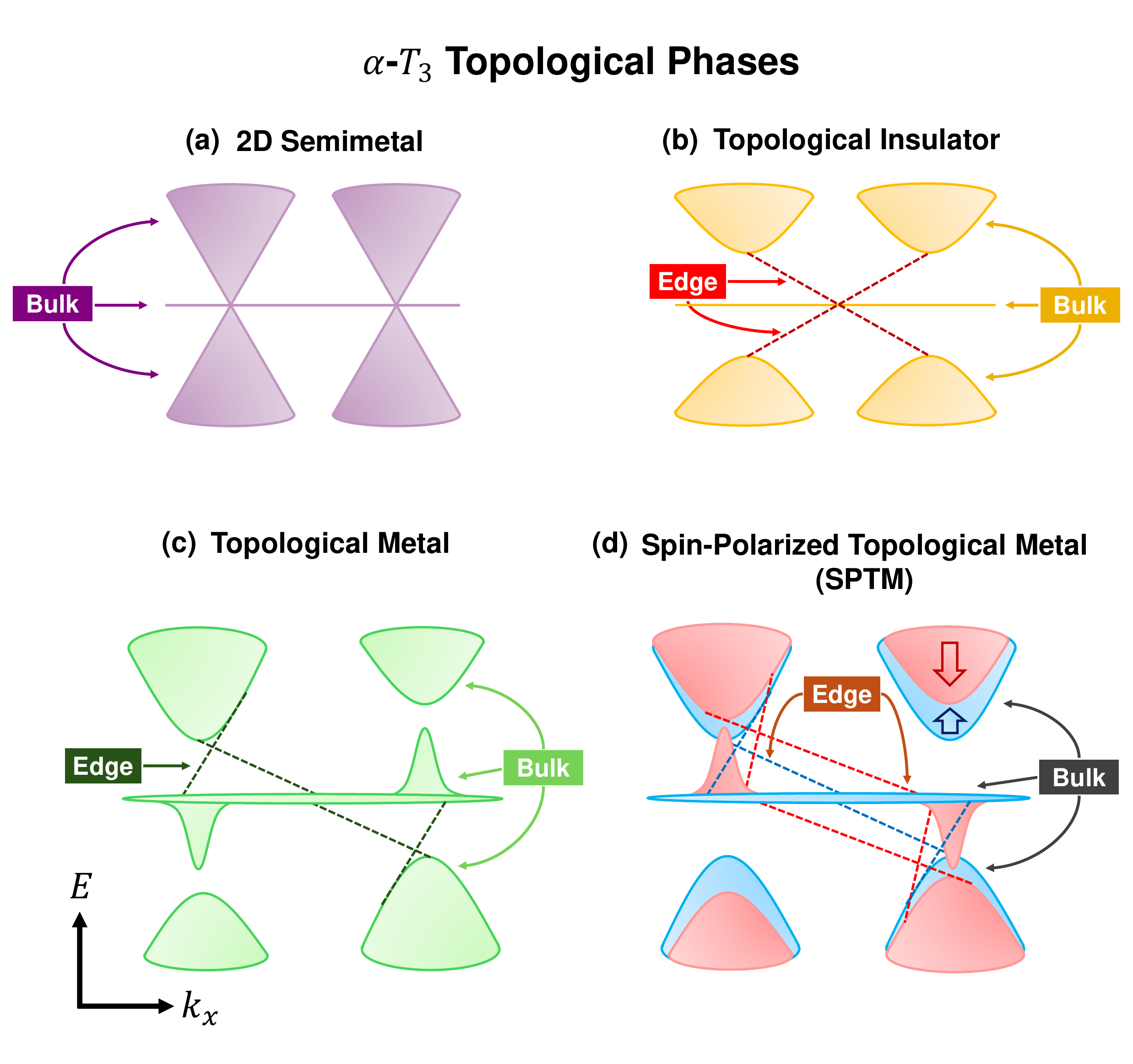}
    \caption{Schematic of the bulk band structure of the $\alpha$-$T_{3}$ topological phases including 
    (a) 2D semimetal \cite{burkov2016topological}, 
    (b) topological insulator \cite{PhysRevB.109.235105, PhysRevB.103.075419} 
    (c) topological metal \cite{PhysRevB.109.235105} 
    and (d) spin-polarized topological metal (SPTM). 
    The tilted dashed lines in (b), (c), and (d) denote the edge states.
    The horizontal solid line in (a) and (b) denotes the flat middle band of the $\alpha$-$T_{3}$ lattice.
    In (c) and (d), the middle band has become dispersive.
    $\downarrow$ and $\uparrow$ in (d) denote spin-down and spin-up respectively.}
    \label{fig:fig1.pdf}
\end{figure}

Studies on quantum spin Hall $\alpha$-$T_{3}$ phase transitions are still ongoing. For instance, a quantum spin quantum anomalous Hall phase is obtained in the presence of magnetization \cite{PhysRevB.103.075419}.
%
%
In the $\alpha$-$T_{3}$ lattice, Floquet driving is shown to induce a phase transition from $\mathcal{C} = 1$ to $\mathcal{C} = 2$ at $\alpha = 1/\sqrt{2}$ \cite{Dey_Bashab}. 
Additionally, Floquet-induced QAHI \cite{PhysRevB.84.235108, PhysRevB.90.115423, mciver2020light} and higher Chern insulator phases \cite{PhysRevResearch.4.033194, PhysRevA.101.043620} can be realized at $\alpha = 0$ (honeycomb) and $\alpha = 1$ (dice) respectively. 
However, the topological properties of the optically driven quantum spin Hall $\alpha$-$T_{3}$ system have yet to be investigated. Crucially, it remains unknown how the topology of the system will be affected by a staggered potential.
%
%
Hence, the interplay between light, Kane-Mele spin-orbit interaction (SOI), $\alpha$, and a staggered potential may offer potentially rich topological properties, broadening the knowledge horizon of condensed matter physics.

\begin{table}[tbp]
\caption{Photoinduced topological phases in the honeycomb, dice and $\alpha$-$T_{3}$ lattices.
%
%
}
\label{Table 1}
\setlength{\tabcolsep}{10pt}
\renewcommand{\arraystretch}{1.2}
\resizebox{0.48\textwidth}{!}{
\begin{tabular}{ccc}
\toprule
\textbf{Lattice} & 
\textbf{Phase(s)}			
& \textbf{Reference(s)} \\
\midrule
Graphene & QAHI & \cite{PhysRevB.84.235108, PhysRevB.90.115423, mciver2020light} \\
\midrule
Dice & Higher Chern Insulator & \cite{PhysRevResearch.4.033194, PhysRevA.101.043620} \\
\midrule
& Higher Chern Insulator &  \\
$\alpha$-$T_{3}$ & Higher Spin Chern Insulator & Current Work \\
& SPTM  &  \\
\bottomrule
\end{tabular}}
\end{table}

In this work, we theoretically investigate Floquet-engineered topological phase transitions in the quantum spin Hall $\alpha$-$T_{3}$ system driven by an off-resonant circularly polarized light (CPL).
In addition to QAHI and QSHI with higher charge Chern and spin Chern numbers respectively, an unprecedented \emph{spin-polarized topological metal} (SPTM) phase is also obtained 
[Fig. \ref{fig:fig1.pdf} (d)], which is distinct from semimetal \cite{burkov2016topological}, topological insulator \cite{Haldane, PhysRevLett.95.226801}, and topological metal \cite{PhysRevB.109.235105} [Figs.\ref{fig:fig1.pdf} (a)-(c)] studied in $\alpha$-$T_{3}$-based systems.
In contrast to the semimetallic (gapless without edge states) and topological insulating (gapped with in-gap edge states) phases,
in the SPTM phase, the bulk topological band gap of one spin sub-band overlaps with the other gapless spin sub-band 
despite any choice of the Fermi level around the band gap.
Hence, although there is a topological gap localized in each valley, the system is gapless, manifesting as a metallic phase.
Compared with the recently proposed  topological metallic phase due to the interplay between the Haldane and modified Haldane models \cite{PhysRevB.109.235105}, the spin polarization of the SPTM phase provides the additional spin degree of freedom. Its unique spin selectivity enriches the family of phases based on the $\alpha$-$T_{3}$ lattice.
Furthermore, the staggered sublattice potential breaks the system's \textcolor{black}{symmetry with respect to the center of energy-momentum plane}, leading to the system topology being dependent on the number of occupied bands.
%
%
Conventionally, the topological invariants characterizing the $\alpha$-$T_{3}$ lattice will always remain \emph{unchanged} regardless of whether we include only the bottom band or both the bottom and middle bands when computing the topological invariants, because the middle band has been topologically trivial. However, once the \textcolor{black}{symmetry with respect to the center of energy-momentum plane} is broken, it is possible to obtain \emph{different} values of the topological invariants either by including or excluding the middle band
in the computation. 
Table \ref{Table 1}
summarizes the photoinduced topological phases in the honeycomb, dice and $\alpha$-$T_{3}$ lattices.
Our work paves the way towards the realization of new possible topological phases in the $\alpha$-$T_{3}$ lattice by using Floquet engineering.

The remainder of this paper is organized as follows. 
In Sec. \ref{Model And Formalism}, the formalism is presented which includes the effective static Floquet Hamiltonian for the periodically driven quantum spin Hall $\alpha$-$T_{3}$ system, topological invariant and the corresponding Hall conductance. 
With this formalism, the phase diagrams, Hall conductance and zigzag nanoribbon band structures are investigated with and without a staggered sublattice potential in Secs. \ref{Spin-Polarized Topological Metal (SPTM)} and 
\ref{Middle-Band Occupancy Dependent Topology}.
The conclusion is given in Sec. \ref{Conclusion}. 

\begin{figure}[tbp]
    \centering
    \includegraphics[width = 0.55\textwidth]{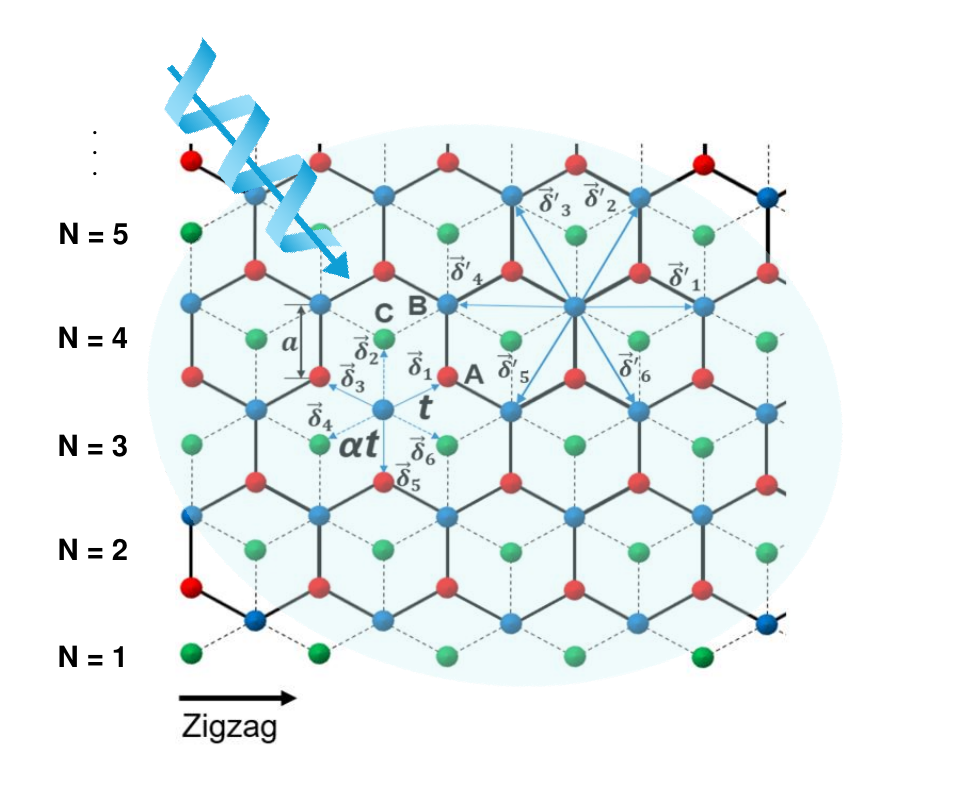}
    \caption{Schematic of the $\alpha$-$T_{3}$ lattice with zigzag edge irradiated by an off-resonant circularly polarized light. 
    $\bm{\delta}_{n}$ and  $\bm{\delta}_{n}^{\prime}$ ($n = 1$, $2$, $3$, $4$, $5$, $6$) denote the nearest-neighbour (NN) and next-nearest-neighbour (NNN) vectors pointing from the B sites respectively \cite{PhysRevB.109.235105}. 
    The A, B, and C sites are colored red, blue and green respectively.
    $\alpha$, which appears in `$\alpha t$' in the figure,
    is the interpolation parameter between the honeycomb ($\alpha = 0$) and dice ($\alpha = 1$) lattices.}
    \label{fig:fig2.pdf}
\end{figure}
\section{Model And Formalism}
\label{Model And Formalism}

\subsection{$\alpha$-$T_{3}$ Kane-Mele Model}
\label{lattice model}
The $\alpha$-$T_{3}$ lattice illustrated schematically in Fig. \ref{fig:fig2.pdf} with the intrinsic SOI is described by a Hamiltonian \cite{PhysRevB.103.075419}:
\begin{eqnarray}
    \mathcal{H} &=& -t\sum _{ \langle ij \rangle \sigma }c_{i\sigma }^{\dag }c_{j\sigma } 
    - \alpha t \sum _{ \langle jk \rangle \sigma }c_{j\sigma }^{\dag }c_{k\sigma } \nonumber \\
    && +\frac{i\lambda }{3\sqrt{3}}\sum _{ \langle  \langle ij \rangle  \rangle \sigma \sigma ^{ \prime }}\nu _{ij}c_{i\sigma }^{\dag }\sigma ^{z}_{\sigma \sigma^{\prime}}c_{j\sigma ^{ \prime }} \nonumber \\
    && +\frac{i \alpha \lambda}{3\sqrt{3}}\sum _{ \langle  \langle jk \rangle  \rangle \sigma \sigma ^{ \prime }}\nu _{kj}c_{j\sigma }^{\dag }\sigma^{z}_{\sigma \sigma^{\prime}}c_{k\sigma^{\prime}} \nonumber \\
    && + H.c.
    \label{real-space Hamiltonian}
\end{eqnarray}
Here, $c_{i\sigma }^{\dag} (c_{i\sigma})$ is the fermionic creation (annihilation) operator with spin polarization $\sigma$ acting at the $i$th site. 
The summation of $\langle ij\rangle $ ($\langle \langle ij\rangle \rangle $) runs over all the nearest (next-nearest)-neighbor sites, $H.c.$ denotes the  Hermitian conjugate whereas $\sigma_{z}$ is the Pauli-Z matrix of spin. 
The nearest-neighbor (NN) and next-nearest-neighbor (NNN) vectors with respect to the B sites are denoted by $\bm{\delta}_{n}$ and $\bm{\delta}_{n}^{\prime}$ ($n = 1$, $2$, $3$, $4$, $5$, $6$) respectively \cite{PhysRevB.109.235105}.
The first (second) term describes the A-B (B-C) NN hopping with strength $t$ or $( \alpha t)$ respectively. 
$\alpha \in[0,1]$ is the parameter which interpolates between the honeycomb ($\alpha = 0$) and dice ($\alpha = 1$) lattices.
The third (fourth) term describes the Kane-Mele A-B-A (B-C-B) NNN hopping with SOI strength $\lambda$ or $(\alpha \lambda)$ respectively.
$\nu_{ij} = +1$ $(-1)$ denotes the anticlockwise (clockwise) NNN hopping with respect to the positive $z$-axis which is perpendicular to the lattice plane. 
In contrast to the honeycomb lattice where an electron crosses from one sublattice to the other to hop to an NNN site (e.g., the path undertaken by an electron hopping from a B site to another B site is B-A-B), there are instead two possible paths for an electron to hop from a B site to another B site in the $\alpha$-$T_{3}$ lattice (i.e., B-A-B and B-C-B with hopping strengths $\lambda$ and $\alpha \lambda$ respectively). 

After a Fourier transformation of Eq. (\ref{real-space Hamiltonian}), the spin-dependent $k$-space Hamiltonian in the basis of $\left( |A_{k}^{\uparrow}\rangle,|B_{k}^{\uparrow}\rangle ,|C_{k}^{\uparrow}\rangle, |A_{k}^{\downarrow}\rangle,|B_{k}^{\downarrow}\rangle ,|C_{k}^{\downarrow}\rangle \right) ^{\top}$ is 

%
\begin{equation}
     \mathcal{\hat{H}}(\bm{k}) =
     \begin{bmatrix}
       \mathcal{\hat{H}}^{+}(\bm{k}) & 0 
     \\ 0 &   \mathcal{\hat{H}}^{-}(\bm{k})
     \end{bmatrix}
     \text{,} 
     \label{k-space Hamiltonian}
\end{equation}
and 
\begin{equation}
    \mathcal{\hat{H}}^{\sigma = \pm 1}(\bm{k}) = 
    \begin{bmatrix} g_{- }^{\sigma}(\bm{k}) & f^\dag\left(\bm{k}\right)
    & 0 \\ f\left(\bm{k}\right) & g_{+ }^{\sigma}(\bm{k}) + \alpha g_{- }^{\sigma}(\bm{k}) & \alpha
    f^\dag\left(\bm{k}\right) \\ 0 & \alpha f\left(\bm{k}\right) & \alpha
    g_{+ }^{\sigma}(\bm{k})  \end{bmatrix}\text{,} 
    \label{spin Hamiltonian}
\end{equation}
where $\bm{k}=\left( k_{x},k_{y}\right) $,
\begin{equation}
    f\left(\bm{k}\right) = -t \sum_{n = 2, 4}^{6}
    e^{i\bm{k}\cdot \bm{\delta}_n} \text{,}
    \label{conventional B-A NN hopping}
\end{equation}
results from the conventional B-A NN hopping  and  
\begin{equation}
    g_{\gamma=\pm 1}^{\sigma}(\bm{k}) =  \frac{i\sigma \lambda}{3\sqrt{3}} 
    \sum_{n = 1}^{6} 
    e^{i[(-1)^{n}\gamma \pi/2  
    + \bm{k}\cdot \bm{\delta}_n^{\prime}]}     
    \text{,}
    \label{Kane-Mele NNN hopping}
\end{equation}
results from the Kane-Mele NNN hopping.

\subsection{Floquet Hamiltonian}
\label{off-resonant driving field} 
We consider the irradiation of the $\alpha$-$T_{3}$ lattice by a beam of CPL incidentally normal to the lattice plane [Fig. \ref{fig:fig2.pdf}] which is described by the vector potential, $\bm{A}(t) = E_{0}(\cos \omega t, \zeta \sin \omega t)/\omega$. 
$E_{0}$ is the electromagnetic field amplitude, $\omega = 2\pi/T$ is the light frequency and $\zeta = +1$ $(-1)$ denotes the right (left)-handed CPL.
In the following, the driving amplitude is normalized as a dimensionless variable, $A_{0} =  a e E_{0}/ (\hbar \omega)$ where $\hbar = e = 1$, $\zeta$ is chosen to be $+1$
and $a$ is the graphene lattice constant set to $1$.
The minimal coupling between $\bm{A}(t)$ and Eq. (\ref{k-space Hamiltonian}) is described by the Peierls substitution, $\bm{k} \rightarrow \bm{k} + e\bm{A}(t)/\hbar$ \cite{peierls1933theorie}, resulting in the time-dependent Hamiltonian, $\mathcal{\hat{H}}^{\sigma}(\bm{k}, t) = \mathcal{\hat{H}}^{\sigma} \left[\bm{k} + e\bm{A}(t)/\hbar \right]$ which satisfies the time periodicity condition: $\mathcal{\hat{H}}^{\sigma}(\bm{k}, t) = \mathcal{\hat{H}}^{\sigma}(\bm{k}, t + T)$. 

According to the Floquet-Bloch theory \cite{PhysRevX.4.031027, PhysRevA.68.013820, PhysRevA.7.2203, PhysRev.138.B979}, such a time-periodic Hamiltonian can be expressed as a discrete Fourier series, $\mathcal{\hat{H}}^{\sigma}(\bm{k}, t) = \sum_{n}\mathcal{\hat{H}}^{\sigma}_{n}(\bm{k})e^{-in \omega t}$ where 
\begin{equation}
    \mathcal{\hat{H}}^{\sigma}_{n}(\bm{k}) = \frac{1}{T} \int_{0}^{T} \mathcal{\hat{H}}^{\sigma}(\bm{k}, t)e^{-in \omega t} dt \text{,}
    \label{Fourier components}
\end{equation}
are the Fourier components of $\mathcal{\hat{H}}^{\sigma}(\bm{k}, t)$. 
In the off-resonant limit $(A_{0}^{2}/\omega \ll 1)$ \cite{Wang_2014}, the optically dressed system is described by the following effective static Floquet Hamiltonian \cite{PhysRevB.91.125105}: 
%
\begin{equation}
    \mathcal{\hat{H}}^{\sigma}_{\text{eff}}(\bm{k}) = \mathcal{\hat{H}}^{\sigma}_{0}(\bm{k}) + 
    \sum_{n > 0}\frac{\left[\mathcal{\hat{H}}^{\sigma}_{+n}(\bm{k}) ,\mathcal{\hat{H}}^{\sigma}_{-n}(\bm{k})\right] }{n \hbar \omega} + \hat{O}\left(\frac{1}{\omega^{2}}\right) 
    \text{.}
    \label{effective static Floquet Hamiltonian}
\end{equation}
\begin{figure}[tbp]
    \centering
    \includegraphics[width = 0.48\textwidth]{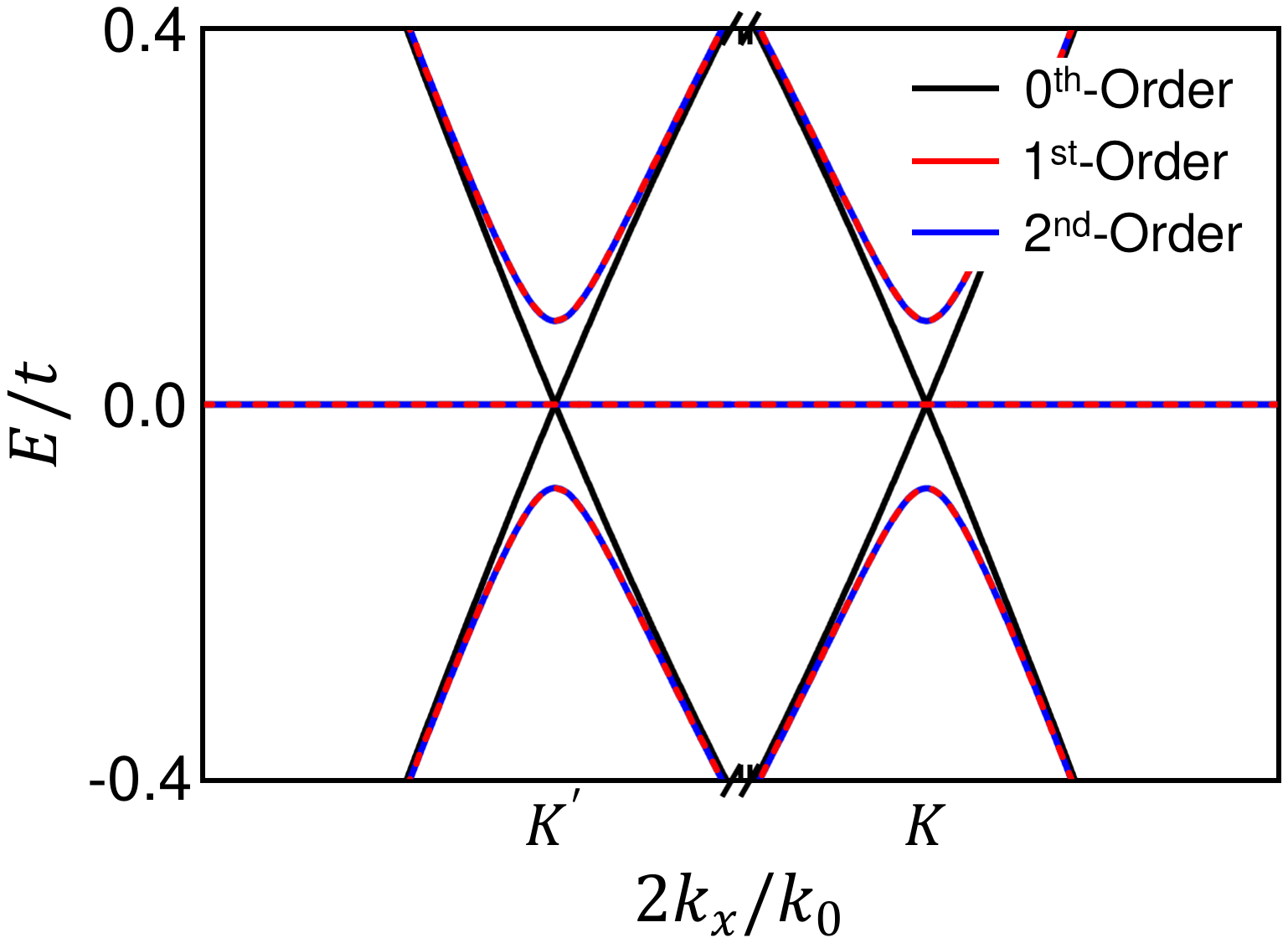}
    \caption{Bulk band structures of the optically dressed quantum spin Hall $\alpha$-$T_{3}$ system 
    with truncation orders $n = 0$, $1$ and $2$. $\lambda = \alpha = 0$. To satisfy the off-resonant condition, $A_{0} = 0.2$. $\omega = t$ such that it is greater than the SOI-induced band gap. $k_{0} = 2\pi/(\sqrt{3}a)$ where $a = 1$ is the graphene lattice constant.}
    \label{fig:fig3.pdf}
\end{figure}
Figure \ref{fig:fig3.pdf} depicts the bulk band structures of the optically dressed quantum spin Hall $\alpha$-$T_{3}$ system obtained by numerically solving the eigenvalue problem of Eq. (\ref{effective static Floquet Hamiltonian}) truncated to various orders ($n = 0$, $1$ and $2$). It can be observed that the bulk band structure converges at $n = 1$.
This shows the first-order $(n = 1)$ truncation of the commutator in Eq. (\ref{effective static Floquet Hamiltonian}) is sufficiently accurate in the off-resonant limit and the higher-order $(n > 1)$ terms can be neglected \cite{PhysRevB.109.085148, Wang_2014}. 
In other words, the light-matter interaction can be effectively described by a single virtual photon absorption-emission process \cite{PhysRevB.84.235108, Fu_2022, FU20173499}. 
\textcolor{black}{Generally, Floquet-driven systems are studied via methods such as the Floquet propagator and Kubo formula 
\cite{PhysRevLett.109.010601, PhysRevB.97.035422}. 
Nevertheless, the quasi-static method chosen here is valid in the high-frequency limit.}
After some calculations, we obtain, to leading order,
\begin{equation}
     \mathcal{\hat{H}}^{\sigma}_{\text{eff}}(\bm{k}) \approx \mathcal{\hat{H}}^{\sigma}_{0}(\bm{k}) + \Delta \mathcal{\hat{H}}_{\text{eff}}(\bm{k})
     \text{,}
    \label{effective Hamiltonian}
\end{equation}
where the first term, $\mathcal{\hat{H}}^{\sigma}_{0}(\bm{k})$ is given as
\begin{equation}
    \mathcal{\hat{H}}^{\sigma}_{0}(\bm{k}) = \begin{bmatrix} g_{0_{-}}^{\sigma}(\bm{k}) & f^\dag_{0}\left(\bm{k}\right)
    & 0 \\ f_{0}\left(\bm{k}\right) & g_{0_{+}}^{\sigma}(\bm{k}) + \alpha g_{0_{-}}^{\sigma}(\bm{k}) & \alpha
    f^\dag_{0}\left(\bm{k}\right) \\ 0 & \alpha f_{0}\left(\bm{k}\right) & \alpha
    g_{0_{+}}^{\sigma}(\bm{k})  \end{bmatrix}
    \text{,} 
    \label{h_{0}}
\end{equation}
with
\begin{equation}
    f_{0}(\bm{k}) = J_{0}(A_{0}a) f(\bm{k})
    \text{,}
    \label{f_{0}}
\end{equation}
and 
\begin{equation}
    g_{0_{\gamma}}^{\sigma}(\bm{k}) = J_{0}(A_{0}\sqrt{3}a) g_{\gamma}^{\sigma}(\bm{k}) 
    \text{,}
    \label{g_{0_{gamma}}^{sigma}}
\end{equation}
whereas the second term, $\Delta \mathcal{\hat{H}}_{\text{eff}}(\bm{k})$ is given as 
\begin{equation}
    \hspace*{-1cm}
    \Delta \mathcal{\hat{H}}_{\text{eff}}(\bm{k}) = \begin{bmatrix} g_{1_{+}}(\bm{k}) & 0
    & 0 \\ 0 & 
    g_{1_{-}}(\bm{k}) + \alpha^{2} g_{1_{+}}(\bm{k}) & 0 \\ 0 & 0 & \alpha^{2}
    g_{1_{-}}(\bm{k}) \end{bmatrix}
    \text{,}
    \label{delta h_{eff}}
\end{equation}
with 
%
\begin{equation}
    g_{1_{\gamma}}(\bm{k}) = \frac{9 t^{2}}{i\sigma \lambda \hbar \omega} J_{1}^{2}(A_{0}a) g_{\gamma}^{\sigma}(\bm{k}) 
    \text{.}
\end{equation}
Here, $J_{n}(x) = \frac{1}{2\pi}
\int^{\pi}_{-\pi} e^{i(nt - x\sin t)} dt $ is the $n^{\text{th}}$ order Bessel function of the first kind.

The effect of the off-resonant CPL on the system is two-fold. 
(i) The hopping strengths are reduced to Eqs. (\ref{f_{0}}) and  (\ref{g_{0_{gamma}}^{sigma}}).
(ii) Most importantly, a Dirac-mass-like term is induced [Eq. (\ref{delta h_{eff}})] 
which can be considered as the light-induced Haldane term in the $\alpha$-$T_3$ lattice \cite{PhysRevB.109.235105, PhysRevB.84.235108}.




\subsection{Low-Energy Effective Model}
\begin{figure*}[tbp]
    \centering
    \includegraphics[width = \textwidth]{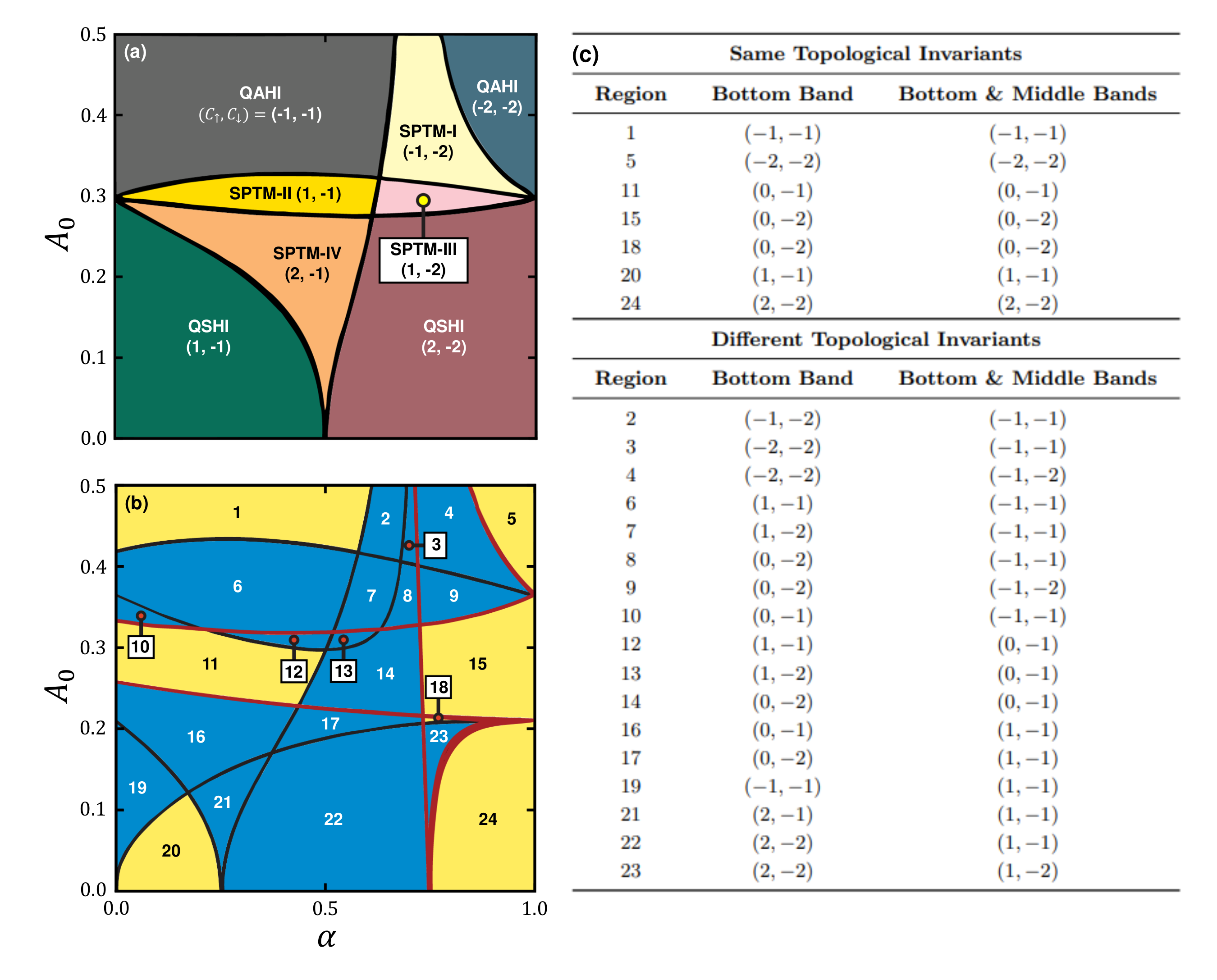}
    \caption{ ($\mathcal{C}_{\uparrow}$,$\mathcal{C}_{\downarrow}$) phase diagram in the $\alpha$-$A_{0}$ plane (a) without and (b) with an A-C staggered sublattice potential, 
    (c) $(\mathcal{C}_{\uparrow}$, $\mathcal{C}_{\downarrow})$ values for (b) obtained by considering only the bottom band or both the bottom and middle bands.
    For (a), $\mathcal{C}_{\uparrow}$ and $\mathcal{C}_{\downarrow}$ are indicated in the figure whereas for (b), they are listed in (c) for clarity purposes.
    In (b), the yellow (blue) color denotes band occupancy number independence (dependence) of $\mathcal{C}_{\uparrow}$ and $\mathcal{C}_{\downarrow}$. For both (a) and (b), $\lambda = 0.2t$ and $\omega = t$. For (b), $M = 0.1t$ whereas the black and red
    phase boundaries are obtained via $\Delta E_{1} = 0$ and $\Delta E_{2} = 0$ respectively .}
    \label{fig:fig4.pdf}
\end{figure*}

A topological phase transition is typically associated with the band gap closing-reopening process \cite{ren2016topological}. 
The gap closing indicates the phase boundary in a phase diagram [Fig. \ref{fig:fig4.pdf}] which can be obtained via the low-energy effective model.

The electronic states in the vicinity of 
the $K$ and $K^{\prime}$ points otherwise known as the $\bm{K}_{\eta = \pm 1} = [\eta 4\pi/ (3\sqrt{3}a), 0]$ points are described by the following low-energy effective Dirac-like Hamiltonian \cite{PhysRevB.109.235105}:
\begin{equation}
     \mathcal{H}^{\eta \sigma}\left( \bm{q}\right) = \begin{bmatrix} i\sigma \lambda \eta & \tilde{q} & 0 \\
    \tilde{q}^{\dag} &
    -i\sigma \lambda \eta + i\alpha \sigma \lambda \eta   
    & \alpha \tilde{q} \\
    0 & \alpha \tilde{q}^{\dag } & -i\alpha \sigma \lambda \eta
    \end{bmatrix}
    \text{,}  
    \label{low-energy effective Dirac-like Hamiltonian}
\end{equation}
where 
$\bm{q}=\left( q_{x},q_{y}\right) = \bm{k} - \bm{K}_{\eta}$ 
, $\tilde{q} =\hbar v_{f}(\eta q_{x} -iq_{y})$, $\hbar v_f = 3at/2$. 
By coupling Eq. (\ref{low-energy effective Dirac-like Hamiltonian}) with the CPL and reapplying the procedure in Sec. \ref{off-resonant driving field}, we obtain a low-energy effective Floquet Hamiltonian which reads
%
%
\begin{equation}
\mathcal{H}_{\text{eff}}^{\eta \sigma}\left( \bm{q}\right) = \mathcal{H}^{\eta \sigma}\left( \bm{q}\right) 
    + \Delta \mathcal{H}_{\text{eff}}^{\eta} 
    \text{,}
    \label{low-energy effective Floquet Hamiltonian}
\end{equation}
with
\begin{equation}
     \Delta \mathcal{H}_{\text{eff}}^{\eta}  = \begin{bmatrix}
      -\beta & 0 & 0 \\ 0 & 
      \beta (1 -\alpha ^{2}) & 0 \\
      0 & 0 & 
      \alpha^{2} \beta
     \end{bmatrix}
     \text{,}
\end{equation}
and $ \beta = \eta A_{0}^{2} \hbar v_{f}^{2}  / \omega $. 
Alternatively, Eq. (\ref{low-energy effective Floquet Hamiltonian}) can also be obtained by performing the low-energy approximation on Eq. (\ref{effective Hamiltonian}).

Equation (\ref{low-energy effective Floquet Hamiltonian}) can be solved analytically via the secular equation, $\det[\mathcal{H}_{\text{eff}}^{\eta \sigma}\left( \bm{q}\right) - E( \bm{q})] = 0$ but the eigenvalues are too complicated to be presented in their exact form here. 
Since we are mainly concerned with the band gap, we shall focus on the $\bm{K}_{\eta}$ points where $\tilde{q} = 0$.  
Then, Eq. (\ref{low-energy effective Floquet Hamiltonian}) becomes a diagonal matrix and the eigenvalues correspond to the diagonal elements as follows:
\begin{subequations}
\begin{equation}
   E_{1} = i\sigma \lambda \eta  - \beta
    \text{,} 
    \label{E_{1}}
\end{equation}
\begin{equation}        
   E_{2} = -i\sigma \lambda \eta + i\alpha \sigma \lambda \eta + \beta (1 -\alpha^{2})
    \text{,} 
    \label{E_{2}}
\end{equation}  
\begin{equation}    
   E_{3} = -i\alpha \sigma \lambda \eta
   + \alpha^{2} \beta
   \text{.} 
   \label{E_{3}}
\end{equation}
\end{subequations}
$E_{1}, E_{2}$ and $E_{3}$ are the band edges where a gap closing corresponds to a band touching between any two of them. 
Using the diagonal elements, we define the following two energy gaps:
%
%
\begin{subequations}
\begin{equation}
    \Delta E_{1} = \left\vert 
    E_{middle} - E_{bottom} \right\vert
     \text{,}
     \label{Delta E_{1}}
\end{equation}
\begin{equation}
    \Delta E_{2} = \left\vert 
    E_{top} - E_{middle} \right\vert
    \text{,}  
    \label{Delta E_{2}}
\end{equation}
\end{subequations}
where $E_{bottom} = \textbf{min}[E_1, E_{2}, E_{3}]$, $E_{top} = \textbf{max}[E_1, E_{2}, E_{3}]$ and $E_{bottom} < E_{middle} < E_{top}$.
Based on the band gap closing-reopening condition, the phase boundary is described precisely by $\Delta E_{1} = \Delta E_{2} = 0$ as shown in Fig. \ref{fig:fig4.pdf}. 

\subsection{Topological Invariant}
\label{topological invariant}
Both the charge $(\mathcal{C})$ and spin $(\mathcal{C}_{s})$ Chern numbers are employed to characterize the topological phases of our optically dressed system, which are defined as follows:
\begin{subequations}
\begin{equation}
    \mathcal{C} = \mathcal{C}_{\uparrow} + \mathcal{C}_{\downarrow}
    \text{,}
    \label{charge_chern}
\end{equation}
\begin{equation}
    \mathcal{C}_{s} = \frac{1}{2} (\mathcal{C}_{\uparrow} - \mathcal{C}_{\downarrow})
    \text{,}
    \label{spin_chern}
\end{equation}
\end{subequations}
where $\mathcal{C}_{\uparrow (\downarrow)}$ is the spin-dependent Chern number defined as \cite{RevModPhys.82.1959},
\begin{equation}
 \mathcal{C}_{\sigma}=\frac{1}{2\pi} \sum_{n\in occ.}\iint_{\text{BZ}}\Omega ^{\sigma}_{n}\left( \bm{k}\right) dk_{x} dk_{y}
    \text{,}
    \label{spin-dependent Chern number}    
\end{equation}
with the summation running over all occupied bands and the integration is performed over the first Brillouin zone (BZ).
The spin-dependent Berry curvature is
\begin{equation}
    \Omega^{\sigma}_{n}(\bm{k}) = i\sum_{m \neq n}\frac{\text{Im} (\langle \psi _{\bm{k}, n}^{\sigma
    }|v_{x}|\psi _{\bm{k}, m}^{\sigma}\rangle \langle \psi _{\bm{k}, m}^{\sigma}|v_{y}|\psi
    _{\bm{k}, n}^{\sigma}\rangle)}{( E_{\bm{k}, n}^{\sigma} - E_{\bm{k}, m}^{\sigma}) ^{2}}
    \text{,}
    \label{berry curvature}
\end{equation}
where the velocity operator $\bm{v} =\partial \mathcal{\hat{H}}^{\sigma}_{\text{eff}} (\bm{k}) /\partial \bm{k}$, $E_{\bm{k}, n}^{\sigma}$ and $|\psi^{\sigma}_{\bm{k}, n}\rangle$ are the $n^{\text{th}}$ 
eigenvalue and eigenvector of $\mathcal{\hat{H}}^{\sigma}_{\text{eff}}(\bm{k})$ [Eq. (\ref{effective Hamiltonian})] respectively.

Notably, the choice of the occupied bands in Eq. (\ref{spin-dependent Chern number}) is crucial in determining the topological invariants of the current system.
For $2$-band systems such as graphene, it is natural to take the bottom band as the sole occupied band.
However, the $\alpha$-$T_{3}$ band structure consists of three bands, namely the bottom, middle, and top bands. 
Hence, there are two choices of the occupied band(s): (i) the bottom band only 
%
%
or (ii) both the bottom and middle bands.
Normally, the topological invariants of the $\alpha$-$T_{3}$ lattice will be unaffected by the two choices. 
However, as we will demonstrate in Sec. \ref{Middle-Band Occupancy Dependent Topology}, 
once the \textcolor{black}{symmetry with respect to the center of energy-momentum plane} is broken by a staggered potential, it is possible to obtain different values of the topological invariants depending on which of the two choices we make. 

\subsection{Hall Conductance}
\label{hall conductance}
The topological invariants $\mathcal{C}$ and $\mathcal{C}_{s}$
can be measured by the charge $(G_{xy})$ and spin $(G_{xy}^{s})$ Hall conductance respectively which are \cite{PhysRevB.100.245430, PhysRevB.94.085410} 
\begin{subequations}
    \begin{equation}
        G_{xy} = G^{\uparrow}_{xy} + G^{\downarrow}_{xy}
    \text{,}
    \label{G_{xy}}
    \end{equation}
    \begin{equation}
        G_{xy}^{s} = G^{\uparrow}_{xy} - G^{\downarrow}_{xy}
    \text{.}
    \label{G_{xy}^{s}}
    \end{equation}
\end{subequations}
Here, $G^{\sigma}_{xy}$ is the spin-dependent Hall conductance \cite{FU20173499} 
\begin{equation}
    G^{\sigma}_{xy} = G_{0} \sum_{n} 
    \iint_{\text{BZ}}  
    \Omega ^{\sigma}_{n}\left( \bm{k}\right)
    f^{\sigma}_{\boldsymbol{k}, n} dk_{x} dk_{y}
    \text{,}
\label{kubo_formula}    
\end{equation}    
where $G_{0} = e^{2}/(2\pi h)$ is the conductance unit, $f^{\sigma}_{\bm{k}, n} = [1 +e^{(E^{\sigma}_{\bm{k}, n} - E_{F})/(k_{B}T)}]^{-1}$ is the Fermi-Dirac distribution of the states with energy $E^{\sigma}_{\bm{k}, n}$ at Fermi energy $E_{F}$ and temperature $T$. 
\textcolor{black}{In addition to the Hall conductance, the spin-dependent transport in a junction \cite{PhysRevLett.123.206601} and the Ruderman-Kittel-Kasuya-Yosida (RKKY) interaction 
\cite{PhysRevB.107.165147, PhysRevB.109.205149} can also be utilized to determine the Floquet-engineered topological phase transitions.}

\begin{figure*}[tbp]
    \centering
    \includegraphics[width = \textwidth]{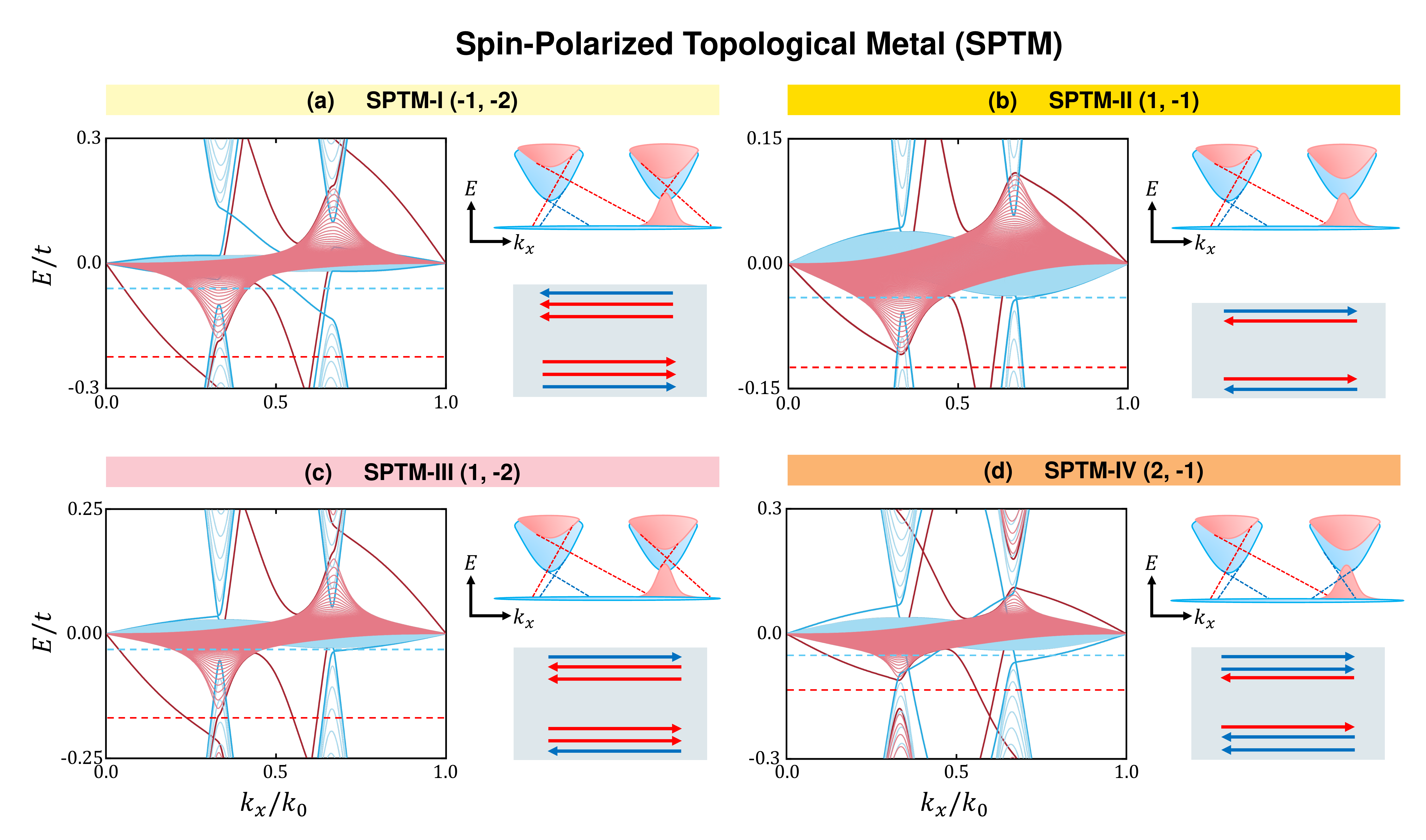}
    \caption{Spin-polarized zigzag nanoribbon band structure of (a) SPTM-\uppercase\expandafter{\romannumeral 1\relax}, (b) SPTM-\uppercase\expandafter{\romannumeral 2\relax}, (c) SPTM-\uppercase\expandafter{\romannumeral 3\relax} and (d) SPTM-\uppercase\expandafter{\romannumeral 4\relax} 
    at $(\alpha$, $A_{0})$ = $(0.74$, $0.38)$, $(0.41$,  $0.30)$, $(0.68$, $0.30)$ and $(0.47$, $0.21)$ respectively.
    For each band structure, a schematic version of it and edge states in a sample strip are depicted. 
    The spin-down and spin-up components of each band structure are colored in red and blue respectively. 
    The respective $(\mathcal{C}_{\uparrow}$, $\mathcal{C}_{\downarrow})$ values are indicated in the subfigure headings. 
    Each zigzag nanoribbon band structure is modelled by using Eq. (\ref {effective Hamiltonian}).
    For each band structure, the horizontal dashed lines denote the spin-specific Fermi levels.}
    \label{fig:fig5.pdf}
\end{figure*}


Hereafter, the NN hopping strength serves as the energy unit ($t = 1$), and the SOI strength, $\lambda$ is set to $0.2t$.
The light frequency, $\omega$ is set to $t$ which is greater than the SOI-induced band gap.
The unit of $k_{x}$ is $k_{0} = 2\pi/ (\sqrt{3} a)$ where $a$ is the graphene lattice constant set to $1$. 
The range of values of $\alpha$ is $0 \le \alpha \le 1$ as the $\alpha$-$T_{3}$ lattice interpolates between the honeycomb $(\alpha=0)$ and dice $(\alpha=1)$ lattices.
On the other hand, the range of values of $A_{0}$ is $0 \le A_{0} \le 0.5$ which satisfies the off-resonant condition.

The concept of bulk-edge correspondence asserts that topological phases possess edge states that are protected by non-trivial bulk topological invariants \cite{PhysRevLett.71.3697, Hatsugai_2, Halperin, traverso2024emerging}. 
Therefore, the appearance of the Chern numbers corresponds to the edge states, indicating the topological phases of the system.
The band structure of the optically dressed quantum spin Hall $\alpha$-$T_{3}$ zigzag nanoribbon \cite{Fujita, Nakada} is obtained by considering periodic boundary conditions along the direction with zigzag edge and open boundary condition along the perpendicular direction [Fig. \ref{fig:fig2.pdf}]. 
The sites along the perpendicular direction are labelled as A\textsubscript{1}, B\textsubscript{1}, C\textsubscript{1}, A\textsubscript{2}, B\textsubscript{2}, C\textsubscript{2},..., A\textsubscript{N}, B\textsubscript{N}, C\textsubscript{N}, etc. 
For the $\alpha$-$T_{3}$ lattice, there are 4 possible zigzag terminations namely C-C, A-C, B-C, and A-B terminations. 
In our work, we demonstrated the A-C termination because similar results are obtained for the other terminations.
Additionally, the results obtained from the zigzag termination are clearer in demonstrating the Floquet engineering phase transitions as compared to the armchair termination.

\section{Spin-Polarized Topological Metal (SPTM)}
\label{Spin-Polarized Topological Metal (SPTM)}
In this section, we discuss the various topological phases of the optically dressed quantum spin Hall $\alpha$-$T_{3}$ system which is depicted in the 
($\mathcal{C}_{\uparrow}$,$\mathcal{C}_{\downarrow}$) phase diagram of Fig. \ref{fig:fig4.pdf} (a). 
The phase boundary of Fig. \ref{fig:fig4.pdf} (a) is described by the condition that $\Delta E_{1} = 0$. 
Note that the phase boundary can also be equivalently described by the condition that $\Delta E_{2} = 0$.

By tuning the values of $\alpha$ and $A_{0}$, a total of 8 topological phases are manifested which can be distinguished into 3 groups. 
The first group is the QAHI phase with higher charge Chern numbers, consisting of 2 regions indexed by $(\mathcal{C}_{\uparrow}$, $\mathcal{C}_{\downarrow}) = (-1, -1)$ and $(-2, -2)$ or equivalently $(\mathcal{C}$, $\mathcal{C}_{s}) = (-2, 0)$ and $(-4, 0)$.
The second group is the QSHI phase consisting of 2 regions as well indexed by $(\mathcal{C}_{\uparrow}$, $\mathcal{C}_{\downarrow}) = (1, -1)$ and $(2, -2)$ or equivalently $(\mathcal{C}$, $\mathcal{C}_{s}) = (0, 1)$ and $(0, 2)$, making the latter a QSHI with a higher spin Chern number.
\textcolor{black}{For $\alpha = 0$, the topological phases depicted in Fig. \ref{fig:fig4.pdf}(a) recover the results demonstrated in Ref. \cite{PhysRevLett.110.026603} for $E_{z} = 0$.}

More significantly, the third group corresponds to four SPTM phases labeled as 
SPTM-\uppercase\expandafter{\romannumeral 1\relax}, \uppercase\expandafter{\romannumeral 2\relax}, \uppercase\expandafter{\romannumeral 3\relax} and \uppercase\expandafter{\romannumeral 4\relax} where they are indexed by $(\mathcal{C}_{\uparrow}$, $\mathcal{C}_{\downarrow}) = (-1, -2)$, $(1, -1)$, $(1, -2)$ and $(2, -1)$ respectively.
Figures \ref{fig:fig5.pdf} (a)-(d) depict the spin-resolved band structures of the 4 SPTM phases in the zigzag nanoribbon. 
\textcolor{black}{The dispersionless spin-degenerate middle bands become dispersive and spin-polarized due to the Kane-Mele SOI and their response to the optical driving in a spin-dependent manner.}
For each of the band structures, a schematic version of it and edge states in a sample strip are included to clearly visualize the physical aspect of $\mathcal{C}_{\uparrow}$ and  $\mathcal{C}_{\downarrow}$, the main features distinguishing the 4 SPTM phases. 
For example, SPTM-\uppercase\expandafter{\romannumeral 1\relax} has $\mathcal{C}_{\uparrow}$ $(\mathcal{C}_{\downarrow})$ = $-1$ $(-2)$ which corresponds to $1$ spin-up ($2$ spin-down) chiral edge state(s) propagating in the same direction.

The spin-polarized zigzag nanoribbon band structure and Hall conductance of SPTM-\uppercase\expandafter{\romannumeral 4\relax} are shown in Fig. \ref{fig:fig6.pdf} (a).
Based on the zigzag nanoribbon band structure, we can observe that the system is gapless where spin-polarized bulk and edge states coexist with each other.  
Alternatively, the system is gapped for each spin component. 
Within the spin-up (spin-down) band gap, the Fermi level crosses 2 (1) chiral edge states which manifest as plateaus in the spin-up (spin-down) Hall conductance at $G_{xy}^{\uparrow}/G_{0} = 2$ $(G_{xy}^{\downarrow}/G_{0} = -1)$. 
%
%
On the other hand, neither the charge nor spin Hall conductance exhibits plateau 
owing to the zero band gap of the system.

\textcolor{black}{Besides the Hall conductance, the various SPTM phases can be distinguished by using spin- and angle-resolved photoemission spectroscopy \cite{jozwiak2016spin} and the spin filter effect via a ferromagnetic tunneling junction \cite{PhysRevB.69.241203}.}
\textcolor{black}{However, a general direct detection of the edge states could be hindered by the bulk states. Generally, the mean free path of the bulk states is much less than that of the edge states, making the latter more robust against disorders. Hence, the contribution of the bulk current can be suppressed by doping a long tunnelling junction such as the Josephson junction \cite{PhysRevLett.124.156601}. Consequently, the current in the junction is mainly contributed by the edge states because the bulk carriers have been strongly scattered by disorders.}

\textcolor{black}{The SPTM phase is an unprecedented topological phase realized in the $\alpha-T_{3}$ lattice. Being spin-polarized, it represents a new potential platform for spintronic \cite{RevModPhys.76.323} applications such as spin transport \cite{vzelezny2018spin}, spin filter \cite{10.1063/1.3537965}, spin transistor \cite{doi:10.1126/science.1195816} and spin valve \cite{https://doi.org/10.1002/adma.202209113}. Besides, it may be useful for probing and controlling the inverse spin galvanic effect \cite{WANG2020126429}. Furthermore, it could also be utilized to investigate the RKKY interaction \cite{PhysRevB.101.235162, PhysRevB.109.205149, Duan_2022}}.

\section{Middle-Band Occupancy Dependent Topology}
\label{Middle-Band Occupancy Dependent Topology}
In this section, we consider modifying the band gap of the quantum spin Hall $\alpha$-$T_{3}$ system. 
Since the band gap is determined by a Dirac mass consisting of $E_{1}, E_{2}$ and $E_{3}$ [Eqs. (\ref{E_{1}})-(\ref{E_{3}})], this means the eigenvalues have to be modified. 
This can be achieved by adding an A-C staggered sublattice potential \cite{ye2020quantum, PhysRevB.103.075419, romhanyi2015hall}:
\begin{equation}
    \mathcal{H}_{\text{st}}  = M \begin{bmatrix}
       1 & 0 & 0 \\ 0 & 
      0 & 0 \\
      0 & 0 & 
       -1
     \end{bmatrix} 
    \text{,}
    \label{A-C staggered
sublattice potential}
\end{equation}
to Eq. (\ref{spin Hamiltonian}) and the resulting band edges are modified as 
\begin{equation}
   E_{1}^{\prime} = E_{1} + M \text{;} \hspace{0.6cm} E_{2}^{\prime} = E_{2} \text{;} \hspace{0.6cm} E_{3}^{\prime} = E_{3} - M 
   \text{.}  
\end{equation}
\textcolor{black}{Experimentally, the on-site energy can be controlled by tuning the laser beam intensities in systems of cold fermionic atoms \cite{Rizzi_Matteo}, thereby generating the A-C staggered sublattice potential. Another possible experimental platform is the \ch{SrTiO_{3}/SrIrO_{3}/SrTiO_{3}} trilayer heterostructure grown along the $(111)$ direction
\cite{PhysRevB.84.241103}. Since the planes of the three sublattices are different, a substrate can be applied to generate different potentials at the three sublattices, namely $u_A$, $u_B$ and $u_C$. $u_B$ can be redefined such that it is the zero-potential reference point.}
The same results can be obtained if Eq. (\ref{A-C staggered
sublattice potential}) is added directly to
Eq. (\ref{effective Hamiltonian}) because the addition does not affect the commutator.
Overall, the topological phases manifested by the system with $\mathcal{H}_{\text{st}}$ are similar to those of the preceding scenario, namely the QAHI, QSHI, and SPTM phases.

\begin{figure}[tbp!]
    \centering
    \includegraphics[width = 0.48\textwidth]{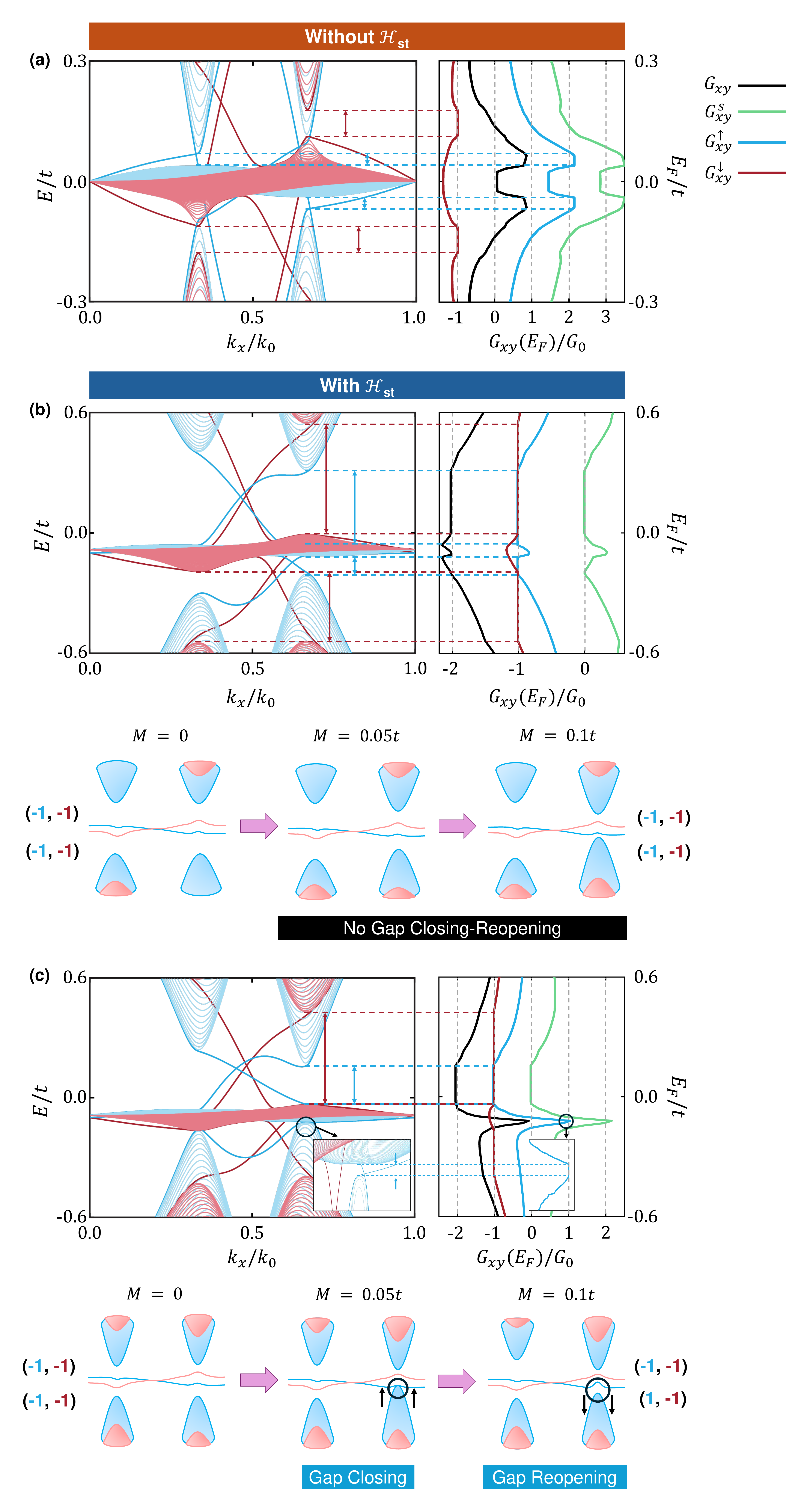}
    \caption{Spin-polarized zigzag nanoribbon band structure and Hall conductance 
    for (a) SPTM-\uppercase\expandafter{\romannumeral 4\relax}, regions (b) $1$ and (c) $6$ at $(\alpha$, $A_{0}) = (0.47$, $0.21)$, $(0.3$, $0.47)$ and $(0.25$, $0.38)$ respectively.
    Schematic of the evolution of the $\alpha$-$T_{3}$ bulk band structure from $M = 0$ to $M = 0.1t$ is included for regions 1 and 6.
    At $M = 0$ and $M = 0.1t$, the 2 sets of $(\mathcal{C}_{\uparrow}$, $\mathcal{C}_{\downarrow})$ indicate the bottom-middle and middle-top in-gap edge states.   
    The spin-down and spin-up components are shown in red and blue respectively. 
    The values of the parameters are $\lambda = 0.2t$, $\omega = t$ and $M = 0.1t$. 
    The total number of unit cells, N along the perpendicular direction of the zigzag chain is $80$.  
    The thermal energy scale, $k_{B}T$ is set at $10^{-6}t$ in calculating the Hall conductance.
    For $ 0.5 \le t \le 3$ eV, this corresponds to $ 5.8 \le T \le   34.8 $ mK.} 
    \label{fig:fig6.pdf}
\end{figure}

The topological properties of the system become dependent on the number of occupied bands as $\mathcal{H}_{\text{st}}$ breaks the \textcolor{black}{symmetry with respect to the center of energy-momentum plane}.
Conventionally, the topological invariants of the system are unchanged regardless of whether the bottom band or both the bottom and middle bands are considered when computing the topological invariants. However, once the \textcolor{black}{symmetry with respect to the center of energy-momentum plane} is broken, it is now possible to obtain different values of the topological invariants either by including or excluding the middle band in the computation. 
%
%
Consequently, a doping-induced phase transition is feasible where the Fermi level can be modulated via doping. 
\textcolor{black}{It is worth noting that the phenomenon is preserved even when (i) the Kane-Mele SOI is absent or
(ii) the alternative staggered sublattice potential, $\pm U$ in Ref. \cite{ye2020quantum} is introduced.}
Notably, besides having important non-topological properties, the nontrivial topology of the middle band, which is essentially a quasi-flat band, makes the $\alpha$-$T_{3}$ lattice a potential platform for investigating the physics of fractional Chern insulators \cite{PhysRevB.86.201101, PhysRevLett.132.126401, PhysRevLett.131.013804}. 
Figure \ref{fig:fig4.pdf} (c) shows $(\mathcal{C}_{\uparrow}$, $\mathcal{C}_{\downarrow})$ values obtained by considering (i) only the bottom band and (ii) both the bottom and middle bands.
For certain cases as indicated by the blue regions in Fig. \ref{fig:fig4.pdf} (b), a different $(\mathcal{C}_{\uparrow}, \mathcal{C}_{\downarrow})$ is obtained depending on whether the middle band is included in the summation in Eq. (\ref{spin-dependent Chern number}).
Unlike the preceding scenario where the phase boundary is equivalently described by $\Delta E_{1} = \Delta E_{2} = 0$, here each vanishing condition describes a different phase boundary and consequently a different phase diagram which are merged in Fig. \ref{fig:fig4.pdf} (b). 


In order to investigate how $\mathcal{H}_{\text{st}}$ affects
the system topology, we choose to discuss and compare the spin-polarized zigzag nanoribbon band structures and Hall conductance of regions $1$ (band occupancy number-independent) and $6$ (band occupancy number-dependent).
Figures \ref{fig:fig6.pdf} (b) and (c) depict the spin-polarized zigzag nanoribbon band structures and Hall conductance of regions $1$ and $6$ respectively. Firstly, it can be noticed that for both regions, the system is gapless (gapped) between the middle and bottom (top and middle) bands.  
The band asymmetry depicted in both Figs. \ref{fig:fig6.pdf} (b) and (c),
in contrast to Fig. \ref{fig:fig6.pdf} (a),
implies the breaking of \textcolor{black}{symmetry with respect to the center of energy-momentum plane} by $\mathcal{H}_{\text{st}}$.  

For region $1$, the system is gapped for each spin component between the middle-bottom and top-middle bands. Within the spin-specific band gaps, the Fermi level crosses one chiral edge state for each spin component which manifests as plateaus in their respective Hall conductance at $G_{xy}^{\uparrow}/G_{0} = G_{xy}^{\downarrow}/G_{0} = -1$.
The charge and spin Hall conductance plateau at $G_{xy}/G_{0} = -2$ and $G_{xy}^{s}/G_{0} = 0$ respectively for $E_{F} > 0$ due to the presence of the top-middle band gap.

For region $6$, the system's spin-down component has both the middle-bottom and top-middle band gaps where the Fermi level crosses one chiral edge state. 
Accordingly, the spin-down Hall conductance plateaus at $G_{xy}^{\downarrow}/G_{0} = -1$ for both $E_{F} < 0$ and $E_{F} > 0$. 	
On the other hand, the spin-up component has a sizeable top-middle band gap and only a very small middle-bottom band gap where the Fermi level also crosses 1 chiral edge state. Consequently, 
the spin-up Hall conductance only shows a small plateau at $G_{xy}^{\uparrow}/G_{0} = 1$ for $E_{F} < 0$.
For $E_{F} > 0$, the spin-up Hall conductance plateaus at $G_{xy}^{\uparrow}/G_{0} = -1$. 
Similarly, the charge and spin Hall conductance plateau at $G_{xy}/G_{0} = -2$ and $G_{xy}^{s}/G_{0} = 0$ respectively for $E_{F} > 0$. 

In Figs. \ref{fig:fig6.pdf} (b) and (c), a schematic of the $\alpha$-$T_{3}$ bulk band structure evolution from $M = 0$ to $M = 0.1t$ is also shown. In region 1, no gap closing-reopening occurs as $M$ varies. In contrast, in region 6, only the spin-up middle-bottom band gap closes and reopens due to band asymmetry
caused by breaking the \textcolor{black}{symmetry with respect to the center of energy-momentum plane}. As a result, the middle-bottom band topology differs from the middle-top band topology. 
This makes the system topology depend on the number of occupied bands.

To estimate the driving frequency and amplitude required to realize the topological phases reported, we assume the $\alpha$-$T_{3}$ NN hopping strength to be in the range $ 0.5 \le t \le 3$ eV.
To satisfy the off-resonant condition, we set the energy of the incident photon, $\hbar \omega = t$ , obtaining the driving frequency range $7.60 \times 10^{14} \le \omega \le 4.56 \times 10^{15} $ Hz. 
To achieve $A_{0} = 0.5$, the driving amplitude range required is $ 1.02 \times 10^{9} \le E_{0} \le 6.10 \times 10^{9} $ V m$^{-1}$.
The estimated driving frequency and amplitude are experimentally possible in the near future \cite{mciver2020light, chen2020observing}.

\section{conclusion}
\label{Conclusion}
In summary, we have theoretically investigated the possible photoinduced topological phases of the quantum spin Hall $\alpha$-$T_{3}$ system driven by an off-resonant circularly polarized light. 
By tuning the values of $\alpha$ $(0 \le \alpha \le 1)$ and $A_{0}$ $(0 \le A_{0} \le 0.5)$, a total of three types of topological phases are manifested by the $\alpha$-$T_{3}$ lattice, namely the QAHI, QSHI, and SPTM phases where the last phase is previously undiscussed for the $\alpha$-$T_{3}$ lattice. 
Within the SPTM phase, the system is gapless where the spin-polarized bulk and edge states coexist with each other.
Next, we introduce an A-C staggered
sublattice potential, $\mathcal{H}_{\text{st}}$
into the quantum spin Hall $\alpha$-$T_{3}$ system which breaks the \textcolor{black}{symmetry with respect to the center of energy-momentum plane}. 
This leads to the system topology being dependent on the number of occupied bands.
The results of our work highlights the importance of Floquet engineering in realizing new topological phases in the $\alpha$-$T_{3}$ lattice.
We expect that more phases can be realized when we include the Rashba spin-orbit coupling (RSOC) \cite{Lin_Fu, PhysRevB.84.195430, PhysRevResearch.6.043108}, modified Haldane flux \cite{PhysRevB.109.235105}, strain engineering \cite{Sun_Junsong}, electron-phonon coupling \cite{PhysRevB.110.045426} and a bilayer architecture \cite{PhysRevB.109.165118}.

\begin{acknowledgments}
This work is supported by the Singapore Ministry of Education (MOE) Academic Research Fund (AcRF) Tier 2 Grant (MOE-T2EP50221-0019). K. W. L. is supported by the SUTD PhD Scholarship. 
M. J. A. Calderon is supported by the SUTD Honours and Research Programme (SHARP). 
C. H. L. acknowledges support 
by Singapore’s NRF Quantum Engineering grant NRF2021-QEP2-02-P09 and Singapore’s MOE Tier-II grant Proposal ID: T2EP50222-0003.
X.-L. Y. is supported by the Natural Science Foundation of Guangdong Province (Grant No. 2023A1515011852).
The computational work for this article was partially performed on resources of the National Supercomputing Centre, Singapore (https://www.nscc.sg).
\end{acknowledgments}


\begin{thebibliography}{111}%
\makeatletter
\providecommand \@ifxundefined [1]{%
 \@ifx{#1\undefined}
}%
\providecommand \@ifnum [1]{%
 \ifnum #1\expandafter \@firstoftwo
 \else \expandafter \@secondoftwo
 \fi
}%
\providecommand \@ifx [1]{%
 \ifx #1\expandafter \@firstoftwo
 \else \expandafter \@secondoftwo
 \fi
}%
\providecommand \natexlab [1]{#1}%
\providecommand \enquote  [1]{``#1''}%
\providecommand \bibnamefont  [1]{#1}%
\providecommand \bibfnamefont [1]{#1}%
\providecommand \citenamefont [1]{#1}%
\providecommand \href@noop [0]{\@secondoftwo}%
\providecommand \href [0]{\begingroup \@sanitize@url \@href}%
\providecommand \@href[1]{\@@startlink{#1}\@@href}%
\providecommand \@@href[1]{\endgroup#1\@@endlink}%
\providecommand \@sanitize@url [0]{\catcode `\\12\catcode `\$12\catcode
  `\&12\catcode `\#12\catcode `\^12\catcode `\_12\catcode `\%12\relax}%
\providecommand \@@startlink[1]{}%
\providecommand \@@endlink[0]{}%
\providecommand \url  [0]{\begingroup\@sanitize@url \@url }%
\providecommand \@url [1]{\endgroup\@href {#1}{\urlprefix }}%
\providecommand \urlprefix  [0]{URL }%
\providecommand \Eprint [0]{\href }%
\providecommand \doibase [0]{https://doi.org/}%
\providecommand \selectlanguage [0]{\@gobble}%
\providecommand \bibinfo  [0]{\@secondoftwo}%
\providecommand \bibfield  [0]{\@secondoftwo}%
\providecommand \translation [1]{[#1]}%
\providecommand \BibitemOpen [0]{}%
\providecommand \bibitemStop [0]{}%
\providecommand \bibitemNoStop [0]{.\EOS\space}%
\providecommand \EOS [0]{\spacefactor3000\relax}%
\providecommand \BibitemShut  [1]{\csname bibitem#1\endcsname}%
\let\auto@bib@innerbib\@empty
\bibitem [{\citenamefont {Hasan}\ and\ \citenamefont
  {Kane}(2010)}]{RevModPhys.82.3045}%
  \BibitemOpen
  \bibfield  {author} {\bibinfo {author} {\bibfnamefont {M.~Z.}\ \bibnamefont
  {Hasan}}\ and\ \bibinfo {author} {\bibfnamefont {C.~L.}\ \bibnamefont
  {Kane}},\ }\bibfield  {title} {\bibinfo {title} {{Colloquium: Topological
  insulators}},\ }\href {https://doi.org/10.1103/RevModPhys.82.3045} {\bibfield
   {journal} {\bibinfo  {journal} {Rev. Mod. Phys.}\ }\textbf {\bibinfo
  {volume} {82}},\ \bibinfo {pages} {3045} (\bibinfo {year}
  {2010})}\BibitemShut {NoStop}%
\bibitem [{\citenamefont {Qi}\ and\ \citenamefont
  {Zhang}(2011)}]{RevModPhys.83.1057}%
  \BibitemOpen
  \bibfield  {author} {\bibinfo {author} {\bibfnamefont {X.-L.}\ \bibnamefont
  {Qi}}\ and\ \bibinfo {author} {\bibfnamefont {S.-C.}\ \bibnamefont {Zhang}},\
  }\bibfield  {title} {\bibinfo {title} {{Topological insulators and
  superconductors}},\ }\href {https://doi.org/10.1103/RevModPhys.83.1057}
  {\bibfield  {journal} {\bibinfo  {journal} {Rev. Mod. Phys.}\ }\textbf
  {\bibinfo {volume} {83}},\ \bibinfo {pages} {1057} (\bibinfo {year}
  {2011})}\BibitemShut {NoStop}%
\bibitem [{\citenamefont {Ren}\ \emph {et~al.}(2016)\citenamefont {Ren},
  \citenamefont {Qiao},\ and\ \citenamefont {Niu}}]{ren2016topological}%
  \BibitemOpen
  \bibfield  {author} {\bibinfo {author} {\bibfnamefont {Y.}~\bibnamefont
  {Ren}}, \bibinfo {author} {\bibfnamefont {Z.}~\bibnamefont {Qiao}},\ and\
  \bibinfo {author} {\bibfnamefont {Q.}~\bibnamefont {Niu}},\ }\bibfield
  {title} {\bibinfo {title} {{Topological phases in two-dimensional materials:
  a review}},\ }\href {https://doi.org/10.1088/0034-4885/79/6/066501}
  {\bibfield  {journal} {\bibinfo  {journal} {Rep. Prog. Phys.}\ }\textbf
  {\bibinfo {volume} {79}},\ \bibinfo {pages} {066501} (\bibinfo {year}
  {2016})}\BibitemShut {NoStop}%
\bibitem [{\citenamefont {Chan}\ \emph {et~al.}(2024)\citenamefont {Chan},
  \citenamefont {Fu}, \citenamefont {Ang},\ and\ \citenamefont
  {Ang}}]{Chan_2024}%
  \BibitemOpen
  \bibfield  {author} {\bibinfo {author} {\bibfnamefont {W.~J.}\ \bibnamefont
  {Chan}}, \bibinfo {author} {\bibfnamefont {P.-H.}\ \bibnamefont {Fu}},
  \bibinfo {author} {\bibfnamefont {L.~K.}\ \bibnamefont {Ang}},\ and\ \bibinfo
  {author} {\bibfnamefont {Y.~S.}\ \bibnamefont {Ang}},\ }\bibfield  {title}
  {\bibinfo {title} {Designing edge states from fractional polarization
  insulators},\ }\href {https://doi.org/10.1088/1361-648X/ad581d} {\bibfield
  {journal} {\bibinfo  {journal} {J. Phys.: Condens. Matter}\ }\textbf
  {\bibinfo {volume} {36}},\ \bibinfo {pages} {385401} (\bibinfo {year}
  {2024})}\BibitemShut {NoStop}%
\bibitem [{\citenamefont {Kane}\ and\ \citenamefont
  {Mele}(2005)}]{PhysRevLett.95.226801}%
  \BibitemOpen
  \bibfield  {author} {\bibinfo {author} {\bibfnamefont {C.~L.}\ \bibnamefont
  {Kane}}\ and\ \bibinfo {author} {\bibfnamefont {E.~J.}\ \bibnamefont
  {Mele}},\ }\bibfield  {title} {\bibinfo {title} {{Quantum Spin Hall Effect in
  Graphene}},\ }\href {https://doi.org/10.1103/PhysRevLett.95.226801}
  {\bibfield  {journal} {\bibinfo  {journal} {Phys. Rev. Lett.}\ }\textbf
  {\bibinfo {volume} {95}},\ \bibinfo {pages} {226801} (\bibinfo {year}
  {2005})}\BibitemShut {NoStop}%
\bibitem [{\citenamefont {König}\ \emph {et~al.}(2007)\citenamefont {König},
  \citenamefont {Wiedmann}, \citenamefont {Brüne}, \citenamefont {Roth},
  \citenamefont {Buhmann}, \citenamefont {Molenkamp}, \citenamefont {Qi},\ and\
  \citenamefont {Zhang}}]{doi:10.1126/science.1148047}%
  \BibitemOpen
  \bibfield  {author} {\bibinfo {author} {\bibfnamefont {M.}~\bibnamefont
  {König}}, \bibinfo {author} {\bibfnamefont {S.}~\bibnamefont {Wiedmann}},
  \bibinfo {author} {\bibfnamefont {C.}~\bibnamefont {Brüne}}, \bibinfo
  {author} {\bibfnamefont {A.}~\bibnamefont {Roth}}, \bibinfo {author}
  {\bibfnamefont {H.}~\bibnamefont {Buhmann}}, \bibinfo {author} {\bibfnamefont
  {L.~W.}\ \bibnamefont {Molenkamp}}, \bibinfo {author} {\bibfnamefont {X.-L.}\
  \bibnamefont {Qi}},\ and\ \bibinfo {author} {\bibfnamefont {S.-C.}\
  \bibnamefont {Zhang}},\ }\bibfield  {title} {\bibinfo {title} {{Quantum Spin
  Hall Insulator State in HgTe Quantum Wells}},\ }\href
  {https://doi.org/10.1126/science.1148047} {\bibfield  {journal} {\bibinfo
  {journal} {Science}\ }\textbf {\bibinfo {volume} {318}},\ \bibinfo {pages}
  {766} (\bibinfo {year} {2007})}\BibitemShut {NoStop}%
\bibitem [{\citenamefont {K{\"o}nig}\ \emph {et~al.}(2008)\citenamefont
  {K{\"o}nig}, \citenamefont {Buhmann}, \citenamefont {W.~Molenkamp},
  \citenamefont {Hughes}, \citenamefont {Liu}, \citenamefont {Qi},\ and\
  \citenamefont {Zhang}}]{konig2008quantum}%
  \BibitemOpen
  \bibfield  {author} {\bibinfo {author} {\bibfnamefont {M.}~\bibnamefont
  {K{\"o}nig}}, \bibinfo {author} {\bibfnamefont {H.}~\bibnamefont {Buhmann}},
  \bibinfo {author} {\bibfnamefont {L.}~\bibnamefont {W.~Molenkamp}}, \bibinfo
  {author} {\bibfnamefont {T.}~\bibnamefont {Hughes}}, \bibinfo {author}
  {\bibfnamefont {C.-X.}\ \bibnamefont {Liu}}, \bibinfo {author} {\bibfnamefont
  {X.-L.}\ \bibnamefont {Qi}},\ and\ \bibinfo {author} {\bibfnamefont {S.-C.}\
  \bibnamefont {Zhang}},\ }\bibfield  {title} {\bibinfo {title} {{The quantum
  spin Hall effect: theory and experiment}},\ }\href
  {https://doi.org/https://doi.org/10.1143/JPSJ.77.031007} {\bibfield
  {journal} {\bibinfo  {journal} {J. Phys. Soc. Jpn.}\ }\textbf {\bibinfo
  {volume} {77}},\ \bibinfo {pages} {031007} (\bibinfo {year}
  {2008})}\BibitemShut {NoStop}%
\bibitem [{\citenamefont {Br{\"u}ne}\ \emph {et~al.}(2012)\citenamefont
  {Br{\"u}ne}, \citenamefont {Roth}, \citenamefont {Buhmann}, \citenamefont
  {Hankiewicz}, \citenamefont {Molenkamp}, \citenamefont {Maciejko},
  \citenamefont {Qi},\ and\ \citenamefont {Zhang}}]{brune2012spin}%
  \BibitemOpen
  \bibfield  {author} {\bibinfo {author} {\bibfnamefont {C.}~\bibnamefont
  {Br{\"u}ne}}, \bibinfo {author} {\bibfnamefont {A.}~\bibnamefont {Roth}},
  \bibinfo {author} {\bibfnamefont {H.}~\bibnamefont {Buhmann}}, \bibinfo
  {author} {\bibfnamefont {E.~M.}\ \bibnamefont {Hankiewicz}}, \bibinfo
  {author} {\bibfnamefont {L.~W.}\ \bibnamefont {Molenkamp}}, \bibinfo {author}
  {\bibfnamefont {J.}~\bibnamefont {Maciejko}}, \bibinfo {author}
  {\bibfnamefont {X.-L.}\ \bibnamefont {Qi}},\ and\ \bibinfo {author}
  {\bibfnamefont {S.-C.}\ \bibnamefont {Zhang}},\ }\bibfield  {title} {\bibinfo
  {title} {{Spin polarization of the quantum spin Hall edge states}},\ }\href
  {https://doi.org/https://doi.org/10.1038/nphys2322} {\bibfield  {journal}
  {\bibinfo  {journal} {Nat. Phys.}\ }\textbf {\bibinfo {volume} {8}},\
  \bibinfo {pages} {485} (\bibinfo {year} {2012})}\BibitemShut {NoStop}%
\bibitem [{\citenamefont {Roth}\ \emph {et~al.}(2009)\citenamefont {Roth},
  \citenamefont {Brüne}, \citenamefont {Buhmann}, \citenamefont {Molenkamp},
  \citenamefont {Maciejko}, \citenamefont {Qi},\ and\ \citenamefont
  {Zhang}}]{doi:10.1126/science.1174736}%
  \BibitemOpen
  \bibfield  {author} {\bibinfo {author} {\bibfnamefont {A.}~\bibnamefont
  {Roth}}, \bibinfo {author} {\bibfnamefont {C.}~\bibnamefont {Brüne}},
  \bibinfo {author} {\bibfnamefont {H.}~\bibnamefont {Buhmann}}, \bibinfo
  {author} {\bibfnamefont {L.~W.}\ \bibnamefont {Molenkamp}}, \bibinfo {author}
  {\bibfnamefont {J.}~\bibnamefont {Maciejko}}, \bibinfo {author}
  {\bibfnamefont {X.-L.}\ \bibnamefont {Qi}},\ and\ \bibinfo {author}
  {\bibfnamefont {S.-C.}\ \bibnamefont {Zhang}},\ }\bibfield  {title} {\bibinfo
  {title} {{Nonlocal Transport in the Quantum Spin Hall State}},\ }\href
  {https://doi.org/10.1126/science.1174736} {\bibfield  {journal} {\bibinfo
  {journal} {Science}\ }\textbf {\bibinfo {volume} {325}},\ \bibinfo {pages}
  {294} (\bibinfo {year} {2009})}\BibitemShut {NoStop}%
\bibitem [{\citenamefont {Knez}\ \emph {et~al.}(2011)\citenamefont {Knez},
  \citenamefont {Du},\ and\ \citenamefont {Sullivan}}]{PhysRevLett.107.136603}%
  \BibitemOpen
  \bibfield  {author} {\bibinfo {author} {\bibfnamefont {I.}~\bibnamefont
  {Knez}}, \bibinfo {author} {\bibfnamefont {R.-R.}\ \bibnamefont {Du}},\ and\
  \bibinfo {author} {\bibfnamefont {G.}~\bibnamefont {Sullivan}},\ }\bibfield
  {title} {\bibinfo {title} {{Evidence for Helical Edge Modes in Inverted
  $\mathrm{InAs}/\mathrm{GaSb}$ Quantum Wells}},\ }\href
  {https://doi.org/10.1103/PhysRevLett.107.136603} {\bibfield  {journal}
  {\bibinfo  {journal} {Phys. Rev. Lett.}\ }\textbf {\bibinfo {volume} {107}},\
  \bibinfo {pages} {136603} (\bibinfo {year} {2011})}\BibitemShut {NoStop}%
\bibitem [{\citenamefont {Young}\ \emph {et~al.}(2014)\citenamefont {Young},
  \citenamefont {Sanchez-Yamagishi}, \citenamefont {Hunt}, \citenamefont
  {Choi}, \citenamefont {Watanabe}, \citenamefont {Taniguchi}, \citenamefont
  {Ashoori},\ and\ \citenamefont {Jarillo-Herrero}}]{young2014tunable}%
  \BibitemOpen
  \bibfield  {author} {\bibinfo {author} {\bibfnamefont {A.}~\bibnamefont
  {Young}}, \bibinfo {author} {\bibfnamefont {J.}~\bibnamefont
  {Sanchez-Yamagishi}}, \bibinfo {author} {\bibfnamefont {B.}~\bibnamefont
  {Hunt}}, \bibinfo {author} {\bibfnamefont {S.}~\bibnamefont {Choi}}, \bibinfo
  {author} {\bibfnamefont {K.}~\bibnamefont {Watanabe}}, \bibinfo {author}
  {\bibfnamefont {T.}~\bibnamefont {Taniguchi}}, \bibinfo {author}
  {\bibfnamefont {R.}~\bibnamefont {Ashoori}},\ and\ \bibinfo {author}
  {\bibfnamefont {P.}~\bibnamefont {Jarillo-Herrero}},\ }\bibfield  {title}
  {\bibinfo {title} {{Tunable symmetry breaking and helical edge transport in a
  graphene quantum spin Hall state}},\ }\href
  {https://doi.org/https://doi.org/10.1038/nature12800} {\bibfield  {journal}
  {\bibinfo  {journal} {Nature}\ }\textbf {\bibinfo {volume} {505}},\ \bibinfo
  {pages} {528} (\bibinfo {year} {2014})}\BibitemShut {NoStop}%
\bibitem [{\citenamefont {Fei}\ \emph {et~al.}(2017)\citenamefont {Fei},
  \citenamefont {Palomaki}, \citenamefont {Wu}, \citenamefont {Zhao},
  \citenamefont {Cai}, \citenamefont {Sun}, \citenamefont {Nguyen},
  \citenamefont {Finney}, \citenamefont {Xu},\ and\ \citenamefont
  {Cobden}}]{fei2017edge}%
  \BibitemOpen
  \bibfield  {author} {\bibinfo {author} {\bibfnamefont {Z.}~\bibnamefont
  {Fei}}, \bibinfo {author} {\bibfnamefont {T.}~\bibnamefont {Palomaki}},
  \bibinfo {author} {\bibfnamefont {S.}~\bibnamefont {Wu}}, \bibinfo {author}
  {\bibfnamefont {W.}~\bibnamefont {Zhao}}, \bibinfo {author} {\bibfnamefont
  {X.}~\bibnamefont {Cai}}, \bibinfo {author} {\bibfnamefont {B.}~\bibnamefont
  {Sun}}, \bibinfo {author} {\bibfnamefont {P.}~\bibnamefont {Nguyen}},
  \bibinfo {author} {\bibfnamefont {J.}~\bibnamefont {Finney}}, \bibinfo
  {author} {\bibfnamefont {X.}~\bibnamefont {Xu}},\ and\ \bibinfo {author}
  {\bibfnamefont {D.~H.}\ \bibnamefont {Cobden}},\ }\bibfield  {title}
  {\bibinfo {title} {{Edge conduction in monolayer WTe2}},\ }\href
  {https://doi.org/https://doi.org/10.1038/nphys4091} {\bibfield  {journal}
  {\bibinfo  {journal} {Nat. Phys.}\ }\textbf {\bibinfo {volume} {13}},\
  \bibinfo {pages} {677} (\bibinfo {year} {2017})}\BibitemShut {NoStop}%
\bibitem [{\citenamefont {Wu}\ \emph {et~al.}(2018)\citenamefont {Wu},
  \citenamefont {Fatemi}, \citenamefont {Gibson}, \citenamefont {Watanabe},
  \citenamefont {Taniguchi}, \citenamefont {Cava},\ and\ \citenamefont
  {Jarillo-Herrero}}]{doi:10.1126/science.aan6003}%
  \BibitemOpen
  \bibfield  {author} {\bibinfo {author} {\bibfnamefont {S.}~\bibnamefont
  {Wu}}, \bibinfo {author} {\bibfnamefont {V.}~\bibnamefont {Fatemi}}, \bibinfo
  {author} {\bibfnamefont {Q.~D.}\ \bibnamefont {Gibson}}, \bibinfo {author}
  {\bibfnamefont {K.}~\bibnamefont {Watanabe}}, \bibinfo {author}
  {\bibfnamefont {T.}~\bibnamefont {Taniguchi}}, \bibinfo {author}
  {\bibfnamefont {R.~J.}\ \bibnamefont {Cava}},\ and\ \bibinfo {author}
  {\bibfnamefont {P.}~\bibnamefont {Jarillo-Herrero}},\ }\bibfield  {title}
  {\bibinfo {title} {{Observation of the quantum spin Hall effect up to 100
  kelvin in a monolayer crystal}},\ }\href
  {https://doi.org/10.1126/science.aan6003} {\bibfield  {journal} {\bibinfo
  {journal} {Science}\ }\textbf {\bibinfo {volume} {359}},\ \bibinfo {pages}
  {76} (\bibinfo {year} {2018})}\BibitemShut {NoStop}%
\bibitem [{\citenamefont {Fu}\ \emph {et~al.}(2024)\citenamefont {Fu},
  \citenamefont {Xu}, \citenamefont {Yang}, \citenamefont {Lee}, \citenamefont
  {Ang},\ and\ \citenamefont {Liu}}]{PhysRevApplied.21.054057}%
  \BibitemOpen
  \bibfield  {author} {\bibinfo {author} {\bibfnamefont {P.-H.}\ \bibnamefont
  {Fu}}, \bibinfo {author} {\bibfnamefont {Y.}~\bibnamefont {Xu}}, \bibinfo
  {author} {\bibfnamefont {S.~A.}\ \bibnamefont {Yang}}, \bibinfo {author}
  {\bibfnamefont {C.~H.}\ \bibnamefont {Lee}}, \bibinfo {author} {\bibfnamefont
  {Y.~S.}\ \bibnamefont {Ang}},\ and\ \bibinfo {author} {\bibfnamefont {J.-F.}\
  \bibnamefont {Liu}},\ }\bibfield  {title} {\bibinfo {title} {{Field-effect
  Josephson diode via asymmetric spin-momentum locking states}},\ }\href
  {https://doi.org/10.1103/PhysRevApplied.21.054057} {\bibfield  {journal}
  {\bibinfo  {journal} {Phys. Rev. Appl.}\ }\textbf {\bibinfo {volume} {21}},\
  \bibinfo {pages} {054057} (\bibinfo {year} {2024})}\BibitemShut {NoStop}%
\bibitem [{\citenamefont {Zhou}\ \emph {et~al.}(2008)\citenamefont {Zhou},
  \citenamefont {Lu}, \citenamefont {Chu}, \citenamefont {Shen},\ and\
  \citenamefont {Niu}}]{PhysRevLett.101.246807}%
  \BibitemOpen
  \bibfield  {author} {\bibinfo {author} {\bibfnamefont {B.}~\bibnamefont
  {Zhou}}, \bibinfo {author} {\bibfnamefont {H.-Z.}\ \bibnamefont {Lu}},
  \bibinfo {author} {\bibfnamefont {R.-L.}\ \bibnamefont {Chu}}, \bibinfo
  {author} {\bibfnamefont {S.-Q.}\ \bibnamefont {Shen}},\ and\ \bibinfo
  {author} {\bibfnamefont {Q.}~\bibnamefont {Niu}},\ }\bibfield  {title}
  {\bibinfo {title} {{Finite Size Effects on Helical Edge States in a Quantum
  Spin-Hall System}},\ }\href {https://doi.org/10.1103/PhysRevLett.101.246807}
  {\bibfield  {journal} {\bibinfo  {journal} {Phys. Rev. Lett.}\ }\textbf
  {\bibinfo {volume} {101}},\ \bibinfo {pages} {246807} (\bibinfo {year}
  {2008})}\BibitemShut {NoStop}%
\bibitem [{\citenamefont {Wang}\ \emph
  {et~al.}(2012{\natexlab{a}})\citenamefont {Wang}, \citenamefont {Hu},\ and\
  \citenamefont {Guo}}]{PhysRevB.85.241402}%
  \BibitemOpen
  \bibfield  {author} {\bibinfo {author} {\bibfnamefont {X.-F.}\ \bibnamefont
  {Wang}}, \bibinfo {author} {\bibfnamefont {Y.}~\bibnamefont {Hu}},\ and\
  \bibinfo {author} {\bibfnamefont {H.}~\bibnamefont {Guo}},\ }\bibfield
  {title} {\bibinfo {title} {{Robustness of helical edge states in topological
  insulators}},\ }\href {https://doi.org/10.1103/PhysRevB.85.241402} {\bibfield
   {journal} {\bibinfo  {journal} {Phys. Rev. B}\ }\textbf {\bibinfo {volume}
  {85}},\ \bibinfo {pages} {241402} (\bibinfo {year}
  {2012}{\natexlab{a}})}\BibitemShut {NoStop}%
\bibitem [{\citenamefont {Lü}\ and\ \citenamefont {Liu}(2024)}]{Lü_2024}%
  \BibitemOpen
  \bibfield  {author} {\bibinfo {author} {\bibfnamefont {X.-L.}\ \bibnamefont
  {Lü}}\ and\ \bibinfo {author} {\bibfnamefont {J.-F.}\ \bibnamefont {Liu}},\
  }\bibfield  {title} {\bibinfo {title} {Generation and edge-state transitions
  of pseudohelical edge state based on side potentials in graphene},\ }\href
  {https://doi.org/10.1088/1367-2630/ad6eae} {\bibfield  {journal} {\bibinfo
  {journal} {New J. Phys.}\ }\textbf {\bibinfo {volume} {26}},\ \bibinfo
  {pages} {093023} (\bibinfo {year} {2024})}\BibitemShut {NoStop}%
\bibitem [{\citenamefont {Dai}\ \emph {et~al.}(2024)\citenamefont {Dai},
  \citenamefont {Fu}, \citenamefont {Ang},\ and\ \citenamefont
  {Chen}}]{PhysRevB.110.195409}%
  \BibitemOpen
  \bibfield  {author} {\bibinfo {author} {\bibfnamefont {X.}~\bibnamefont
  {Dai}}, \bibinfo {author} {\bibfnamefont {P.-H.}\ \bibnamefont {Fu}},
  \bibinfo {author} {\bibfnamefont {Y.~S.}\ \bibnamefont {Ang}},\ and\ \bibinfo
  {author} {\bibfnamefont {Q.}~\bibnamefont {Chen}},\ }\bibfield  {title}
  {\bibinfo {title} {{Two-dimensional Weyl nodal-line semimetal and antihelical
  edge states in a modified Kane-Mele model}},\ }\href
  {https://doi.org/10.1103/PhysRevB.110.195409} {\bibfield  {journal} {\bibinfo
   {journal} {Phys. Rev. B}\ }\textbf {\bibinfo {volume} {110}},\ \bibinfo
  {pages} {195409} (\bibinfo {year} {2024})}\BibitemShut {NoStop}%
\bibitem [{\citenamefont {Wang}\ and\ \citenamefont
  {Liu}(2021)}]{PhysRevB.103.075419}%
  \BibitemOpen
  \bibfield  {author} {\bibinfo {author} {\bibfnamefont {J.}~\bibnamefont
  {Wang}}\ and\ \bibinfo {author} {\bibfnamefont {J.-F.}\ \bibnamefont {Liu}},\
  }\bibfield  {title} {\bibinfo {title} {{Quantum spin Hall phase transition in
  the $\alpha$-$T_{3}$ lattice}},\ }\href
  {https://doi.org/10.1103/PhysRevB.103.075419} {\bibfield  {journal} {\bibinfo
   {journal} {Phys. Rev. B}\ }\textbf {\bibinfo {volume} {103}},\ \bibinfo
  {pages} {075419} (\bibinfo {year} {2021})}\BibitemShut {NoStop}%
\bibitem [{\citenamefont {Vidal}\ \emph {et~al.}(1998)\citenamefont {Vidal},
  \citenamefont {Mosseri},\ and\ \citenamefont {Dou\ifmmode~\mbox{\c{c}}\else
  \c{c}\fi{}ot}}]{PhysRevLett.81.5888}%
  \BibitemOpen
  \bibfield  {author} {\bibinfo {author} {\bibfnamefont {J.}~\bibnamefont
  {Vidal}}, \bibinfo {author} {\bibfnamefont {R.}~\bibnamefont {Mosseri}},\
  and\ \bibinfo {author} {\bibfnamefont {B.}~\bibnamefont
  {Dou\ifmmode~\mbox{\c{c}}\else \c{c}\fi{}ot}},\ }\bibfield  {title} {\bibinfo
  {title} {{Aharonov-Bohm Cages in Two-Dimensional Structures}},\ }\href
  {https://doi.org/10.1103/PhysRevLett.81.5888} {\bibfield  {journal} {\bibinfo
   {journal} {Phys. Rev. Lett.}\ }\textbf {\bibinfo {volume} {81}},\ \bibinfo
  {pages} {5888} (\bibinfo {year} {1998})}\BibitemShut {NoStop}%
\bibitem [{\citenamefont {Raoux}\ \emph {et~al.}(2014)\citenamefont {Raoux},
  \citenamefont {Morigi}, \citenamefont {Fuchs}, \citenamefont {Pi\'echon},\
  and\ \citenamefont {Montambaux}}]{PhysRevLett.112.026402}%
  \BibitemOpen
  \bibfield  {author} {\bibinfo {author} {\bibfnamefont {A.}~\bibnamefont
  {Raoux}}, \bibinfo {author} {\bibfnamefont {M.}~\bibnamefont {Morigi}},
  \bibinfo {author} {\bibfnamefont {J.-N.}\ \bibnamefont {Fuchs}}, \bibinfo
  {author} {\bibfnamefont {F.}~\bibnamefont {Pi\'echon}},\ and\ \bibinfo
  {author} {\bibfnamefont {G.}~\bibnamefont {Montambaux}},\ }\bibfield  {title}
  {\bibinfo {title} {{From Dia- to Paramagnetic Orbital Susceptibility of
  Massless Fermions}},\ }\href {https://doi.org/10.1103/PhysRevLett.112.026402}
  {\bibfield  {journal} {\bibinfo  {journal} {Phys. Rev. Lett.}\ }\textbf
  {\bibinfo {volume} {112}},\ \bibinfo {pages} {026402} (\bibinfo {year}
  {2014})}\BibitemShut {NoStop}%
\bibitem [{\citenamefont {Illes}\ \emph {et~al.}(2015)\citenamefont {Illes},
  \citenamefont {Carbotte},\ and\ \citenamefont {Nicol}}]{PhysRevB.92.245410}%
  \BibitemOpen
  \bibfield  {author} {\bibinfo {author} {\bibfnamefont {E.}~\bibnamefont
  {Illes}}, \bibinfo {author} {\bibfnamefont {J.~P.}\ \bibnamefont
  {Carbotte}},\ and\ \bibinfo {author} {\bibfnamefont {E.~J.}\ \bibnamefont
  {Nicol}},\ }\bibfield  {title} {\bibinfo {title} {{Hall quantization and
  optical conductivity evolution with variable Berry phase in the
  $\alpha$-$T_{3}$ model}},\ }\href
  {https://doi.org/10.1103/PhysRevB.92.245410} {\bibfield  {journal} {\bibinfo
  {journal} {Phys. Rev. B}\ }\textbf {\bibinfo {volume} {92}},\ \bibinfo
  {pages} {245410} (\bibinfo {year} {2015})}\BibitemShut {NoStop}%
\bibitem [{\citenamefont {Biswas}\ and\ \citenamefont {Ghosh}(2016)}]{Biswas}%
  \BibitemOpen
  \bibfield  {author} {\bibinfo {author} {\bibfnamefont {T.}~\bibnamefont
  {Biswas}}\ and\ \bibinfo {author} {\bibfnamefont {T.~K.}\ \bibnamefont
  {Ghosh}},\ }\bibfield  {title} {\bibinfo {title} {{Magnetotransport
  properties of the $\alpha$-$T_{3}$ model}},\ }\href
  {https://doi.org/10.1088/0953-8984/28/49/495302} {\bibfield  {journal}
  {\bibinfo  {journal} {J. Phys. Condens. Matter}\ }\textbf {\bibinfo {volume}
  {28}},\ \bibinfo {pages} {495302} (\bibinfo {year} {2016})}\BibitemShut
  {NoStop}%
\bibitem [{\citenamefont {Tamang}\ and\ \citenamefont
  {Biswas}(2023)}]{Tamang_Lakpa}%
  \BibitemOpen
  \bibfield  {author} {\bibinfo {author} {\bibfnamefont {L.}~\bibnamefont
  {Tamang}}\ and\ \bibinfo {author} {\bibfnamefont {T.}~\bibnamefont
  {Biswas}},\ }\bibfield  {title} {\bibinfo {title} {{Probing topological
  signatures in an optically driven $\alpha$-$T_{3}$ lattice}},\ }\href
  {https://doi.org/10.1103/PhysRevB.107.085408} {\bibfield  {journal} {\bibinfo
   {journal} {Phys. Rev. B}\ }\textbf {\bibinfo {volume} {107}},\ \bibinfo
  {pages} {085408} (\bibinfo {year} {2023})}\BibitemShut {NoStop}%
\bibitem [{\citenamefont {Tamang}\ \emph {et~al.}(2024)\citenamefont {Tamang},
  \citenamefont {Verma},\ and\ \citenamefont {Biswas}}]{PhysRevB.110.165426}%
  \BibitemOpen
  \bibfield  {author} {\bibinfo {author} {\bibfnamefont {L.}~\bibnamefont
  {Tamang}}, \bibinfo {author} {\bibfnamefont {S.}~\bibnamefont {Verma}},\ and\
  \bibinfo {author} {\bibfnamefont {T.}~\bibnamefont {Biswas}},\ }\bibfield
  {title} {\bibinfo {title} {{Orbital magnetization senses the topological
  phase transition in a spin-orbit coupled
  $\ensuremath{\alpha}\text{\ensuremath{-}}{T}_{3}$ system}},\ }\href
  {https://doi.org/10.1103/PhysRevB.110.165426} {\bibfield  {journal} {\bibinfo
   {journal} {Phys. Rev. B}\ }\textbf {\bibinfo {volume} {110}},\ \bibinfo
  {pages} {165426} (\bibinfo {year} {2024})}\BibitemShut {NoStop}%
\bibitem [{\citenamefont {Lin}\ \emph {et~al.}(2023)\citenamefont {Lin},
  \citenamefont {Tan}, \citenamefont {Fu},\ and\ \citenamefont {Liu}}]{Lin_Fu}%
  \BibitemOpen
  \bibfield  {author} {\bibinfo {author} {\bibfnamefont {S.-Q.}\ \bibnamefont
  {Lin}}, \bibinfo {author} {\bibfnamefont {H.}~\bibnamefont {Tan}}, \bibinfo
  {author} {\bibfnamefont {P.-H.}\ \bibnamefont {Fu}},\ and\ \bibinfo {author}
  {\bibfnamefont {J.-F.}\ \bibnamefont {Liu}},\ }\bibfield  {title} {\bibinfo
  {title} {{Interaction-driven Chern insulating phases in the $\alpha$-$T_{3}$
  lattice with Rashba spin-orbit coupling}},\ }\href
  {https://www.sciencedirect.com/science/article/pii/S2589004223016231}
  {\bibfield  {journal} {\bibinfo  {journal} {iScience}\ }\textbf {\bibinfo
  {volume} {26}} (\bibinfo {year} {2023})}\BibitemShut {NoStop}%
\bibitem [{\citenamefont {Pizzi}\ \emph {et~al.}(2023)\citenamefont {Pizzi},
  \citenamefont {Gorlach}, \citenamefont {Rivera}, \citenamefont {Nunnenkamp},\
  and\ \citenamefont {Kaminer}}]{pizzi2023light}%
  \BibitemOpen
  \bibfield  {author} {\bibinfo {author} {\bibfnamefont {A.}~\bibnamefont
  {Pizzi}}, \bibinfo {author} {\bibfnamefont {A.}~\bibnamefont {Gorlach}},
  \bibinfo {author} {\bibfnamefont {N.}~\bibnamefont {Rivera}}, \bibinfo
  {author} {\bibfnamefont {A.}~\bibnamefont {Nunnenkamp}},\ and\ \bibinfo
  {author} {\bibfnamefont {I.}~\bibnamefont {Kaminer}},\ }\bibfield  {title}
  {\bibinfo {title} {{Light emission from strongly driven many-body systems}},\
  }\href {https://doi.org/https://doi.org/10.1038/s41567-022-01910-7}
  {\bibfield  {journal} {\bibinfo  {journal} {Nat. Phys.}\ }\textbf {\bibinfo
  {volume} {19}},\ \bibinfo {pages} {551} (\bibinfo {year} {2023})}\BibitemShut
  {NoStop}%
\bibitem [{\citenamefont {T\"orm\"a}(2023)}]{PhysRevLett.131.240001}%
  \BibitemOpen
  \bibfield  {author} {\bibinfo {author} {\bibfnamefont {P.}~\bibnamefont
  {T\"orm\"a}},\ }\bibfield  {title} {\bibinfo {title} {{Essay: Where Can
  Quantum Geometry Lead Us?}},\ }\href
  {https://doi.org/10.1103/PhysRevLett.131.240001} {\bibfield  {journal}
  {\bibinfo  {journal} {Phys. Rev. Lett.}\ }\textbf {\bibinfo {volume} {131}},\
  \bibinfo {pages} {240001} (\bibinfo {year} {2023})}\BibitemShut {NoStop}%
\bibitem [{\citenamefont {Mera}\ and\ \citenamefont
  {Mitscherling}(2022)}]{PhysRevB.106.165133}%
  \BibitemOpen
  \bibfield  {author} {\bibinfo {author} {\bibfnamefont {B.}~\bibnamefont
  {Mera}}\ and\ \bibinfo {author} {\bibfnamefont {J.}~\bibnamefont
  {Mitscherling}},\ }\bibfield  {title} {\bibinfo {title} {{Nontrivial quantum
  geometry of degenerate flat bands}},\ }\href
  {https://doi.org/10.1103/PhysRevB.106.165133} {\bibfield  {journal} {\bibinfo
   {journal} {Phys. Rev. B}\ }\textbf {\bibinfo {volume} {106}},\ \bibinfo
  {pages} {165133} (\bibinfo {year} {2022})}\BibitemShut {NoStop}%
\bibitem [{\citenamefont {\ifmmode \check{Z}\else
  \v{Z}\fi{}uti\ifmmode~\acute{c}\else \'{c}\fi{}}\ \emph
  {et~al.}(2004)\citenamefont {\ifmmode \check{Z}\else
  \v{Z}\fi{}uti\ifmmode~\acute{c}\else \'{c}\fi{}}, \citenamefont {Fabian},\
  and\ \citenamefont {Das~Sarma}}]{RevModPhys.76.323}%
  \BibitemOpen
  \bibfield  {author} {\bibinfo {author} {\bibfnamefont {I.}~\bibnamefont
  {\ifmmode \check{Z}\else \v{Z}\fi{}uti\ifmmode~\acute{c}\else \'{c}\fi{}}},
  \bibinfo {author} {\bibfnamefont {J.}~\bibnamefont {Fabian}},\ and\ \bibinfo
  {author} {\bibfnamefont {S.}~\bibnamefont {Das~Sarma}},\ }\bibfield  {title}
  {\bibinfo {title} {{Spintronics: Fundamentals and applications}},\ }\href
  {https://doi.org/10.1103/RevModPhys.76.323} {\bibfield  {journal} {\bibinfo
  {journal} {Rev. Mod. Phys.}\ }\textbf {\bibinfo {volume} {76}},\ \bibinfo
  {pages} {323} (\bibinfo {year} {2004})}\BibitemShut {NoStop}%
\bibitem [{\citenamefont {Di~Sante}\ \emph {et~al.}(2023)\citenamefont
  {Di~Sante}, \citenamefont {Bigi}, \citenamefont {Eck}, \citenamefont
  {Enzner}, \citenamefont {Consiglio}, \citenamefont {Pokharel}, \citenamefont
  {Carrara}, \citenamefont {Orgiani}, \citenamefont {Polewczyk}, \citenamefont
  {Fujii} \emph {et~al.}}]{di2023flat}%
  \BibitemOpen
  \bibfield  {author} {\bibinfo {author} {\bibfnamefont {D.}~\bibnamefont
  {Di~Sante}}, \bibinfo {author} {\bibfnamefont {C.}~\bibnamefont {Bigi}},
  \bibinfo {author} {\bibfnamefont {P.}~\bibnamefont {Eck}}, \bibinfo {author}
  {\bibfnamefont {S.}~\bibnamefont {Enzner}}, \bibinfo {author} {\bibfnamefont
  {A.}~\bibnamefont {Consiglio}}, \bibinfo {author} {\bibfnamefont
  {G.}~\bibnamefont {Pokharel}}, \bibinfo {author} {\bibfnamefont
  {P.}~\bibnamefont {Carrara}}, \bibinfo {author} {\bibfnamefont
  {P.}~\bibnamefont {Orgiani}}, \bibinfo {author} {\bibfnamefont
  {V.}~\bibnamefont {Polewczyk}}, \bibinfo {author} {\bibfnamefont
  {J.}~\bibnamefont {Fujii}}, \emph {et~al.},\ }\bibfield  {title} {\bibinfo
  {title} {{Flat band separation and robust spin Berry curvature in bilayer
  kagome metals}},\ }\href
  {https://doi.org/https://doi.org/10.1038/s41567-023-02053-z} {\bibfield
  {journal} {\bibinfo  {journal} {Nat. Phys.}\ }\textbf {\bibinfo {volume}
  {19}},\ \bibinfo {pages} {1135} (\bibinfo {year} {2023})}\BibitemShut
  {NoStop}%
\bibitem [{\citenamefont {Xue}\ and\ \citenamefont
  {Lee}(2024)}]{xue2024topologicallyprotectednegativeentanglement}%
  \BibitemOpen
  \bibfield  {author} {\bibinfo {author} {\bibfnamefont {W.-T.}\ \bibnamefont
  {Xue}}\ and\ \bibinfo {author} {\bibfnamefont {C.~H.}\ \bibnamefont {Lee}},\
  }\href {https://arxiv.org/abs/2403.03259} {\bibinfo {title} {{Topologically
  protected negative entanglement}}} (\bibinfo {year} {2024}),\ \Eprint
  {https://arxiv.org/abs/2403.03259} {arXiv:2403.03259 [quant-ph]} \BibitemShut
  {NoStop}%
\bibitem [{\citenamefont {Lee}\ \emph {et~al.}(2016)\citenamefont {Lee},
  \citenamefont {Arovas},\ and\ \citenamefont {Thomale}}]{PhysRevB.93.155155}%
  \BibitemOpen
  \bibfield  {author} {\bibinfo {author} {\bibfnamefont {C.~H.}\ \bibnamefont
  {Lee}}, \bibinfo {author} {\bibfnamefont {D.~P.}\ \bibnamefont {Arovas}},\
  and\ \bibinfo {author} {\bibfnamefont {R.}~\bibnamefont {Thomale}},\
  }\bibfield  {title} {\bibinfo {title} {{Band flatness optimization through
  complex analysis}},\ }\href {https://doi.org/10.1103/PhysRevB.93.155155}
  {\bibfield  {journal} {\bibinfo  {journal} {Phys. Rev. B}\ }\textbf {\bibinfo
  {volume} {93}},\ \bibinfo {pages} {155155} (\bibinfo {year}
  {2016})}\BibitemShut {NoStop}%
\bibitem [{\citenamefont {Lee}\ \emph {et~al.}(2017)\citenamefont {Lee},
  \citenamefont {Claassen},\ and\ \citenamefont
  {Thomale}}]{PhysRevB.96.165150}%
  \BibitemOpen
  \bibfield  {author} {\bibinfo {author} {\bibfnamefont {C.~H.}\ \bibnamefont
  {Lee}}, \bibinfo {author} {\bibfnamefont {M.}~\bibnamefont {Claassen}},\ and\
  \bibinfo {author} {\bibfnamefont {R.}~\bibnamefont {Thomale}},\ }\bibfield
  {title} {\bibinfo {title} {{Band structure engineering of ideal fractional
  Chern insulators}},\ }\href {https://doi.org/10.1103/PhysRevB.96.165150}
  {\bibfield  {journal} {\bibinfo  {journal} {Phys. Rev. B}\ }\textbf {\bibinfo
  {volume} {96}},\ \bibinfo {pages} {165150} (\bibinfo {year}
  {2017})}\BibitemShut {NoStop}%
\bibitem [{\citenamefont {Li}\ \emph {et~al.}(2021)\citenamefont {Li},
  \citenamefont {Kumar}, \citenamefont {Sun},\ and\ \citenamefont
  {Lin}}]{PhysRevResearch.3.L032070}%
  \BibitemOpen
  \bibfield  {author} {\bibinfo {author} {\bibfnamefont {H.}~\bibnamefont
  {Li}}, \bibinfo {author} {\bibfnamefont {U.}~\bibnamefont {Kumar}}, \bibinfo
  {author} {\bibfnamefont {K.}~\bibnamefont {Sun}},\ and\ \bibinfo {author}
  {\bibfnamefont {S.-Z.}\ \bibnamefont {Lin}},\ }\bibfield  {title} {\bibinfo
  {title} {{Spontaneous fractional Chern insulators in transition metal
  dichalcogenide moir\'e superlattices}},\ }\href
  {https://doi.org/10.1103/PhysRevResearch.3.L032070} {\bibfield  {journal}
  {\bibinfo  {journal} {Phys. Rev. Res.}\ }\textbf {\bibinfo {volume} {3}},\
  \bibinfo {pages} {L032070} (\bibinfo {year} {2021})}\BibitemShut {NoStop}%
\bibitem [{\citenamefont {Wu}\ \emph {et~al.}(2012)\citenamefont {Wu},
  \citenamefont {Jain},\ and\ \citenamefont {Sun}}]{PhysRevB.86.165129}%
  \BibitemOpen
  \bibfield  {author} {\bibinfo {author} {\bibfnamefont {Y.-H.}\ \bibnamefont
  {Wu}}, \bibinfo {author} {\bibfnamefont {J.~K.}\ \bibnamefont {Jain}},\ and\
  \bibinfo {author} {\bibfnamefont {K.}~\bibnamefont {Sun}},\ }\bibfield
  {title} {\bibinfo {title} {{Adiabatic continuity between Hofstadter and Chern
  insulator states}},\ }\href {https://doi.org/10.1103/PhysRevB.86.165129}
  {\bibfield  {journal} {\bibinfo  {journal} {Phys. Rev. B}\ }\textbf {\bibinfo
  {volume} {86}},\ \bibinfo {pages} {165129} (\bibinfo {year}
  {2012})}\BibitemShut {NoStop}%
\bibitem [{\citenamefont {Liu}\ \emph {et~al.}(2012)\citenamefont {Liu},
  \citenamefont {Bergholtz}, \citenamefont {Fan},\ and\ \citenamefont
  {L\"auchli}}]{PhysRevLett.109.186805}%
  \BibitemOpen
  \bibfield  {author} {\bibinfo {author} {\bibfnamefont {Z.}~\bibnamefont
  {Liu}}, \bibinfo {author} {\bibfnamefont {E.~J.}\ \bibnamefont {Bergholtz}},
  \bibinfo {author} {\bibfnamefont {H.}~\bibnamefont {Fan}},\ and\ \bibinfo
  {author} {\bibfnamefont {A.~M.}\ \bibnamefont {L\"auchli}},\ }\bibfield
  {title} {\bibinfo {title} {{Fractional Chern Insulators in Topological Flat
  Bands with Higher Chern Number}},\ }\href
  {https://doi.org/10.1103/PhysRevLett.109.186805} {\bibfield  {journal}
  {\bibinfo  {journal} {Phys. Rev. Lett.}\ }\textbf {\bibinfo {volume} {109}},\
  \bibinfo {pages} {186805} (\bibinfo {year} {2012})}\BibitemShut {NoStop}%
\bibitem [{\citenamefont {Chang}\ \emph {et~al.}(2013)\citenamefont {Chang},
  \citenamefont {Zhang}, \citenamefont {Feng}, \citenamefont {Shen},
  \citenamefont {Zhang}, \citenamefont {Guo}, \citenamefont {Li}, \citenamefont
  {Ou}, \citenamefont {Wei}, \citenamefont {Wang}, \citenamefont {Ji},
  \citenamefont {Feng}, \citenamefont {Ji}, \citenamefont {Chen}, \citenamefont
  {Jia}, \citenamefont {Dai}, \citenamefont {Fang}, \citenamefont {Zhang},
  \citenamefont {He}, \citenamefont {Wang}, \citenamefont {Lu}, \citenamefont
  {Ma},\ and\ \citenamefont {Xue}}]{Qi-Kun}%
  \BibitemOpen
  \bibfield  {author} {\bibinfo {author} {\bibfnamefont {C.-Z.}\ \bibnamefont
  {Chang}}, \bibinfo {author} {\bibfnamefont {J.}~\bibnamefont {Zhang}},
  \bibinfo {author} {\bibfnamefont {X.}~\bibnamefont {Feng}}, \bibinfo {author}
  {\bibfnamefont {J.}~\bibnamefont {Shen}}, \bibinfo {author} {\bibfnamefont
  {Z.}~\bibnamefont {Zhang}}, \bibinfo {author} {\bibfnamefont
  {M.}~\bibnamefont {Guo}}, \bibinfo {author} {\bibfnamefont {K.}~\bibnamefont
  {Li}}, \bibinfo {author} {\bibfnamefont {Y.}~\bibnamefont {Ou}}, \bibinfo
  {author} {\bibfnamefont {P.}~\bibnamefont {Wei}}, \bibinfo {author}
  {\bibfnamefont {L.-L.}\ \bibnamefont {Wang}}, \bibinfo {author}
  {\bibfnamefont {Z.-Q.}\ \bibnamefont {Ji}}, \bibinfo {author} {\bibfnamefont
  {Y.}~\bibnamefont {Feng}}, \bibinfo {author} {\bibfnamefont {S.}~\bibnamefont
  {Ji}}, \bibinfo {author} {\bibfnamefont {X.}~\bibnamefont {Chen}}, \bibinfo
  {author} {\bibfnamefont {J.}~\bibnamefont {Jia}}, \bibinfo {author}
  {\bibfnamefont {X.}~\bibnamefont {Dai}}, \bibinfo {author} {\bibfnamefont
  {Z.}~\bibnamefont {Fang}}, \bibinfo {author} {\bibfnamefont {S.-C.}\
  \bibnamefont {Zhang}}, \bibinfo {author} {\bibfnamefont {K.}~\bibnamefont
  {He}}, \bibinfo {author} {\bibfnamefont {Y.}~\bibnamefont {Wang}}, \bibinfo
  {author} {\bibfnamefont {L.}~\bibnamefont {Lu}}, \bibinfo {author}
  {\bibfnamefont {X.-C.}\ \bibnamefont {Ma}},\ and\ \bibinfo {author}
  {\bibfnamefont {Q.-K.}\ \bibnamefont {Xue}},\ }\bibfield  {title} {\bibinfo
  {title} {{Experimental Observation of the Quantum Anomalous Hall Effect in a
  Magnetic Topological Insulator}},\ }\href
  {https://doi.org/10.1126/science.1234414} {\bibfield  {journal} {\bibinfo
  {journal} {Science}\ }\textbf {\bibinfo {volume} {340}},\ \bibinfo {pages}
  {167} (\bibinfo {year} {2013})}\BibitemShut {NoStop}%
\bibitem [{\citenamefont {Haldane}(1988)}]{Haldane}%
  \BibitemOpen
  \bibfield  {author} {\bibinfo {author} {\bibfnamefont {F.~D.~M.}\
  \bibnamefont {Haldane}},\ }\bibfield  {title} {\bibinfo {title} {{Model for a
  Quantum Hall Effect without Landau Levels: Condensed-Matter Realization of
  the "Parity Anomaly"}},\ }\href {https://doi.org/10.1103/PhysRevLett.61.2015}
  {\bibfield  {journal} {\bibinfo  {journal} {Phys. Rev. Lett.}\ }\textbf
  {\bibinfo {volume} {61}},\ \bibinfo {pages} {2015} (\bibinfo {year}
  {1988})}\BibitemShut {NoStop}%
\bibitem [{\citenamefont {Hatsugai}(1993{\natexlab{a}})}]{PhysRevLett.71.3697}%
  \BibitemOpen
  \bibfield  {author} {\bibinfo {author} {\bibfnamefont {Y.}~\bibnamefont
  {Hatsugai}},\ }\bibfield  {title} {\bibinfo {title} {{Chern number and edge
  states in the integer quantum Hall effect}},\ }\href
  {https://doi.org/10.1103/PhysRevLett.71.3697} {\bibfield  {journal} {\bibinfo
   {journal} {Phys. Rev. Lett.}\ }\textbf {\bibinfo {volume} {71}},\ \bibinfo
  {pages} {3697} (\bibinfo {year} {1993}{\natexlab{a}})}\BibitemShut {NoStop}%
\bibitem [{\citenamefont {Thouless}\ \emph {et~al.}(1982)\citenamefont
  {Thouless}, \citenamefont {Kohmoto}, \citenamefont {Nightingale},\ and\
  \citenamefont {den Nijs}}]{PhysRevLett.49.405}%
  \BibitemOpen
  \bibfield  {author} {\bibinfo {author} {\bibfnamefont {D.~J.}\ \bibnamefont
  {Thouless}}, \bibinfo {author} {\bibfnamefont {M.}~\bibnamefont {Kohmoto}},
  \bibinfo {author} {\bibfnamefont {M.~P.}\ \bibnamefont {Nightingale}},\ and\
  \bibinfo {author} {\bibfnamefont {M.}~\bibnamefont {den Nijs}},\ }\bibfield
  {title} {\bibinfo {title} {{Quantized Hall Conductance in a Two-Dimensional
  Periodic Potential}},\ }\href {https://doi.org/10.1103/PhysRevLett.49.405}
  {\bibfield  {journal} {\bibinfo  {journal} {Phys. Rev. Lett.}\ }\textbf
  {\bibinfo {volume} {49}},\ \bibinfo {pages} {405} (\bibinfo {year}
  {1982})}\BibitemShut {NoStop}%
\bibitem [{\citenamefont {Oka}\ and\ \citenamefont
  {Kitamura}(2019)}]{oka2019floquet}%
  \BibitemOpen
  \bibfield  {author} {\bibinfo {author} {\bibfnamefont {T.}~\bibnamefont
  {Oka}}\ and\ \bibinfo {author} {\bibfnamefont {S.}~\bibnamefont {Kitamura}},\
  }\bibfield  {title} {\bibinfo {title} {{Floquet Engineering of Quantum
  Materials}},\ }\href
  {https://doi.org/https://doi.org/10.1146/annurev-conmatphys-031218-013423}
  {\bibfield  {journal} {\bibinfo  {journal} {Annu. Rev. Condens. Matter
  Phys.}\ }\textbf {\bibinfo {volume} {10}},\ \bibinfo {pages} {387} (\bibinfo
  {year} {2019})}\BibitemShut {NoStop}%
\bibitem [{\citenamefont {Goldman}\ and\ \citenamefont
  {Dalibard}(2014)}]{PhysRevX.4.031027}%
  \BibitemOpen
  \bibfield  {author} {\bibinfo {author} {\bibfnamefont {N.}~\bibnamefont
  {Goldman}}\ and\ \bibinfo {author} {\bibfnamefont {J.}~\bibnamefont
  {Dalibard}},\ }\bibfield  {title} {\bibinfo {title} {{Periodically Driven
  Quantum Systems: Effective Hamiltonians and Engineered Gauge Fields}},\
  }\href {https://doi.org/10.1103/PhysRevX.4.031027} {\bibfield  {journal}
  {\bibinfo  {journal} {Phys. Rev. X}\ }\textbf {\bibinfo {volume} {4}},\
  \bibinfo {pages} {031027} (\bibinfo {year} {2014})}\BibitemShut {NoStop}%
\bibitem [{\citenamefont {Rahav}\ \emph {et~al.}(2003)\citenamefont {Rahav},
  \citenamefont {Gilary},\ and\ \citenamefont {Fishman}}]{PhysRevA.68.013820}%
  \BibitemOpen
  \bibfield  {author} {\bibinfo {author} {\bibfnamefont {S.}~\bibnamefont
  {Rahav}}, \bibinfo {author} {\bibfnamefont {I.}~\bibnamefont {Gilary}},\ and\
  \bibinfo {author} {\bibfnamefont {S.}~\bibnamefont {Fishman}},\ }\bibfield
  {title} {\bibinfo {title} {{Effective Hamiltonians for periodically driven
  systems}},\ }\href {https://doi.org/10.1103/PhysRevA.68.013820} {\bibfield
  {journal} {\bibinfo  {journal} {Phys. Rev. A}\ }\textbf {\bibinfo {volume}
  {68}},\ \bibinfo {pages} {013820} (\bibinfo {year} {2003})}\BibitemShut
  {NoStop}%
\bibitem [{\citenamefont {Sambe}(1973)}]{PhysRevA.7.2203}%
  \BibitemOpen
  \bibfield  {author} {\bibinfo {author} {\bibfnamefont {H.}~\bibnamefont
  {Sambe}},\ }\bibfield  {title} {\bibinfo {title} {{Steady States and
  Quasienergies of a Quantum-Mechanical System in an Oscillating Field}},\
  }\href {https://doi.org/10.1103/PhysRevA.7.2203} {\bibfield  {journal}
  {\bibinfo  {journal} {Phys. Rev. A}\ }\textbf {\bibinfo {volume} {7}},\
  \bibinfo {pages} {2203} (\bibinfo {year} {1973})}\BibitemShut {NoStop}%
\bibitem [{\citenamefont {Shirley}(1965)}]{PhysRev.138.B979}%
  \BibitemOpen
  \bibfield  {author} {\bibinfo {author} {\bibfnamefont {J.~H.}\ \bibnamefont
  {Shirley}},\ }\bibfield  {title} {\bibinfo {title} {{Solution of the
  Schr\"odinger Equation with a Hamiltonian Periodic in Time}},\ }\href
  {https://doi.org/10.1103/PhysRev.138.B979} {\bibfield  {journal} {\bibinfo
  {journal} {Phys. Rev.}\ }\textbf {\bibinfo {volume} {138}},\ \bibinfo {pages}
  {B979} (\bibinfo {year} {1965})}\BibitemShut {NoStop}%
\bibitem [{\citenamefont {Wan}\ \emph {et~al.}(2024)\citenamefont {Wan},
  \citenamefont {Ning}, \citenamefont {Xu}, \citenamefont {Wang},\ and\
  \citenamefont {Zheng}}]{PhysRevB.109.085148}%
  \BibitemOpen
  \bibfield  {author} {\bibinfo {author} {\bibfnamefont {X.}~\bibnamefont
  {Wan}}, \bibinfo {author} {\bibfnamefont {Z.}~\bibnamefont {Ning}}, \bibinfo
  {author} {\bibfnamefont {D.-H.}\ \bibnamefont {Xu}}, \bibinfo {author}
  {\bibfnamefont {R.}~\bibnamefont {Wang}},\ and\ \bibinfo {author}
  {\bibfnamefont {B.}~\bibnamefont {Zheng}},\ }\bibfield  {title} {\bibinfo
  {title} {{Photoinduced high-Chern-number quantum anomalous Hall effect from
  higher-order topological insulators}},\ }\href
  {https://doi.org/10.1103/PhysRevB.109.085148} {\bibfield  {journal} {\bibinfo
   {journal} {Phys. Rev. B}\ }\textbf {\bibinfo {volume} {109}},\ \bibinfo
  {pages} {085148} (\bibinfo {year} {2024})}\BibitemShut {NoStop}%
\bibitem [{\citenamefont {Titum}\ \emph {et~al.}(2015)\citenamefont {Titum},
  \citenamefont {Lindner}, \citenamefont {Rechtsman},\ and\ \citenamefont
  {Refael}}]{PhysRevLett.114.056801}%
  \BibitemOpen
  \bibfield  {author} {\bibinfo {author} {\bibfnamefont {P.}~\bibnamefont
  {Titum}}, \bibinfo {author} {\bibfnamefont {N.~H.}\ \bibnamefont {Lindner}},
  \bibinfo {author} {\bibfnamefont {M.~C.}\ \bibnamefont {Rechtsman}},\ and\
  \bibinfo {author} {\bibfnamefont {G.}~\bibnamefont {Refael}},\ }\bibfield
  {title} {\bibinfo {title} {{Disorder-Induced Floquet Topological
  Insulators}},\ }\href {https://doi.org/10.1103/PhysRevLett.114.056801}
  {\bibfield  {journal} {\bibinfo  {journal} {Phys. Rev. Lett.}\ }\textbf
  {\bibinfo {volume} {114}},\ \bibinfo {pages} {056801} (\bibinfo {year}
  {2015})}\BibitemShut {NoStop}%
\bibitem [{\citenamefont {Liu}\ \emph {et~al.}(2024{\natexlab{a}})\citenamefont
  {Liu}, \citenamefont {Chen},\ and\ \citenamefont
  {Zhou}}]{PhysRevB.109.125303}%
  \BibitemOpen
  \bibfield  {author} {\bibinfo {author} {\bibfnamefont {Z.-R.}\ \bibnamefont
  {Liu}}, \bibinfo {author} {\bibfnamefont {R.}~\bibnamefont {Chen}},\ and\
  \bibinfo {author} {\bibfnamefont {B.}~\bibnamefont {Zhou}},\ }\bibfield
  {title} {\bibinfo {title} {{Four-dimensional Floquet topological insulator
  with an emergent second Chern number}},\ }\href
  {https://doi.org/10.1103/PhysRevB.109.125303} {\bibfield  {journal} {\bibinfo
   {journal} {Phys. Rev. B}\ }\textbf {\bibinfo {volume} {109}},\ \bibinfo
  {pages} {125303} (\bibinfo {year} {2024}{\natexlab{a}})}\BibitemShut
  {NoStop}%
\bibitem [{\citenamefont {Liu}\ \emph {et~al.}(2024{\natexlab{b}})\citenamefont
  {Liu}, \citenamefont {Chen},\ and\ \citenamefont {Zhou}}]{liu2024tuning}%
  \BibitemOpen
  \bibfield  {author} {\bibinfo {author} {\bibfnamefont {Z.-R.}\ \bibnamefont
  {Liu}}, \bibinfo {author} {\bibfnamefont {R.}~\bibnamefont {Chen}},\ and\
  \bibinfo {author} {\bibfnamefont {B.}~\bibnamefont {Zhou}},\ }\bibfield
  {title} {\bibinfo {title} {{Tuning Second Chern Number in a Four-Dimensional
  Topological Insulator by High-Frequency Time-Periodic Driving}},\ }\href
  {https://doi.org/10.1088/0256-307X/41/4/047102} {\bibfield  {journal}
  {\bibinfo  {journal} {Chinese Phys. Lett.}\ }\textbf {\bibinfo {volume}
  {41}},\ \bibinfo {pages} {047102} (\bibinfo {year}
  {2024}{\natexlab{b}})}\BibitemShut {NoStop}%
\bibitem [{\citenamefont {Chan}\ \emph {et~al.}(2016)\citenamefont {Chan},
  \citenamefont {Oh}, \citenamefont {Han},\ and\ \citenamefont
  {Lee}}]{PhysRevB.94.121106}%
  \BibitemOpen
  \bibfield  {author} {\bibinfo {author} {\bibfnamefont {C.-K.}\ \bibnamefont
  {Chan}}, \bibinfo {author} {\bibfnamefont {Y.-T.}\ \bibnamefont {Oh}},
  \bibinfo {author} {\bibfnamefont {J.~H.}\ \bibnamefont {Han}},\ and\ \bibinfo
  {author} {\bibfnamefont {P.~A.}\ \bibnamefont {Lee}},\ }\bibfield  {title}
  {\bibinfo {title} {{Type-II Weyl cone transitions in driven semimetals}},\
  }\href {https://doi.org/10.1103/PhysRevB.94.121106} {\bibfield  {journal}
  {\bibinfo  {journal} {Phys. Rev. B}\ }\textbf {\bibinfo {volume} {94}},\
  \bibinfo {pages} {121106} (\bibinfo {year} {2016})}\BibitemShut {NoStop}%
\bibitem [{\citenamefont {Bonasera}\ \emph {et~al.}(2022)\citenamefont
  {Bonasera}, \citenamefont {Zhang}, \citenamefont {Privitera},\ and\
  \citenamefont {Pellegrino}}]{PhysRevB.106.195115}%
  \BibitemOpen
  \bibfield  {author} {\bibinfo {author} {\bibfnamefont {F.}~\bibnamefont
  {Bonasera}}, \bibinfo {author} {\bibfnamefont {S.-B.}\ \bibnamefont {Zhang}},
  \bibinfo {author} {\bibfnamefont {L.}~\bibnamefont {Privitera}},\ and\
  \bibinfo {author} {\bibfnamefont {F.~M.~D.}\ \bibnamefont {Pellegrino}},\
  }\bibfield  {title} {\bibinfo {title} {{Tunable interface states between
  Floquet-Weyl semimetals}},\ }\href
  {https://doi.org/10.1103/PhysRevB.106.195115} {\bibfield  {journal} {\bibinfo
   {journal} {Phys. Rev. B}\ }\textbf {\bibinfo {volume} {106}},\ \bibinfo
  {pages} {195115} (\bibinfo {year} {2022})}\BibitemShut {NoStop}%
\bibitem [{\citenamefont {Wang}\ \emph {et~al.}(2014)\citenamefont {Wang},
  \citenamefont {Wang}, \citenamefont {Shen}, \citenamefont {Sheng},\ and\
  \citenamefont {Xing}}]{Wang_2014}%
  \BibitemOpen
  \bibfield  {author} {\bibinfo {author} {\bibfnamefont {R.}~\bibnamefont
  {Wang}}, \bibinfo {author} {\bibfnamefont {B.}~\bibnamefont {Wang}}, \bibinfo
  {author} {\bibfnamefont {R.}~\bibnamefont {Shen}}, \bibinfo {author}
  {\bibfnamefont {L.}~\bibnamefont {Sheng}},\ and\ \bibinfo {author}
  {\bibfnamefont {D.~Y.}\ \bibnamefont {Xing}},\ }\bibfield  {title} {\bibinfo
  {title} {{Floquet Weyl semimetal induced by off-resonant light}},\ }\href
  {https://doi.org/10.1209/0295-5075/105/17004} {\bibfield  {journal} {\bibinfo
   {journal} {EPL}\ }\textbf {\bibinfo {volume} {105}},\ \bibinfo {pages}
  {17004} (\bibinfo {year} {2014})}\BibitemShut {NoStop}%
\bibitem [{\citenamefont {Fu}\ \emph {et~al.}(2017)\citenamefont {Fu},
  \citenamefont {Duan}, \citenamefont {Wang},\ and\ \citenamefont
  {Chen}}]{FU20173499}%
  \BibitemOpen
  \bibfield  {author} {\bibinfo {author} {\bibfnamefont {P.-H.}\ \bibnamefont
  {Fu}}, \bibinfo {author} {\bibfnamefont {H.-J.}\ \bibnamefont {Duan}},
  \bibinfo {author} {\bibfnamefont {R.-Q.}\ \bibnamefont {Wang}},\ and\
  \bibinfo {author} {\bibfnamefont {H.}~\bibnamefont {Chen}},\ }\bibfield
  {title} {\bibinfo {title} {{Phase transitions in three-dimensional Dirac
  semimetal induced by off-resonant circularly polarized light}},\ }\href
  {https://doi.org/https://doi.org/10.1016/j.physleta.2017.08.055} {\bibfield
  {journal} {\bibinfo  {journal} {Phys. Lett. A}\ }\textbf {\bibinfo {volume}
  {381}},\ \bibinfo {pages} {3499} (\bibinfo {year} {2017})}\BibitemShut
  {NoStop}%
\bibitem [{\citenamefont {Fu}\ \emph {et~al.}(2021)\citenamefont {Fu},
  \citenamefont {Lv}, \citenamefont {Yu}, \citenamefont {Liu},\ and\
  \citenamefont {Wu}}]{Fu_2022}%
  \BibitemOpen
  \bibfield  {author} {\bibinfo {author} {\bibfnamefont {P.-H.}\ \bibnamefont
  {Fu}}, \bibinfo {author} {\bibfnamefont {Q.}~\bibnamefont {Lv}}, \bibinfo
  {author} {\bibfnamefont {X.-L.}\ \bibnamefont {Yu}}, \bibinfo {author}
  {\bibfnamefont {J.-F.}\ \bibnamefont {Liu}},\ and\ \bibinfo {author}
  {\bibfnamefont {J.}~\bibnamefont {Wu}},\ }\bibfield  {title} {\bibinfo
  {title} {{The generation of switchable polarized currents in nodal ring
  semimetals using high-frequency periodic driving}},\ }\href
  {https://doi.org/10.1088/1361-648X/ac37db} {\bibfield  {journal} {\bibinfo
  {journal} {J. Phys. Condens. Matter}\ }\textbf {\bibinfo {volume} {34}},\
  \bibinfo {pages} {075401} (\bibinfo {year} {2021})}\BibitemShut {NoStop}%
\bibitem [{\citenamefont {Li}\ \emph {et~al.}(2019)\citenamefont {Li},
  \citenamefont {Wang}, \citenamefont {Deng}, \citenamefont {Duan},
  \citenamefont {Fu}, \citenamefont {Wang}, \citenamefont {Sheng},\ and\
  \citenamefont {Xing}}]{PhysRevLett.123.206601}%
  \BibitemOpen
  \bibfield  {author} {\bibinfo {author} {\bibfnamefont {X.-S.}\ \bibnamefont
  {Li}}, \bibinfo {author} {\bibfnamefont {C.}~\bibnamefont {Wang}}, \bibinfo
  {author} {\bibfnamefont {M.-X.}\ \bibnamefont {Deng}}, \bibinfo {author}
  {\bibfnamefont {H.-J.}\ \bibnamefont {Duan}}, \bibinfo {author}
  {\bibfnamefont {P.-H.}\ \bibnamefont {Fu}}, \bibinfo {author} {\bibfnamefont
  {R.-Q.}\ \bibnamefont {Wang}}, \bibinfo {author} {\bibfnamefont
  {L.}~\bibnamefont {Sheng}},\ and\ \bibinfo {author} {\bibfnamefont {D.~Y.}\
  \bibnamefont {Xing}},\ }\bibfield  {title} {\bibinfo {title} {{Photon-Induced
  Weyl Half-Metal Phase and Spin Filter Effect from Topological Dirac
  Semimetals}},\ }\href {https://doi.org/10.1103/PhysRevLett.123.206601}
  {\bibfield  {journal} {\bibinfo  {journal} {Phys. Rev. Lett.}\ }\textbf
  {\bibinfo {volume} {123}},\ \bibinfo {pages} {206601} (\bibinfo {year}
  {2019})}\BibitemShut {NoStop}%
\bibitem [{\citenamefont {Cayao}\ \emph {et~al.}(2021)\citenamefont {Cayao},
  \citenamefont {Triola},\ and\ \citenamefont
  {Black-Schaffer}}]{PhysRevB.103.104505}%
  \BibitemOpen
  \bibfield  {author} {\bibinfo {author} {\bibfnamefont {J.}~\bibnamefont
  {Cayao}}, \bibinfo {author} {\bibfnamefont {C.}~\bibnamefont {Triola}},\ and\
  \bibinfo {author} {\bibfnamefont {A.~M.}\ \bibnamefont {Black-Schaffer}},\
  }\bibfield  {title} {\bibinfo {title} {{Floquet engineering bulk
  odd-frequency superconducting pairs}},\ }\href
  {https://doi.org/10.1103/PhysRevB.103.104505} {\bibfield  {journal} {\bibinfo
   {journal} {Phys. Rev. B}\ }\textbf {\bibinfo {volume} {103}},\ \bibinfo
  {pages} {104505} (\bibinfo {year} {2021})}\BibitemShut {NoStop}%
\bibitem [{\citenamefont {Berdakin}\ \emph {et~al.}(2021)\citenamefont
  {Berdakin}, \citenamefont {Rodr{\'\i}guez-Mena},\ and\ \citenamefont
  {Foa~Torres}}]{berdakin2021spin}%
  \BibitemOpen
  \bibfield  {author} {\bibinfo {author} {\bibfnamefont {M.}~\bibnamefont
  {Berdakin}}, \bibinfo {author} {\bibfnamefont {E.~A.}\ \bibnamefont
  {Rodr{\'\i}guez-Mena}},\ and\ \bibinfo {author} {\bibfnamefont {L.~E.}\
  \bibnamefont {Foa~Torres}},\ }\bibfield  {title} {\bibinfo {title}
  {{Spin-polarized tunable photocurrents}},\ }\href
  {https://doi.org/https://doi.org/10.1021/acs.nanolett.1c00420} {\bibfield
  {journal} {\bibinfo  {journal} {Nano Lett.}\ }\textbf {\bibinfo {volume}
  {21}},\ \bibinfo {pages} {3177} (\bibinfo {year} {2021})}\BibitemShut
  {NoStop}%
\bibitem [{\citenamefont {Kuhn}\ \emph {et~al.}(2024)\citenamefont {Kuhn},
  \citenamefont {Sothmann},\ and\ \citenamefont {Cayao}}]{PhysRevB.109.134517}%
  \BibitemOpen
  \bibfield  {author} {\bibinfo {author} {\bibfnamefont {T.}~\bibnamefont
  {Kuhn}}, \bibinfo {author} {\bibfnamefont {B.}~\bibnamefont {Sothmann}},\
  and\ \bibinfo {author} {\bibfnamefont {J.}~\bibnamefont {Cayao}},\ }\bibfield
   {title} {\bibinfo {title} {{Floquet engineering Higgs dynamics in
  time-periodic superconductors}},\ }\href
  {https://doi.org/10.1103/PhysRevB.109.134517} {\bibfield  {journal} {\bibinfo
   {journal} {Phys. Rev. B}\ }\textbf {\bibinfo {volume} {109}},\ \bibinfo
  {pages} {134517} (\bibinfo {year} {2024})}\BibitemShut {NoStop}%
\bibitem [{\citenamefont {Fu}\ \emph {et~al.}(2022)\citenamefont {Fu},
  \citenamefont {Xu}, \citenamefont {Yu}, \citenamefont {Liu},\ and\
  \citenamefont {Wu}}]{PhysRevB.105.064503}%
  \BibitemOpen
  \bibfield  {author} {\bibinfo {author} {\bibfnamefont {P.-H.}\ \bibnamefont
  {Fu}}, \bibinfo {author} {\bibfnamefont {Y.}~\bibnamefont {Xu}}, \bibinfo
  {author} {\bibfnamefont {X.-L.}\ \bibnamefont {Yu}}, \bibinfo {author}
  {\bibfnamefont {J.-F.}\ \bibnamefont {Liu}},\ and\ \bibinfo {author}
  {\bibfnamefont {J.}~\bibnamefont {Wu}},\ }\bibfield  {title} {\bibinfo
  {title} {{Electrically modulated Josephson junction of light-dressed
  topological insulators}},\ }\href
  {https://doi.org/10.1103/PhysRevB.105.064503} {\bibfield  {journal} {\bibinfo
   {journal} {Phys. Rev. B}\ }\textbf {\bibinfo {volume} {105}},\ \bibinfo
  {pages} {064503} (\bibinfo {year} {2022})}\BibitemShut {NoStop}%
\bibitem [{\citenamefont {Lee}\ \emph {et~al.}(2018)\citenamefont {Lee},
  \citenamefont {Ho}, \citenamefont {Yang}, \citenamefont {Gong},\ and\
  \citenamefont {Papi\ifmmode~\acute{c}\else
  \'{c}\fi{}}}]{PhysRevLett.121.237401}%
  \BibitemOpen
  \bibfield  {author} {\bibinfo {author} {\bibfnamefont {C.~H.}\ \bibnamefont
  {Lee}}, \bibinfo {author} {\bibfnamefont {W.~W.}\ \bibnamefont {Ho}},
  \bibinfo {author} {\bibfnamefont {B.}~\bibnamefont {Yang}}, \bibinfo {author}
  {\bibfnamefont {J.}~\bibnamefont {Gong}},\ and\ \bibinfo {author}
  {\bibfnamefont {Z.}~\bibnamefont {Papi\ifmmode~\acute{c}\else \'{c}\fi{}}},\
  }\bibfield  {title} {\bibinfo {title} {{Floquet Mechanism for Non-Abelian
  Fractional Quantum Hall States}},\ }\href
  {https://doi.org/10.1103/PhysRevLett.121.237401} {\bibfield  {journal}
  {\bibinfo  {journal} {Phys. Rev. Lett.}\ }\textbf {\bibinfo {volume} {121}},\
  \bibinfo {pages} {237401} (\bibinfo {year} {2018})}\BibitemShut {NoStop}%
\bibitem [{\citenamefont {Qin}\ \emph {et~al.}(2023)\citenamefont {Qin},
  \citenamefont {Lee},\ and\ \citenamefont {Chen}}]{PhysRevB.108.075435}%
  \BibitemOpen
  \bibfield  {author} {\bibinfo {author} {\bibfnamefont {F.}~\bibnamefont
  {Qin}}, \bibinfo {author} {\bibfnamefont {C.~H.}\ \bibnamefont {Lee}},\ and\
  \bibinfo {author} {\bibfnamefont {R.}~\bibnamefont {Chen}},\ }\bibfield
  {title} {\bibinfo {title} {{Light-induced half-quantized Hall effect and
  axion insulator}},\ }\href {https://doi.org/10.1103/PhysRevB.108.075435}
  {\bibfield  {journal} {\bibinfo  {journal} {Phys. Rev. B}\ }\textbf {\bibinfo
  {volume} {108}},\ \bibinfo {pages} {075435} (\bibinfo {year}
  {2023})}\BibitemShut {NoStop}%
\bibitem [{\citenamefont {Qin}\ \emph {et~al.}(2024)\citenamefont {Qin},
  \citenamefont {Chen},\ and\ \citenamefont {Lee}}]{qin2024light}%
  \BibitemOpen
  \bibfield  {author} {\bibinfo {author} {\bibfnamefont {F.}~\bibnamefont
  {Qin}}, \bibinfo {author} {\bibfnamefont {R.}~\bibnamefont {Chen}},\ and\
  \bibinfo {author} {\bibfnamefont {C.~H.}\ \bibnamefont {Lee}},\ }\bibfield
  {title} {\bibinfo {title} {{Light-enhanced nonlinear Hall effect}},\ }\href
  {https://doi.org/https://doi.org/10.1038/s42005-024-01820-5} {\bibfield
  {journal} {\bibinfo  {journal} {Commun Phys}\ }\textbf {\bibinfo {volume}
  {7}},\ \bibinfo {pages} {368} (\bibinfo {year} {2024})}\BibitemShut {NoStop}%
\bibitem [{\citenamefont {Li}\ \emph {et~al.}(2018)\citenamefont {Li},
  \citenamefont {Lee},\ and\ \citenamefont {Gong}}]{PhysRevLett.121.036401}%
  \BibitemOpen
  \bibfield  {author} {\bibinfo {author} {\bibfnamefont {L.}~\bibnamefont
  {Li}}, \bibinfo {author} {\bibfnamefont {C.~H.}\ \bibnamefont {Lee}},\ and\
  \bibinfo {author} {\bibfnamefont {J.}~\bibnamefont {Gong}},\ }\bibfield
  {title} {\bibinfo {title} {{Realistic Floquet Semimetal with Exotic
  Topological Linkages between Arbitrarily Many Nodal Loops}},\ }\href
  {https://doi.org/10.1103/PhysRevLett.121.036401} {\bibfield  {journal}
  {\bibinfo  {journal} {Phys. Rev. Lett.}\ }\textbf {\bibinfo {volume} {121}},\
  \bibinfo {pages} {036401} (\bibinfo {year} {2018})}\BibitemShut {NoStop}%
\bibitem [{\citenamefont {Yap}\ \emph {et~al.}(2018)\citenamefont {Yap},
  \citenamefont {Zhou}, \citenamefont {Lee},\ and\ \citenamefont
  {Gong}}]{PhysRevB.97.165142}%
  \BibitemOpen
  \bibfield  {author} {\bibinfo {author} {\bibfnamefont {H.~H.}\ \bibnamefont
  {Yap}}, \bibinfo {author} {\bibfnamefont {L.}~\bibnamefont {Zhou}}, \bibinfo
  {author} {\bibfnamefont {C.~H.}\ \bibnamefont {Lee}},\ and\ \bibinfo {author}
  {\bibfnamefont {J.}~\bibnamefont {Gong}},\ }\bibfield  {title} {\bibinfo
  {title} {{Photoinduced half-integer quantized conductance plateaus in
  topological-insulator/superconductor heterostructures}},\ }\href
  {https://doi.org/10.1103/PhysRevB.97.165142} {\bibfield  {journal} {\bibinfo
  {journal} {Phys. Rev. B}\ }\textbf {\bibinfo {volume} {97}},\ \bibinfo
  {pages} {165142} (\bibinfo {year} {2018})}\BibitemShut {NoStop}%
\bibitem [{\citenamefont {Burkov}(2016)}]{burkov2016topological}%
  \BibitemOpen
  \bibfield  {author} {\bibinfo {author} {\bibfnamefont {A.}~\bibnamefont
  {Burkov}},\ }\bibfield  {title} {\bibinfo {title} {{Topological
  semimetals}},\ }\href {https://doi.org/https://doi.org/10.1038/nmat4788}
  {\bibfield  {journal} {\bibinfo  {journal} {Nature Mater}\ }\textbf {\bibinfo
  {volume} {15}},\ \bibinfo {pages} {1145} (\bibinfo {year}
  {2016})}\BibitemShut {NoStop}%
\bibitem [{\citenamefont {Lee}\ \emph {et~al.}(2024)\citenamefont {Lee},
  \citenamefont {Fu},\ and\ \citenamefont {Ang}}]{PhysRevB.109.235105}%
  \BibitemOpen
  \bibfield  {author} {\bibinfo {author} {\bibfnamefont {K.~W.}\ \bibnamefont
  {Lee}}, \bibinfo {author} {\bibfnamefont {P.-H.}\ \bibnamefont {Fu}},\ and\
  \bibinfo {author} {\bibfnamefont {Y.~S.}\ \bibnamefont {Ang}},\ }\bibfield
  {title} {\bibinfo {title} {{Interplay between Haldane and modified Haldane
  models in $\alpha$-$T_{3}$ lattice: Band structures, phase diagrams, and edge
  states}},\ }\href {https://doi.org/10.1103/PhysRevB.109.235105} {\bibfield
  {journal} {\bibinfo  {journal} {Phys. Rev. B}\ }\textbf {\bibinfo {volume}
  {109}},\ \bibinfo {pages} {235105} (\bibinfo {year} {2024})}\BibitemShut
  {NoStop}%
\bibitem [{\citenamefont {Dey}\ and\ \citenamefont {Ghosh}(2019)}]{Dey_Bashab}%
  \BibitemOpen
  \bibfield  {author} {\bibinfo {author} {\bibfnamefont {B.}~\bibnamefont
  {Dey}}\ and\ \bibinfo {author} {\bibfnamefont {T.~K.}\ \bibnamefont
  {Ghosh}},\ }\bibfield  {title} {\bibinfo {title} {{Floquet topological phase
  transition in the $\alpha$-$T_{3}$ lattice}},\ }\href
  {https://doi.org/10.1103/PhysRevB.99.205429} {\bibfield  {journal} {\bibinfo
  {journal} {Phys. Rev. B}\ }\textbf {\bibinfo {volume} {99}},\ \bibinfo
  {pages} {205429} (\bibinfo {year} {2019})}\BibitemShut {NoStop}%
\bibitem [{\citenamefont {Kitagawa}\ \emph {et~al.}(2011)\citenamefont
  {Kitagawa}, \citenamefont {Oka}, \citenamefont {Brataas}, \citenamefont
  {Fu},\ and\ \citenamefont {Demler}}]{PhysRevB.84.235108}%
  \BibitemOpen
  \bibfield  {author} {\bibinfo {author} {\bibfnamefont {T.}~\bibnamefont
  {Kitagawa}}, \bibinfo {author} {\bibfnamefont {T.}~\bibnamefont {Oka}},
  \bibinfo {author} {\bibfnamefont {A.}~\bibnamefont {Brataas}}, \bibinfo
  {author} {\bibfnamefont {L.}~\bibnamefont {Fu}},\ and\ \bibinfo {author}
  {\bibfnamefont {E.}~\bibnamefont {Demler}},\ }\bibfield  {title} {\bibinfo
  {title} {{Transport properties of nonequilibrium systems under the
  application of light: Photoinduced quantum Hall insulators without Landau
  levels}},\ }\href {https://doi.org/10.1103/PhysRevB.84.235108} {\bibfield
  {journal} {\bibinfo  {journal} {Phys. Rev. B}\ }\textbf {\bibinfo {volume}
  {84}},\ \bibinfo {pages} {235108} (\bibinfo {year} {2011})}\BibitemShut
  {NoStop}%
\bibitem [{\citenamefont {Usaj}\ \emph {et~al.}(2014)\citenamefont {Usaj},
  \citenamefont {Perez-Piskunow}, \citenamefont {Foa~Torres},\ and\
  \citenamefont {Balseiro}}]{PhysRevB.90.115423}%
  \BibitemOpen
  \bibfield  {author} {\bibinfo {author} {\bibfnamefont {G.}~\bibnamefont
  {Usaj}}, \bibinfo {author} {\bibfnamefont {P.~M.}\ \bibnamefont
  {Perez-Piskunow}}, \bibinfo {author} {\bibfnamefont {L.~E.~F.}\ \bibnamefont
  {Foa~Torres}},\ and\ \bibinfo {author} {\bibfnamefont {C.~A.}\ \bibnamefont
  {Balseiro}},\ }\bibfield  {title} {\bibinfo {title} {{Irradiated graphene as
  a tunable Floquet topological insulator}},\ }\href
  {https://doi.org/10.1103/PhysRevB.90.115423} {\bibfield  {journal} {\bibinfo
  {journal} {Phys. Rev. B}\ }\textbf {\bibinfo {volume} {90}},\ \bibinfo
  {pages} {115423} (\bibinfo {year} {2014})}\BibitemShut {NoStop}%
\bibitem [{\citenamefont {McIver}\ \emph {et~al.}(2020)\citenamefont {McIver},
  \citenamefont {Schulte}, \citenamefont {Stein}, \citenamefont {Matsuyama},
  \citenamefont {Jotzu}, \citenamefont {Meier},\ and\ \citenamefont
  {Cavalleri}}]{mciver2020light}%
  \BibitemOpen
  \bibfield  {author} {\bibinfo {author} {\bibfnamefont {J.~W.}\ \bibnamefont
  {McIver}}, \bibinfo {author} {\bibfnamefont {B.}~\bibnamefont {Schulte}},
  \bibinfo {author} {\bibfnamefont {F.-U.}\ \bibnamefont {Stein}}, \bibinfo
  {author} {\bibfnamefont {T.}~\bibnamefont {Matsuyama}}, \bibinfo {author}
  {\bibfnamefont {G.}~\bibnamefont {Jotzu}}, \bibinfo {author} {\bibfnamefont
  {G.}~\bibnamefont {Meier}},\ and\ \bibinfo {author} {\bibfnamefont
  {A.}~\bibnamefont {Cavalleri}},\ }\bibfield  {title} {\bibinfo {title}
  {{Light-induced anomalous Hall effect in graphene}},\ }\href
  {https://doi.org/https://doi.org/10.1038/s41567-019-0698-y} {\bibfield
  {journal} {\bibinfo  {journal} {Nat. Phys.}\ }\textbf {\bibinfo {volume}
  {16}},\ \bibinfo {pages} {38} (\bibinfo {year} {2020})}\BibitemShut {NoStop}%
\bibitem [{\citenamefont {Cheng}\ and\ \citenamefont
  {Xianlong}(2022)}]{PhysRevResearch.4.033194}%
  \BibitemOpen
  \bibfield  {author} {\bibinfo {author} {\bibfnamefont {S.}~\bibnamefont
  {Cheng}}\ and\ \bibinfo {author} {\bibfnamefont {G.}~\bibnamefont
  {Xianlong}},\ }\bibfield  {title} {\bibinfo {title} {{Topological Floquet
  bands in a circularly shaken dice lattice}},\ }\href
  {https://doi.org/10.1103/PhysRevResearch.4.033194} {\bibfield  {journal}
  {\bibinfo  {journal} {Phys. Rev. Res.}\ }\textbf {\bibinfo {volume} {4}},\
  \bibinfo {pages} {033194} (\bibinfo {year} {2022})}\BibitemShut {NoStop}%
\bibitem [{\citenamefont {Cheng}\ \emph {et~al.}(2020)\citenamefont {Cheng},
  \citenamefont {Yin}, \citenamefont {Lu}, \citenamefont {He}, \citenamefont
  {Wang},\ and\ \citenamefont {Xianlong}}]{PhysRevA.101.043620}%
  \BibitemOpen
  \bibfield  {author} {\bibinfo {author} {\bibfnamefont {S.}~\bibnamefont
  {Cheng}}, \bibinfo {author} {\bibfnamefont {H.}~\bibnamefont {Yin}}, \bibinfo
  {author} {\bibfnamefont {Z.}~\bibnamefont {Lu}}, \bibinfo {author}
  {\bibfnamefont {C.}~\bibnamefont {He}}, \bibinfo {author} {\bibfnamefont
  {P.}~\bibnamefont {Wang}},\ and\ \bibinfo {author} {\bibfnamefont
  {G.}~\bibnamefont {Xianlong}},\ }\bibfield  {title} {\bibinfo {title}
  {{Predicting large-Chern-number phases in a shaken optical dice lattice}},\
  }\href {https://doi.org/10.1103/PhysRevA.101.043620} {\bibfield  {journal}
  {\bibinfo  {journal} {Phys. Rev. A}\ }\textbf {\bibinfo {volume} {101}},\
  \bibinfo {pages} {043620} (\bibinfo {year} {2020})}\BibitemShut {NoStop}%
\bibitem [{\citenamefont {Peierls}(1933)}]{peierls1933theorie}%
  \BibitemOpen
  \bibfield  {author} {\bibinfo {author} {\bibfnamefont {R.}~\bibnamefont
  {Peierls}},\ }\bibfield  {title} {\bibinfo {title} {{Zur theorie des
  diamagnetismus von leitungselektronen}},\ }\href
  {https://doi.org/https://doi.org/10.1007/BF01342591} {\bibfield  {journal}
  {\bibinfo  {journal} {Z. Physik}\ }\textbf {\bibinfo {volume} {80}},\
  \bibinfo {pages} {763} (\bibinfo {year} {1933})}\BibitemShut {NoStop}%
\bibitem [{\citenamefont {L\'opez}\ \emph {et~al.}(2015)\citenamefont
  {L\'opez}, \citenamefont {Scholz}, \citenamefont {Santos},\ and\
  \citenamefont {Schliemann}}]{PhysRevB.91.125105}%
  \BibitemOpen
  \bibfield  {author} {\bibinfo {author} {\bibfnamefont {A.}~\bibnamefont
  {L\'opez}}, \bibinfo {author} {\bibfnamefont {A.}~\bibnamefont {Scholz}},
  \bibinfo {author} {\bibfnamefont {B.}~\bibnamefont {Santos}},\ and\ \bibinfo
  {author} {\bibfnamefont {J.}~\bibnamefont {Schliemann}},\ }\bibfield  {title}
  {\bibinfo {title} {{Photoinduced pseudospin effects in silicene beyond the
  off-resonant condition}},\ }\href
  {https://doi.org/10.1103/PhysRevB.91.125105} {\bibfield  {journal} {\bibinfo
  {journal} {Phys. Rev. B}\ }\textbf {\bibinfo {volume} {91}},\ \bibinfo
  {pages} {125105} (\bibinfo {year} {2015})}\BibitemShut {NoStop}%
\bibitem [{\citenamefont {Ho}\ and\ \citenamefont
  {Gong}(2012)}]{PhysRevLett.109.010601}%
  \BibitemOpen
  \bibfield  {author} {\bibinfo {author} {\bibfnamefont {D.~Y.~H.}\
  \bibnamefont {Ho}}\ and\ \bibinfo {author} {\bibfnamefont {J.}~\bibnamefont
  {Gong}},\ }\bibfield  {title} {\bibinfo {title} {{Quantized Adiabatic
  Transport In Momentum Space}},\ }\href
  {https://doi.org/10.1103/PhysRevLett.109.010601} {\bibfield  {journal}
  {\bibinfo  {journal} {Phys. Rev. Lett.}\ }\textbf {\bibinfo {volume} {109}},\
  \bibinfo {pages} {010601} (\bibinfo {year} {2012})}\BibitemShut {NoStop}%
\bibitem [{\citenamefont {Chen}\ \emph {et~al.}(2018)\citenamefont {Chen},
  \citenamefont {Du},\ and\ \citenamefont {Fiete}}]{PhysRevB.97.035422}%
  \BibitemOpen
  \bibfield  {author} {\bibinfo {author} {\bibfnamefont {Q.}~\bibnamefont
  {Chen}}, \bibinfo {author} {\bibfnamefont {L.}~\bibnamefont {Du}},\ and\
  \bibinfo {author} {\bibfnamefont {G.~A.}\ \bibnamefont {Fiete}},\ }\bibfield
  {title} {\bibinfo {title} {{Floquet band structure of a semi-Dirac system}},\
  }\href {https://doi.org/10.1103/PhysRevB.97.035422} {\bibfield  {journal}
  {\bibinfo  {journal} {Phys. Rev. B}\ }\textbf {\bibinfo {volume} {97}},\
  \bibinfo {pages} {035422} (\bibinfo {year} {2018})}\BibitemShut {NoStop}%
\bibitem [{\citenamefont {Xiao}\ \emph {et~al.}(2010)\citenamefont {Xiao},
  \citenamefont {Chang},\ and\ \citenamefont {Niu}}]{RevModPhys.82.1959}%
  \BibitemOpen
  \bibfield  {author} {\bibinfo {author} {\bibfnamefont {D.}~\bibnamefont
  {Xiao}}, \bibinfo {author} {\bibfnamefont {M.-C.}\ \bibnamefont {Chang}},\
  and\ \bibinfo {author} {\bibfnamefont {Q.}~\bibnamefont {Niu}},\ }\bibfield
  {title} {\bibinfo {title} {{Berry phase effects on electronic properties}},\
  }\href {https://doi.org/10.1103/RevModPhys.82.1959} {\bibfield  {journal}
  {\bibinfo  {journal} {Rev. Mod. Phys.}\ }\textbf {\bibinfo {volume} {82}},\
  \bibinfo {pages} {1959} (\bibinfo {year} {2010})}\BibitemShut {NoStop}%
\bibitem [{\citenamefont {Matusalem}\ \emph {et~al.}(2019)\citenamefont
  {Matusalem}, \citenamefont {Marques}, \citenamefont {Teles}, \citenamefont
  {Matthes}, \citenamefont {Furthm\"uller},\ and\ \citenamefont
  {Bechstedt}}]{PhysRevB.100.245430}%
  \BibitemOpen
  \bibfield  {author} {\bibinfo {author} {\bibfnamefont {F.}~\bibnamefont
  {Matusalem}}, \bibinfo {author} {\bibfnamefont {M.}~\bibnamefont {Marques}},
  \bibinfo {author} {\bibfnamefont {L.~K.}\ \bibnamefont {Teles}}, \bibinfo
  {author} {\bibfnamefont {L.}~\bibnamefont {Matthes}}, \bibinfo {author}
  {\bibfnamefont {J.}~\bibnamefont {Furthm\"uller}},\ and\ \bibinfo {author}
  {\bibfnamefont {F.}~\bibnamefont {Bechstedt}},\ }\bibfield  {title} {\bibinfo
  {title} {{Quantization of spin Hall conductivity in two-dimensional
  topological insulators versus symmetry and spin-orbit interaction}},\ }\href
  {https://doi.org/10.1103/PhysRevB.100.245430} {\bibfield  {journal} {\bibinfo
   {journal} {Phys. Rev. B}\ }\textbf {\bibinfo {volume} {100}},\ \bibinfo
  {pages} {245430} (\bibinfo {year} {2019})}\BibitemShut {NoStop}%
\bibitem [{\citenamefont {Matthes}\ \emph {et~al.}(2016)\citenamefont
  {Matthes}, \citenamefont {K\"ufner}, \citenamefont {Furthm\"uller},\ and\
  \citenamefont {Bechstedt}}]{PhysRevB.94.085410}%
  \BibitemOpen
  \bibfield  {author} {\bibinfo {author} {\bibfnamefont {L.}~\bibnamefont
  {Matthes}}, \bibinfo {author} {\bibfnamefont {S.}~\bibnamefont {K\"ufner}},
  \bibinfo {author} {\bibfnamefont {J.}~\bibnamefont {Furthm\"uller}},\ and\
  \bibinfo {author} {\bibfnamefont {F.}~\bibnamefont {Bechstedt}},\ }\bibfield
  {title} {\bibinfo {title} {{Intrinsic spin Hall conductivity in one-, two-,
  and three-dimensional trivial and topological systems}},\ }\href
  {https://doi.org/10.1103/PhysRevB.94.085410} {\bibfield  {journal} {\bibinfo
  {journal} {Phys. Rev. B}\ }\textbf {\bibinfo {volume} {94}},\ \bibinfo
  {pages} {085410} (\bibinfo {year} {2016})}\BibitemShut {NoStop}%
\bibitem [{\citenamefont {Duan}\ \emph {et~al.}(2023)\citenamefont {Duan},
  \citenamefont {Wu}, \citenamefont {Deng}, \citenamefont {Wang},\ and\
  \citenamefont {Yang}}]{PhysRevB.107.165147}%
  \BibitemOpen
  \bibfield  {author} {\bibinfo {author} {\bibfnamefont {H.-J.}\ \bibnamefont
  {Duan}}, \bibinfo {author} {\bibfnamefont {Y.-J.}\ \bibnamefont {Wu}},
  \bibinfo {author} {\bibfnamefont {M.-X.}\ \bibnamefont {Deng}}, \bibinfo
  {author} {\bibfnamefont {R.-Q.}\ \bibnamefont {Wang}},\ and\ \bibinfo
  {author} {\bibfnamefont {M.}~\bibnamefont {Yang}},\ }\bibfield  {title}
  {\bibinfo {title} {{Indirect magnetic signals in Weyl semimetals mediated by
  a single Fermi arc}},\ }\href {https://doi.org/10.1103/PhysRevB.107.165147}
  {\bibfield  {journal} {\bibinfo  {journal} {Phys. Rev. B}\ }\textbf {\bibinfo
  {volume} {107}},\ \bibinfo {pages} {165147} (\bibinfo {year}
  {2023})}\BibitemShut {NoStop}%
\bibitem [{\citenamefont {Duan}\ \emph {et~al.}(2024)\citenamefont {Duan},
  \citenamefont {Cai}, \citenamefont {Wei}, \citenamefont {Chen}, \citenamefont
  {Wu}, \citenamefont {Deng}, \citenamefont {Wang},\ and\ \citenamefont
  {Yang}}]{PhysRevB.109.205149}%
  \BibitemOpen
  \bibfield  {author} {\bibinfo {author} {\bibfnamefont {H.-J.}\ \bibnamefont
  {Duan}}, \bibinfo {author} {\bibfnamefont {S.-M.}\ \bibnamefont {Cai}},
  \bibinfo {author} {\bibfnamefont {X.}~\bibnamefont {Wei}}, \bibinfo {author}
  {\bibfnamefont {Y.-C.}\ \bibnamefont {Chen}}, \bibinfo {author}
  {\bibfnamefont {Y.-J.}\ \bibnamefont {Wu}}, \bibinfo {author} {\bibfnamefont
  {M.-X.}\ \bibnamefont {Deng}}, \bibinfo {author} {\bibfnamefont
  {R.}~\bibnamefont {Wang}},\ and\ \bibinfo {author} {\bibfnamefont
  {M.}~\bibnamefont {Yang}},\ }\bibfield  {title} {\bibinfo {title} {{RKKY
  signals characterizing the topological phase transitions in Floquet Dirac
  semimetals}},\ }\href {https://doi.org/10.1103/PhysRevB.109.205149}
  {\bibfield  {journal} {\bibinfo  {journal} {Phys. Rev. B}\ }\textbf {\bibinfo
  {volume} {109}},\ \bibinfo {pages} {205149} (\bibinfo {year}
  {2024})}\BibitemShut {NoStop}%
\bibitem [{\citenamefont {Hatsugai}(1993{\natexlab{b}})}]{Hatsugai_2}%
  \BibitemOpen
  \bibfield  {author} {\bibinfo {author} {\bibfnamefont {Y.}~\bibnamefont
  {Hatsugai}},\ }\bibfield  {title} {\bibinfo {title} {{Edge states in the
  integer quantum Hall effect and the Riemann surface of the Bloch function}},\
  }\href {https://doi.org/10.1103/PhysRevB.48.11851} {\bibfield  {journal}
  {\bibinfo  {journal} {Phys. Rev. B}\ }\textbf {\bibinfo {volume} {48}},\
  \bibinfo {pages} {11851} (\bibinfo {year} {1993}{\natexlab{b}})}\BibitemShut
  {NoStop}%
\bibitem [{\citenamefont {Halperin}(1982)}]{Halperin}%
  \BibitemOpen
  \bibfield  {author} {\bibinfo {author} {\bibfnamefont {B.~I.}\ \bibnamefont
  {Halperin}},\ }\bibfield  {title} {\bibinfo {title} {{Quantized Hall
  conductance, current-carrying edge states, and the existence of extended
  states in a two-dimensional disordered potential}},\ }\href
  {https://doi.org/10.1103/PhysRevB.25.2185} {\bibfield  {journal} {\bibinfo
  {journal} {Phys. Rev. B}\ }\textbf {\bibinfo {volume} {25}},\ \bibinfo
  {pages} {2185} (\bibinfo {year} {1982})}\BibitemShut {NoStop}%
\bibitem [{\citenamefont {Traverso}\ \emph {et~al.}(2024)\citenamefont
  {Traverso}, \citenamefont {Sassetti},\ and\ \citenamefont
  {Traverso~Ziani}}]{traverso2024emerging}%
  \BibitemOpen
  \bibfield  {author} {\bibinfo {author} {\bibfnamefont {S.}~\bibnamefont
  {Traverso}}, \bibinfo {author} {\bibfnamefont {M.}~\bibnamefont {Sassetti}},\
  and\ \bibinfo {author} {\bibfnamefont {N.}~\bibnamefont {Traverso~Ziani}},\
  }\bibfield  {title} {\bibinfo {title} {{Emerging topological bound states in
  Haldane model zigzag nanoribbons}},\ }\href
  {https://doi.org/https://doi.org/10.1038/s41535-023-00615-1} {\bibfield
  {journal} {\bibinfo  {journal} {npj Quantum Mater.}\ }\textbf {\bibinfo
  {volume} {9}},\ \bibinfo {pages} {9} (\bibinfo {year} {2024})}\BibitemShut
  {NoStop}%
\bibitem [{\citenamefont {Fujita}\ \emph {et~al.}(1996)\citenamefont {Fujita},
  \citenamefont {Wakabayashi}, \citenamefont {Nakada},\ and\ \citenamefont
  {Kusakabe}}]{Fujita}%
  \BibitemOpen
  \bibfield  {author} {\bibinfo {author} {\bibfnamefont {M.}~\bibnamefont
  {Fujita}}, \bibinfo {author} {\bibfnamefont {K.}~\bibnamefont {Wakabayashi}},
  \bibinfo {author} {\bibfnamefont {K.}~\bibnamefont {Nakada}},\ and\ \bibinfo
  {author} {\bibfnamefont {K.}~\bibnamefont {Kusakabe}},\ }\bibfield  {title}
  {\bibinfo {title} {{Peculiar localized state at zigzag graphite edge}},\
  }\href {https://doi.org/https://doi.org/10.1143/JPSJ.65.1920} {\bibfield
  {journal} {\bibinfo  {journal} {J. Phys. Soc. Jpn.}\ }\textbf {\bibinfo
  {volume} {65}},\ \bibinfo {pages} {1920} (\bibinfo {year}
  {1996})}\BibitemShut {NoStop}%
\bibitem [{\citenamefont {Nakada}\ \emph {et~al.}(1996)\citenamefont {Nakada},
  \citenamefont {Fujita}, \citenamefont {Dresselhaus},\ and\ \citenamefont
  {Dresselhaus}}]{Nakada}%
  \BibitemOpen
  \bibfield  {author} {\bibinfo {author} {\bibfnamefont {K.}~\bibnamefont
  {Nakada}}, \bibinfo {author} {\bibfnamefont {M.}~\bibnamefont {Fujita}},
  \bibinfo {author} {\bibfnamefont {G.}~\bibnamefont {Dresselhaus}},\ and\
  \bibinfo {author} {\bibfnamefont {M.~S.}\ \bibnamefont {Dresselhaus}},\
  }\bibfield  {title} {\bibinfo {title} {{Edge state in graphene ribbons:
  Nanometer size effect and edge shape dependence}},\ }\href
  {https://doi.org/10.1103/PhysRevB.54.17954} {\bibfield  {journal} {\bibinfo
  {journal} {Phys. Rev. B}\ }\textbf {\bibinfo {volume} {54}},\ \bibinfo
  {pages} {17954} (\bibinfo {year} {1996})}\BibitemShut {NoStop}%
\bibitem [{\citenamefont {Ezawa}(2013)}]{PhysRevLett.110.026603}%
  \BibitemOpen
  \bibfield  {author} {\bibinfo {author} {\bibfnamefont {M.}~\bibnamefont
  {Ezawa}},\ }\bibfield  {title} {\bibinfo {title} {{Photoinduced Topological
  Phase Transition and a Single Dirac-Cone State in Silicene}},\ }\href
  {https://doi.org/10.1103/PhysRevLett.110.026603} {\bibfield  {journal}
  {\bibinfo  {journal} {Phys. Rev. Lett.}\ }\textbf {\bibinfo {volume} {110}},\
  \bibinfo {pages} {026603} (\bibinfo {year} {2013})}\BibitemShut {NoStop}%
\bibitem [{\citenamefont {Jozwiak}\ \emph {et~al.}(2016)\citenamefont
  {Jozwiak}, \citenamefont {Sobota}, \citenamefont {Gotlieb}, \citenamefont
  {Kemper}, \citenamefont {Rotundu}, \citenamefont {Birgeneau}, \citenamefont
  {Hussain}, \citenamefont {Lee}, \citenamefont {Shen},\ and\ \citenamefont
  {Lanzara}}]{jozwiak2016spin}%
  \BibitemOpen
  \bibfield  {author} {\bibinfo {author} {\bibfnamefont {C.}~\bibnamefont
  {Jozwiak}}, \bibinfo {author} {\bibfnamefont {J.~A.}\ \bibnamefont {Sobota}},
  \bibinfo {author} {\bibfnamefont {K.}~\bibnamefont {Gotlieb}}, \bibinfo
  {author} {\bibfnamefont {A.~F.}\ \bibnamefont {Kemper}}, \bibinfo {author}
  {\bibfnamefont {C.~R.}\ \bibnamefont {Rotundu}}, \bibinfo {author}
  {\bibfnamefont {R.~J.}\ \bibnamefont {Birgeneau}}, \bibinfo {author}
  {\bibfnamefont {Z.}~\bibnamefont {Hussain}}, \bibinfo {author} {\bibfnamefont
  {D.-H.}\ \bibnamefont {Lee}}, \bibinfo {author} {\bibfnamefont {Z.-X.}\
  \bibnamefont {Shen}},\ and\ \bibinfo {author} {\bibfnamefont
  {A.}~\bibnamefont {Lanzara}},\ }\bibfield  {title} {\bibinfo {title}
  {{Spin-polarized surface resonances accompanying topological surface state
  formation}},\ }\href {https://doi.org/https://doi.org/10.1038/ncomms13143}
  {\bibfield  {journal} {\bibinfo  {journal} {Nat Commun}\ }\textbf {\bibinfo
  {volume} {7}},\ \bibinfo {pages} {13143} (\bibinfo {year}
  {2016})}\BibitemShut {NoStop}%
\bibitem [{\citenamefont {Santos}\ and\ \citenamefont
  {Moodera}(2004)}]{PhysRevB.69.241203}%
  \BibitemOpen
  \bibfield  {author} {\bibinfo {author} {\bibfnamefont {T.~S.}\ \bibnamefont
  {Santos}}\ and\ \bibinfo {author} {\bibfnamefont {J.~S.}\ \bibnamefont
  {Moodera}},\ }\bibfield  {title} {\bibinfo {title} {{Observation of spin
  filtering with a ferromagnetic $\mathrm{EuO}$ tunnel barrier}},\ }\href
  {https://doi.org/10.1103/PhysRevB.69.241203} {\bibfield  {journal} {\bibinfo
  {journal} {Phys. Rev. B}\ }\textbf {\bibinfo {volume} {69}},\ \bibinfo
  {pages} {241203} (\bibinfo {year} {2004})}\BibitemShut {NoStop}%
\bibitem [{\citenamefont {Li}\ \emph {et~al.}(2020)\citenamefont {Li},
  \citenamefont {Wang}, \citenamefont {Li}, \citenamefont {Zheng},
  \citenamefont {Brinkman}, \citenamefont {Yu},\ and\ \citenamefont
  {Liao}}]{PhysRevLett.124.156601}%
  \BibitemOpen
  \bibfield  {author} {\bibinfo {author} {\bibfnamefont {C.-Z.}\ \bibnamefont
  {Li}}, \bibinfo {author} {\bibfnamefont {A.-Q.}\ \bibnamefont {Wang}},
  \bibinfo {author} {\bibfnamefont {C.}~\bibnamefont {Li}}, \bibinfo {author}
  {\bibfnamefont {W.-Z.}\ \bibnamefont {Zheng}}, \bibinfo {author}
  {\bibfnamefont {A.}~\bibnamefont {Brinkman}}, \bibinfo {author}
  {\bibfnamefont {D.-P.}\ \bibnamefont {Yu}},\ and\ \bibinfo {author}
  {\bibfnamefont {Z.-M.}\ \bibnamefont {Liao}},\ }\bibfield  {title} {\bibinfo
  {title} {{Reducing Electronic Transport Dimension to Topological Hinge States
  by Increasing Geometry Size of Dirac Semimetal Josephson Junctions}},\ }\href
  {https://doi.org/10.1103/PhysRevLett.124.156601} {\bibfield  {journal}
  {\bibinfo  {journal} {Phys. Rev. Lett.}\ }\textbf {\bibinfo {volume} {124}},\
  \bibinfo {pages} {156601} (\bibinfo {year} {2020})}\BibitemShut {NoStop}%
\bibitem [{\citenamefont {{\v{Z}}elezn{\`y}}\ \emph {et~al.}(2018)\citenamefont
  {{\v{Z}}elezn{\`y}}, \citenamefont {Wadley}, \citenamefont {Olejn{\'\i}k},
  \citenamefont {Hoffmann},\ and\ \citenamefont {Ohno}}]{vzelezny2018spin}%
  \BibitemOpen
  \bibfield  {author} {\bibinfo {author} {\bibfnamefont {J.}~\bibnamefont
  {{\v{Z}}elezn{\`y}}}, \bibinfo {author} {\bibfnamefont {P.}~\bibnamefont
  {Wadley}}, \bibinfo {author} {\bibfnamefont {K.}~\bibnamefont
  {Olejn{\'\i}k}}, \bibinfo {author} {\bibfnamefont {A.}~\bibnamefont
  {Hoffmann}},\ and\ \bibinfo {author} {\bibfnamefont {H.}~\bibnamefont
  {Ohno}},\ }\bibfield  {title} {\bibinfo {title} {{Spin transport and spin
  torque in antiferromagnetic devices}},\ }\href
  {https://doi.org/https://doi.org/10.1038/s41567-018-0062-7} {\bibfield
  {journal} {\bibinfo  {journal} {Nature Phys}\ }\textbf {\bibinfo {volume}
  {14}},\ \bibinfo {pages} {220} (\bibinfo {year} {2018})}\BibitemShut
  {NoStop}%
\bibitem [{\citenamefont {Saffarzadeh}\ and\ \citenamefont
  {Farghadan}(2011)}]{10.1063/1.3537965}%
  \BibitemOpen
  \bibfield  {author} {\bibinfo {author} {\bibfnamefont {A.}~\bibnamefont
  {Saffarzadeh}}\ and\ \bibinfo {author} {\bibfnamefont {R.}~\bibnamefont
  {Farghadan}},\ }\bibfield  {title} {\bibinfo {title} {{A spin-filter device
  based on armchair graphene nanoribbons}},\ }\href
  {https://doi.org/10.1063/1.3537965} {\bibfield  {journal} {\bibinfo
  {journal} {Appl. Phys. Lett.}\ }\textbf {\bibinfo {volume} {98}},\ \bibinfo
  {pages} {023106} (\bibinfo {year} {2011})}\BibitemShut {NoStop}%
\bibitem [{\citenamefont {Wunderlich}\ \emph {et~al.}(2010)\citenamefont
  {Wunderlich}, \citenamefont {Park}, \citenamefont {Irvine}, \citenamefont
  {Zârbo}, \citenamefont {Rozkotová}, \citenamefont {Nemec}, \citenamefont
  {Novák}, \citenamefont {Sinova},\ and\ \citenamefont
  {Jungwirth}}]{doi:10.1126/science.1195816}%
  \BibitemOpen
  \bibfield  {author} {\bibinfo {author} {\bibfnamefont {J.}~\bibnamefont
  {Wunderlich}}, \bibinfo {author} {\bibfnamefont {B.-G.}\ \bibnamefont
  {Park}}, \bibinfo {author} {\bibfnamefont {A.~C.}\ \bibnamefont {Irvine}},
  \bibinfo {author} {\bibfnamefont {L.~P.}\ \bibnamefont {Zârbo}}, \bibinfo
  {author} {\bibfnamefont {E.}~\bibnamefont {Rozkotová}}, \bibinfo {author}
  {\bibfnamefont {P.}~\bibnamefont {Nemec}}, \bibinfo {author} {\bibfnamefont
  {V.}~\bibnamefont {Novák}}, \bibinfo {author} {\bibfnamefont
  {J.}~\bibnamefont {Sinova}},\ and\ \bibinfo {author} {\bibfnamefont
  {T.}~\bibnamefont {Jungwirth}},\ }\bibfield  {title} {\bibinfo {title} {{Spin
  Hall Effect Transistor}},\ }\href {https://doi.org/10.1126/science.1195816}
  {\bibfield  {journal} {\bibinfo  {journal} {Science}\ }\textbf {\bibinfo
  {volume} {330}},\ \bibinfo {pages} {1801} (\bibinfo {year}
  {2010})}\BibitemShut {NoStop}%
\bibitem [{\citenamefont {Zhao}\ \emph {et~al.}(2023)\citenamefont {Zhao},
  \citenamefont {Ngaloy}, \citenamefont {Ghosh}, \citenamefont {Ershadrad},
  \citenamefont {Gupta}, \citenamefont {Ali}, \citenamefont {Hoque},
  \citenamefont {Karpiak}, \citenamefont {Khokhriakov}, \citenamefont {Polley},
  \citenamefont {Thiagarajan}, \citenamefont {Kalaboukhov}, \citenamefont
  {Svedlindh}, \citenamefont {Sanyal},\ and\ \citenamefont
  {Dash}}]{https://doi.org/10.1002/adma.202209113}%
  \BibitemOpen
  \bibfield  {author} {\bibinfo {author} {\bibfnamefont {B.}~\bibnamefont
  {Zhao}}, \bibinfo {author} {\bibfnamefont {R.}~\bibnamefont {Ngaloy}},
  \bibinfo {author} {\bibfnamefont {S.}~\bibnamefont {Ghosh}}, \bibinfo
  {author} {\bibfnamefont {S.}~\bibnamefont {Ershadrad}}, \bibinfo {author}
  {\bibfnamefont {R.}~\bibnamefont {Gupta}}, \bibinfo {author} {\bibfnamefont
  {K.}~\bibnamefont {Ali}}, \bibinfo {author} {\bibfnamefont {A.~M.}\
  \bibnamefont {Hoque}}, \bibinfo {author} {\bibfnamefont {B.}~\bibnamefont
  {Karpiak}}, \bibinfo {author} {\bibfnamefont {D.}~\bibnamefont
  {Khokhriakov}}, \bibinfo {author} {\bibfnamefont {C.}~\bibnamefont {Polley}},
  \bibinfo {author} {\bibfnamefont {B.}~\bibnamefont {Thiagarajan}}, \bibinfo
  {author} {\bibfnamefont {A.}~\bibnamefont {Kalaboukhov}}, \bibinfo {author}
  {\bibfnamefont {P.}~\bibnamefont {Svedlindh}}, \bibinfo {author}
  {\bibfnamefont {B.}~\bibnamefont {Sanyal}},\ and\ \bibinfo {author}
  {\bibfnamefont {S.~P.}\ \bibnamefont {Dash}},\ }\bibfield  {title} {\bibinfo
  {title} {{A Room-Temperature Spin-Valve with van der Waals Ferromagnet
  Fe5GeTe2/Graphene Heterostructure}},\ }\href
  {https://doi.org/https://doi.org/10.1002/adma.202209113} {\bibfield
  {journal} {\bibinfo  {journal} {Adv. Mater.}\ }\textbf {\bibinfo {volume}
  {35}},\ \bibinfo {pages} {2209113} (\bibinfo {year} {2023})}\BibitemShut
  {NoStop}%
\bibitem [{\citenamefont {kun Wang}\ \emph {et~al.}(2020)\citenamefont {kun
  Wang}, \citenamefont {Li}, \citenamefont {Duan},\ and\ \citenamefont
  {Wang}}]{WANG2020126429}%
  \BibitemOpen
  \bibfield  {author} {\bibinfo {author} {\bibfnamefont {J.}~\bibnamefont {kun
  Wang}}, \bibinfo {author} {\bibfnamefont {J.-Y.}\ \bibnamefont {Li}},
  \bibinfo {author} {\bibfnamefont {H.-J.}\ \bibnamefont {Duan}},\ and\
  \bibinfo {author} {\bibfnamefont {R.-Q.}\ \bibnamefont {Wang}},\ }\bibfield
  {title} {\bibinfo {title} {{Photoinduced anisotropic edge states and
  signatures of nonequilibrium spin polarization in silicene nanoribbons}},\
  }\href {https://doi.org/https://doi.org/10.1016/j.physleta.2020.126429}
  {\bibfield  {journal} {\bibinfo  {journal} {Phys. Lett. A}\ }\textbf
  {\bibinfo {volume} {384}},\ \bibinfo {pages} {126429} (\bibinfo {year}
  {2020})}\BibitemShut {NoStop}%
\bibitem [{\citenamefont {Oriekhov}\ and\ \citenamefont
  {Gusynin}(2020)}]{PhysRevB.101.235162}%
  \BibitemOpen
  \bibfield  {author} {\bibinfo {author} {\bibfnamefont {D.~O.}\ \bibnamefont
  {Oriekhov}}\ and\ \bibinfo {author} {\bibfnamefont {V.~P.}\ \bibnamefont
  {Gusynin}},\ }\bibfield  {title} {\bibinfo {title} {{RKKY interaction in a
  doped pseudospin-1 fermion system at finite temperature}},\ }\href
  {https://doi.org/10.1103/PhysRevB.101.235162} {\bibfield  {journal} {\bibinfo
   {journal} {Phys. Rev. B}\ }\textbf {\bibinfo {volume} {101}},\ \bibinfo
  {pages} {235162} (\bibinfo {year} {2020})}\BibitemShut {NoStop}%
\bibitem [{\citenamefont {Duan}\ \emph {et~al.}(2022)\citenamefont {Duan},
  \citenamefont {Wu}, \citenamefont {Yang}, \citenamefont {Zheng},
  \citenamefont {Zhu}, \citenamefont {Deng}, \citenamefont {Yang},\ and\
  \citenamefont {Wang}}]{Duan_2022}%
  \BibitemOpen
  \bibfield  {author} {\bibinfo {author} {\bibfnamefont {H.-J.}\ \bibnamefont
  {Duan}}, \bibinfo {author} {\bibfnamefont {Y.-J.}\ \bibnamefont {Wu}},
  \bibinfo {author} {\bibfnamefont {Y.-Y.}\ \bibnamefont {Yang}}, \bibinfo
  {author} {\bibfnamefont {S.-H.}\ \bibnamefont {Zheng}}, \bibinfo {author}
  {\bibfnamefont {C.-Y.}\ \bibnamefont {Zhu}}, \bibinfo {author} {\bibfnamefont
  {M.-X.}\ \bibnamefont {Deng}}, \bibinfo {author} {\bibfnamefont
  {M.}~\bibnamefont {Yang}},\ and\ \bibinfo {author} {\bibfnamefont {R.-Q.}\
  \bibnamefont {Wang}},\ }\bibfield  {title} {\bibinfo {title} {{The prolonged
  decay of RKKY interactions by interplay of relativistic and non-relativistic
  electrons in semi-Dirac semimetals}},\ }\href
  {https://doi.org/10.1088/1367-2630/ac5842} {\bibfield  {journal} {\bibinfo
  {journal} {New J. Phys.}\ }\textbf {\bibinfo {volume} {24}},\ \bibinfo
  {pages} {033029} (\bibinfo {year} {2022})}\BibitemShut {NoStop}%
\bibitem [{\citenamefont {Ye}\ \emph {et~al.}(2020)\citenamefont {Ye},
  \citenamefont {Ke}, \citenamefont {Du}, \citenamefont {Guo},\ and\
  \citenamefont {L{\"u}}}]{ye2020quantum}%
  \BibitemOpen
  \bibfield  {author} {\bibinfo {author} {\bibfnamefont {X.}~\bibnamefont
  {Ye}}, \bibinfo {author} {\bibfnamefont {S.-S.}\ \bibnamefont {Ke}}, \bibinfo
  {author} {\bibfnamefont {X.-W.}\ \bibnamefont {Du}}, \bibinfo {author}
  {\bibfnamefont {Y.}~\bibnamefont {Guo}},\ and\ \bibinfo {author}
  {\bibfnamefont {H.-F.}\ \bibnamefont {L{\"u}}},\ }\bibfield  {title}
  {\bibinfo {title} {{Quantum tunneling in the $\alpha$-$T_{3}$ model with an
  effective mass term}},\ }\href
  {https://doi.org/https://doi.org/10.1007/s10909-020-02440-3} {\bibfield
  {journal} {\bibinfo  {journal} {J Low Temp Phys}\ }\textbf {\bibinfo {volume}
  {199}},\ \bibinfo {pages} {1332} (\bibinfo {year} {2020})}\BibitemShut
  {NoStop}%
\bibitem [{\citenamefont {Romh{\'a}nyi}\ \emph {et~al.}(2015)\citenamefont
  {Romh{\'a}nyi}, \citenamefont {Penc},\ and\ \citenamefont
  {Ganesh}}]{romhanyi2015hall}%
  \BibitemOpen
  \bibfield  {author} {\bibinfo {author} {\bibfnamefont {J.}~\bibnamefont
  {Romh{\'a}nyi}}, \bibinfo {author} {\bibfnamefont {K.}~\bibnamefont {Penc}},\
  and\ \bibinfo {author} {\bibfnamefont {R.}~\bibnamefont {Ganesh}},\
  }\bibfield  {title} {\bibinfo {title} {{Hall effect of triplons in a
  dimerized quantum magnet}},\ }\href
  {https://doi.org/https://doi.org/10.1038/ncomms7805} {\bibfield  {journal}
  {\bibinfo  {journal} {Nat Commun}\ }\textbf {\bibinfo {volume} {6}},\
  \bibinfo {pages} {6805} (\bibinfo {year} {2015})}\BibitemShut {NoStop}%
\bibitem [{\citenamefont {Rizzi}\ \emph {et~al.}(2006)\citenamefont {Rizzi},
  \citenamefont {Cataudella},\ and\ \citenamefont {Fazio}}]{Rizzi_Matteo}%
  \BibitemOpen
  \bibfield  {author} {\bibinfo {author} {\bibfnamefont {M.}~\bibnamefont
  {Rizzi}}, \bibinfo {author} {\bibfnamefont {V.}~\bibnamefont {Cataudella}},\
  and\ \bibinfo {author} {\bibfnamefont {R.}~\bibnamefont {Fazio}},\ }\bibfield
   {title} {\bibinfo {title} {{Phase diagram of the Bose-Hubbard model with
  ${\mathcal{T}}_{3}$ symmetry}},\ }\href
  {https://doi.org/10.1103/PhysRevB.73.144511} {\bibfield  {journal} {\bibinfo
  {journal} {Phys. Rev. B}\ }\textbf {\bibinfo {volume} {73}},\ \bibinfo
  {pages} {144511} (\bibinfo {year} {2006})}\BibitemShut {NoStop}%
\bibitem [{\citenamefont {Wang}\ and\ \citenamefont
  {Ran}(2011)}]{PhysRevB.84.241103}%
  \BibitemOpen
  \bibfield  {author} {\bibinfo {author} {\bibfnamefont {F.}~\bibnamefont
  {Wang}}\ and\ \bibinfo {author} {\bibfnamefont {Y.}~\bibnamefont {Ran}},\
  }\bibfield  {title} {\bibinfo {title} {{Nearly flat band with Chern number
  $C=2$ on the dice lattice}},\ }\href
  {https://doi.org/10.1103/PhysRevB.84.241103} {\bibfield  {journal} {\bibinfo
  {journal} {Phys. Rev. B}\ }\textbf {\bibinfo {volume} {84}},\ \bibinfo
  {pages} {241103} (\bibinfo {year} {2011})}\BibitemShut {NoStop}%
\bibitem [{\citenamefont {Wang}\ \emph
  {et~al.}(2012{\natexlab{b}})\citenamefont {Wang}, \citenamefont {Yao},
  \citenamefont {Gong},\ and\ \citenamefont {Sheng}}]{PhysRevB.86.201101}%
  \BibitemOpen
  \bibfield  {author} {\bibinfo {author} {\bibfnamefont {Y.-F.}\ \bibnamefont
  {Wang}}, \bibinfo {author} {\bibfnamefont {H.}~\bibnamefont {Yao}}, \bibinfo
  {author} {\bibfnamefont {C.-D.}\ \bibnamefont {Gong}},\ and\ \bibinfo
  {author} {\bibfnamefont {D.~N.}\ \bibnamefont {Sheng}},\ }\bibfield  {title}
  {\bibinfo {title} {{Fractional quantum Hall effect in topological flat bands
  with Chern number two}},\ }\href {https://doi.org/10.1103/PhysRevB.86.201101}
  {\bibfield  {journal} {\bibinfo  {journal} {Phys. Rev. B}\ }\textbf {\bibinfo
  {volume} {86}},\ \bibinfo {pages} {201101} (\bibinfo {year}
  {2012}{\natexlab{b}})}\BibitemShut {NoStop}%
\bibitem [{\citenamefont {Al~Ezzi}\ \emph {et~al.}(2024)\citenamefont
  {Al~Ezzi}, \citenamefont {Hu}, \citenamefont {Ariando}, \citenamefont
  {Guinea},\ and\ \citenamefont {Adam}}]{PhysRevLett.132.126401}%
  \BibitemOpen
  \bibfield  {author} {\bibinfo {author} {\bibfnamefont {M.~M.}\ \bibnamefont
  {Al~Ezzi}}, \bibinfo {author} {\bibfnamefont {J.}~\bibnamefont {Hu}},
  \bibinfo {author} {\bibfnamefont {A.}~\bibnamefont {Ariando}}, \bibinfo
  {author} {\bibfnamefont {F.}~\bibnamefont {Guinea}},\ and\ \bibinfo {author}
  {\bibfnamefont {S.}~\bibnamefont {Adam}},\ }\bibfield  {title} {\bibinfo
  {title} {{Topological Flat Bands in Graphene Super-Moir\'e Lattices}},\
  }\href {https://doi.org/10.1103/PhysRevLett.132.126401} {\bibfield  {journal}
  {\bibinfo  {journal} {Phys. Rev. Lett.}\ }\textbf {\bibinfo {volume} {132}},\
  \bibinfo {pages} {126401} (\bibinfo {year} {2024})}\BibitemShut {NoStop}%
\bibitem [{\citenamefont {Xia}\ \emph {et~al.}(2023)\citenamefont {Xia},
  \citenamefont {Liang}, \citenamefont {Tang}, \citenamefont {Song},
  \citenamefont {Xu},\ and\ \citenamefont {Chen}}]{PhysRevLett.131.013804}%
  \BibitemOpen
  \bibfield  {author} {\bibinfo {author} {\bibfnamefont {S.}~\bibnamefont
  {Xia}}, \bibinfo {author} {\bibfnamefont {Y.}~\bibnamefont {Liang}}, \bibinfo
  {author} {\bibfnamefont {L.}~\bibnamefont {Tang}}, \bibinfo {author}
  {\bibfnamefont {D.}~\bibnamefont {Song}}, \bibinfo {author} {\bibfnamefont
  {J.}~\bibnamefont {Xu}},\ and\ \bibinfo {author} {\bibfnamefont
  {Z.}~\bibnamefont {Chen}},\ }\bibfield  {title} {\bibinfo {title} {{Photonic
  Realization of a Generic Type of Graphene Edge States Exhibiting Topological
  Flat Band}},\ }\href {https://doi.org/10.1103/PhysRevLett.131.013804}
  {\bibfield  {journal} {\bibinfo  {journal} {Phys. Rev. Lett.}\ }\textbf
  {\bibinfo {volume} {131}},\ \bibinfo {pages} {013804} (\bibinfo {year}
  {2023})}\BibitemShut {NoStop}%
\bibitem [{\citenamefont {Chen}\ \emph {et~al.}(2020)\citenamefont {Chen},
  \citenamefont {Wang}, \citenamefont {Claassen}, \citenamefont {Moritz},\ and\
  \citenamefont {Devereaux}}]{chen2020observing}%
  \BibitemOpen
  \bibfield  {author} {\bibinfo {author} {\bibfnamefont {Y.}~\bibnamefont
  {Chen}}, \bibinfo {author} {\bibfnamefont {Y.}~\bibnamefont {Wang}}, \bibinfo
  {author} {\bibfnamefont {M.}~\bibnamefont {Claassen}}, \bibinfo {author}
  {\bibfnamefont {B.}~\bibnamefont {Moritz}},\ and\ \bibinfo {author}
  {\bibfnamefont {T.~P.}\ \bibnamefont {Devereaux}},\ }\bibfield  {title}
  {\bibinfo {title} {{Observing photo-induced chiral edge states of graphene
  nanoribbons in pump-probe spectroscopies}},\ }\href
  {https://doi.org/https://doi.org/10.1038/s41535-020-00283-5} {\bibfield
  {journal} {\bibinfo  {journal} {npj Quantum Mater.}\ }\textbf {\bibinfo
  {volume} {5}},\ \bibinfo {pages} {84} (\bibinfo {year} {2020})}\BibitemShut
  {NoStop}%
\bibitem [{\citenamefont {Liu}\ \emph {et~al.}(2011)\citenamefont {Liu},
  \citenamefont {Jiang},\ and\ \citenamefont {Yao}}]{PhysRevB.84.195430}%
  \BibitemOpen
  \bibfield  {author} {\bibinfo {author} {\bibfnamefont {C.-C.}\ \bibnamefont
  {Liu}}, \bibinfo {author} {\bibfnamefont {H.}~\bibnamefont {Jiang}},\ and\
  \bibinfo {author} {\bibfnamefont {Y.}~\bibnamefont {Yao}},\ }\bibfield
  {title} {\bibinfo {title} {{Low-energy effective Hamiltonian involving
  spin-orbit coupling in silicene and two-dimensional germanium and tin}},\
  }\href {https://doi.org/10.1103/PhysRevB.84.195430} {\bibfield  {journal}
  {\bibinfo  {journal} {Phys. Rev. B}\ }\textbf {\bibinfo {volume} {84}},\
  \bibinfo {pages} {195430} (\bibinfo {year} {2011})}\BibitemShut {NoStop}%
\bibitem [{\citenamefont {L\"u}\ \emph {et~al.}(2024)\citenamefont {L\"u},
  \citenamefont {Zhang}, \citenamefont {Fu},\ and\ \citenamefont
  {Liu}}]{PhysRevResearch.6.043108}%
  \BibitemOpen
  \bibfield  {author} {\bibinfo {author} {\bibfnamefont {X.-L.}\ \bibnamefont
  {L\"u}}, \bibinfo {author} {\bibfnamefont {Y.-C.}\ \bibnamefont {Zhang}},
  \bibinfo {author} {\bibfnamefont {P.-H.}\ \bibnamefont {Fu}},\ and\ \bibinfo
  {author} {\bibfnamefont {J.-F.}\ \bibnamefont {Liu}},\ }\bibfield  {title}
  {\bibinfo {title} {{Phase diagrams and topological mixed edge states in
  silicene with intrinsic and extrinsic Rashba effects}},\ }\href
  {https://doi.org/10.1103/PhysRevResearch.6.043108} {\bibfield  {journal}
  {\bibinfo  {journal} {Phys. Rev. Res.}\ }\textbf {\bibinfo {volume} {6}},\
  \bibinfo {pages} {043108} (\bibinfo {year} {2024})}\BibitemShut {NoStop}%
\bibitem [{\citenamefont {Sun}\ \emph {et~al.}(2022)\citenamefont {Sun},
  \citenamefont {Liu}, \citenamefont {Du},\ and\ \citenamefont
  {Guo}}]{Sun_Junsong}%
  \BibitemOpen
  \bibfield  {author} {\bibinfo {author} {\bibfnamefont {J.}~\bibnamefont
  {Sun}}, \bibinfo {author} {\bibfnamefont {T.}~\bibnamefont {Liu}}, \bibinfo
  {author} {\bibfnamefont {Y.}~\bibnamefont {Du}},\ and\ \bibinfo {author}
  {\bibfnamefont {H.}~\bibnamefont {Guo}},\ }\bibfield  {title} {\bibinfo
  {title} {{Strain-induced pseudo magnetic field in the $\alpha$-$T_{3}$
  lattice}},\ }\href {https://doi.org/10.1103/PhysRevB.106.155417} {\bibfield
  {journal} {\bibinfo  {journal} {Phys. Rev. B}\ }\textbf {\bibinfo {volume}
  {106}},\ \bibinfo {pages} {155417} (\bibinfo {year} {2022})}\BibitemShut
  {NoStop}%
\bibitem [{\citenamefont {Islam}\ \emph {et~al.}(2024)\citenamefont {Islam},
  \citenamefont {Bhattacharyya},\ and\ \citenamefont
  {Basu}}]{PhysRevB.110.045426}%
  \BibitemOpen
  \bibfield  {author} {\bibinfo {author} {\bibfnamefont {M.}~\bibnamefont
  {Islam}}, \bibinfo {author} {\bibfnamefont {K.}~\bibnamefont
  {Bhattacharyya}},\ and\ \bibinfo {author} {\bibfnamefont {S.}~\bibnamefont
  {Basu}},\ }\bibfield  {title} {\bibinfo {title} {{Electron-phonon coupling
  induced topological phase transition in an
  $\ensuremath{\alpha}\text{\ensuremath{-}}{T}_{3}$ Haldane-Holstein model}},\
  }\href {https://doi.org/10.1103/PhysRevB.110.045426} {\bibfield  {journal}
  {\bibinfo  {journal} {Phys. Rev. B}\ }\textbf {\bibinfo {volume} {110}},\
  \bibinfo {pages} {045426} (\bibinfo {year} {2024})}\BibitemShut {NoStop}%
\bibitem [{\citenamefont {Parui}\ \emph {et~al.}(2024)\citenamefont {Parui},
  \citenamefont {Ghosh},\ and\ \citenamefont {Chittari}}]{PhysRevB.109.165118}%
  \BibitemOpen
  \bibfield  {author} {\bibinfo {author} {\bibfnamefont {P.}~\bibnamefont
  {Parui}}, \bibinfo {author} {\bibfnamefont {S.}~\bibnamefont {Ghosh}},\ and\
  \bibinfo {author} {\bibfnamefont {B.~L.}\ \bibnamefont {Chittari}},\
  }\bibfield  {title} {\bibinfo {title} {{Topological properties of nearly flat
  bands in bilayer $\alpha$-$T_{3}$ lattice}},\ }\href
  {https://doi.org/10.1103/PhysRevB.109.165118} {\bibfield  {journal} {\bibinfo
   {journal} {Phys. Rev. B}\ }\textbf {\bibinfo {volume} {109}},\ \bibinfo
  {pages} {165118} (\bibinfo {year} {2024})}\BibitemShut {NoStop}%
\end{thebibliography}

%

\end{document}